\documentclass[english,10pt,aps,prd,a4paper,preprintnumbers,floatfix,nofootinbib,showpacs,superscriptaddress,notitlepage,hyperref]{revtex4-2} 
\usepackage{amssymb}
\usepackage{slashed}
\usepackage{amsmath}
\usepackage{mathtools}
\usepackage{braket}
\usepackage{float}
\usepackage[utf8]{inputenc}
\usepackage{graphicx}
\usepackage{subfigure}
\usepackage{rotating}
\usepackage{latexsym}
\usepackage{epic}
\usepackage{eepic}
\usepackage{epsfig}
\usepackage{color}
\usepackage{cancel}
\usepackage{natbib}
\usepackage[colorlinks=true,citecolor=darkred,urlcolor=darkred, pdfborder={0 0 0}]{hyperref}
\usepackage[normalem]{ulem}

\newcommand{\bea}{\begin{eqnarray}}
\newcommand{\eea}{\end{eqnarray}}
\newcommand\dd{\mathrm{d}}

\definecolor{darkred}{rgb}{0.6,0,0}

\definecolor{linkcolor}{rgb}{0,0,0.5}

\begin{document}

\title{
Probing for chiral $Z^\prime$ gauge boson through scattering measurement experiments
}
\author{Kento Asai}
\email{kento@icrr.u-tokyo.ac.jp}
\affiliation{Institute for Cosmic Ray Research (ICRR), The University of Tokyo, Kashiwa, Chiba 277-8582, Japan}
\author{Arindam Das}
\email{arindamdas@oia.hokudai.ac.jp}
\affiliation{Institute for the Advancement of Higher Education, Hokkaido University, Sapporo 060-0817, Japan}
\affiliation{Department of Physics, Hokkaido University, Sapporo 060-0810, Japan}
\author{Jinmian Li}
\email{jmli@scu.edu.cn}
\affiliation{College of Physics, Sichuan University, Chengdu 610065, China}
\author{Takaaki Nomura}
\email{nomura@scu.edu.cn}
\affiliation{College of Physics, Sichuan University, Chengdu 610065, China}
\author{Osamu Seto}
\email{seto@particle.sci.hokudai.ac.jp}
\affiliation{Department of Physics, Hokkaido University, Sapporo 060-0810, Japan}
\vskip .4in
\begin{abstract}\noindent
Motivated by the observation of tiny neutrino mass can not be explained within the framework of Standard Model (SM), 
we consider extra gauge extended scenarios in which tiny neutrino masses are generated through seesaw mechanism. These scenarios are equipped with beyond the standard model (BSM) neutral gauge boson called $Z^\prime$ 
in the general $U(1)_X$ symmetry which is a linear combination of $U(1)_Y$ and $U(1)_{B-L}$. In this case,  left and right handed fermions interact differently with the $Z^\prime$. The $Z^\prime$ gives rise to different processes involving neutrino-nucleon, neutrino-electron, electron-nucleus and electron-muon scattering processes.
By comparing with proton, electron beam-dump experiments data, recast data from searches for the long-lived and dark photon at BaBaR, LHCb and CMS experiments, 
%
the electron and muon $g-2$ data, and the data of the dilepton and dijet searches at the LEP experiment,  we derive bounds on the gauge coupling and the corresponding gauge boson mass for different $U(1)_X$ charges
and evaluate the prospective limits from the future beam-dump scenarios at DUNE, FASER(2) and ILC. We conclude that large parameter regions could be probed by scattering, beam-dump and collider experiments in future.    
\end{abstract}
\vspace{-3cm}
\preprint{EPHOU-23-013}

\maketitle
\setcounter{page}{1}
\setcounter{footnote}{0}
\section{Introduction}
\label{sec:introduction}
Tiny neutrino masses and flavor mixing are important experimental observations \cite{ParticleDataGroup:2020ssz} which motivate us to think beyond the Standard Model (SM). Various cosmological data indicate that nonluminous objects called dark matter capture nearly $0.25$ fraction of the energy budget of the Universe~\cite{Bertone:2004pz,
Planck:2018vyg} which further indicate that an extension of the SM is certain. The origin of tiny neutrino mass can be explained by the seesaw mechanism where the SM is extended by SM-singlet Majorana Right Handed Neutrinos (RHNs)~\cite{Minkowski:1977sc,Yanagida:1979as,Gell-Mann:1979vob,Mohapatra:1979ia,Schechter:1980gr}. In this case, light neutrino mass can be originated by the suppression of heavy mass scale of the RHNs, which introduces a lepton number violation of unit two which is a nice realization of dimension five Weinberg operator \cite{Weinberg:1979sa}. 

This simple but interesting ultraviolet (UV) theory is constructed if the SM is extended by a general $U(1)_X$ gauge group which is a linear combination of $U(1)_Y$ and $U(1)_{B-L}$. In this set-up, three generations of SM-singlet RHNs are introduced to cancel the gauge and mixed gauge-gravity anomalies. 
After the general $U(1)_X$ symmetry is broken by a SM-singlet scalar, the scalar acquires vacuum expectation value (VEV) which lets the RHNs to acquire Majorana masses. Followed by the electroweak symmetry breaking, a Dirac mass term is generated from the Yukawa interaction between the SM lepton and Higgs doublets along with the SM-singlet RHNs. These Majorana and Dirac masses get involved in the seesaw mechanism to generate light neutrino mass and flavor mixing~\cite{Asai:2022zxw}. All these couplings and interactions are protected by general $U(1)_X$ gauge symmetry.

In general, an extra gauged $U(1)$ extension of the SM, a neutral and beyond the SM (BSM) gauge boson, commonly known as the $Z^\prime$, exists and acquires the mass after the extra $U(1)$ symmetry breaking. 
New physics contributions from the $Z^\prime$ are considered to be very well motivated and being studied at low and high energy experiments~\cite{Bauer:2018onh,Ilten:2018crw,Dev:2021qjj,Baruch:2022esd,Greljo:2022dwn,Bernal:2022qba,Boyarsky:2021moj,Ansarifard:2021elw,Aboubrahim:2022qln,
Kling:2022ykt,Kelly:2021jgj,Dev:2021xzd,Chauhan:2022iuh,Coy:2021wfs,
Foguel:2022unm,Campos:2017dgc, Lindner:2018kjo,
AtzoriCorona:2022moj,AtzoriCorona:2022moj,Chakraborty:2021apc,Ismail:2020yqc,Bakhti:2020szu,Cheung:2022kjd,Cheung:2021tmx,Bakhti:2020vfq,Dev:2020drf,Masiero:2020vxk,Nath:2021uqb,delaVega:2021wpx,Araki:2020wkq,Felkl:2023nan,Elpe:2022hqp,DeRomeri:2022twg,Melas:2023olz} from various aspects. 
A remarkable aspect of our general $U(1)_X$ models compared with the others studied previously is that the SM left and right chiral fermions could be charged differently under the $U(1)_X$. 
We will study such a chiral scenario to estimate the limits on the general $U(1)_X$ gauge coupling with respect to the $Z^\prime$ mass for different general $U(1)_X$ charges and compare with a variety of existing bounds for the $Z^\prime$ mass to be probed. 
In this paper, we consider two cases of general $U(1)_X$ extensions of the SM. Those two are solutions of anomaly free conditions and mathematically equally possible. In the first case, three generations of RHNs are universal under the general $U(1)_X$ gauge group having same charge.
The second is the less-studied another interesting possibility where first two generations of the RHNs have general $U(1)_X$ charge $-4$, the third generation RHN has the charge $+5$ and two $SU(2)_L$ Higgs doublet are differently charged under the general $U(1)_X$ gauge group. One of $SU(2)_L$ Higgs doublet couples with the RHNs and charged leptons, which is making the scenario neutrinophilic. In addition to the doublet scalars we introduce three SM-singlet scalars which are differently charged under the general $U(1)_X$ gauge group. We call the second case an alternative general $U(1)_X$ scenario. 
Due to the anomaly cancellation conditions, left and right handed SM fermions must be differently charged under the general $U(1)_X$ gauge group manifesting the chiral nature of the model.

We study the chiral nature of the $Z^\prime$ interactions 
through the $Z^\prime$ mediated neutrino-electron, neutrino-nucleon and electron-muon scattering process at different experiments such as FASER$\nu$ and FASER$\nu2$~\cite{FASER:2018bac,FASER:2019aik,FASER:2021mtu,FASER:2019dxq,FASER:2020gpr,Arguelles:2022tki,Feng:2022inv,Ariga:2022roc}, SND$@$LHC~\cite{Ahdida:2750060,SHiP:2015vad,SNDLHC:2022ihg,SHiP:2020sos}, NA64~\cite{NA64:2022yly,NA64:2016oww,NA64:2017vtt,Banerjee:2019pds,Gninenko:2020hbd,Sieber:2021fue}, JSNS2, COHERENT, and MuonE~\cite{Abbiendi:26774711,Abbiendi:2022oks}. These are the $Z^\prime$ mediated $t$-channel processes and those interaction vertices depending on general $U(1)_X$ charges manifest the chiral nature which has not been studied before in the literature for these experiments. 

%

In order to compare our results with the existing bounds we obtain respective bounds from different scattering and beam dump experiments in chiral scenario. To do that, we estimate bounds from electron/ positron beam dump experiments Orsay~\cite{Davier:1989wz}, NA64~\cite{NA64:2019auh}, KEK~\cite{Beer:1986qr}, E141~\cite{Riordan:1987aw}, E137~\cite{Bjorken:1988as} and E774~\cite{Bross:1989mp} respectively. Studying the bounds from the neutrino-electron scattering experiments from TEXONO~\cite{TEXONO:2009knm,TEXONO:2002pra,TEXONO:2006xds}, BOREXINO~\cite{Borexino:2000uvj,Bellini:2011rx,Cleveland:1998nv,Lande:2003ex,Borexino:2007kvk,Borexino:2008gab} and JSNS2 at J-PARC experiment~\cite{Ajimura:2017fld}, we compare our results for different $U(1)_X$ charges. Depending on the choice of the $Z^\prime$ mass, we compare our bounds for muon neutrino- and muon anti-neutrino-electron scattering from the CHARM-II experiment~\cite{CHARM-II:1991ydz,CHARM-II:1989nic,CHARM-II:1993phx,CHARM-II:1994dzw}. We study neutrino-nucleon scattering to estimate bounds for different $U(1)_X$ charges from the COHERENT experiment~\cite{COHERENT:2018imc,COHERENT:2018imc,Cadeddu:2017etk,Cadeddu:2020nbr,COHERENT:2020iec,COHERENT:2020ybo}. In addition to those, we compare our results with the GEMMA experiment studying neutrino magnetic moment~\cite{Beda:2009kx,Lindner:2018kjo}. We compare the bounds form the dark photon searches at the LHCb experiment~\cite{Foguel:2022unm,Baruch:2022esd}, dark photon searches from the CMS experiment~\cite{CMS:2023slr}, visible and invisible decay of dark photon from BaBar experiment~\cite{BaBar:2014zli,BaBar:2017tiz}, respectively. We compare our results with proton beam dump experiments like Nomad~\cite{NOMAD:2001xxt}, CHARM~\cite{CHARM:1985anb} and $\nu-$cal~\cite{Blumlein:2011mv,Blumlein:2013cua} experiments respectively. We compare our bounds with the dilepton and dijet final states from the LEP-II study~\cite{ALEPH:2006bhb,ALEPH:1997gvm,LEPWorkingGroupforHiggsbosonsearches:2003ing,ALEPH:2005ab}. Finally we estimate bounds on general $U(1)_X$ coupling with respect to $M_{Z^\prime}$ from muon and electron $g-2$ experiments to show the complementarity scenario. 

Our paper is organized as follows. We discuss the models in Sec.~\ref{sec:model}. We calculate constraints on the gauge coupling fro different $U(1)_X$ charges in Sec.~\ref{sec:calculation}. We discuss our results in Sec.~\ref{sec:result} and finally conclude the paper in Sec.~\ref{sec:summary}. 
\section{Model}
\label{sec:model}
A general $U(1)_X$ extension of the SM involves three generations of RHNs to cancel the gauge and mixed gauge-gravity anomalies. As a result, we observe that left and right handed charged fermions in the SM are differently charged under $U(1)$ group. We write down the UV-complete models in the following:
\subsection{Case-I}
We consider a general $U(1)_X$ extension of the SM to investigate the chiral scenario introducing three RHNs $(N_R)$ and an SM-singlet scalar $(\Phi)$ field. Three generations of RHNs are introduced to cancel gauge and mixed gauge-gravity anomalies. The corresponding field content is given in Tab.~\ref{tab1} where general $U(1)_X$ charges, being independent of generations, are written as $\tilde{x}_f$ before anomaly cancellation and $f$ stands for the three generations of quarks $(q_L^\alpha, u_R^\alpha, d_R^\alpha)$ and leptons $(\ell_L^\alpha, e_R^\alpha, N_R^\alpha)$ respectively where $\alpha (=1,~2,~3)$ is the generation index. The gauge and mixed gauge-gravity anomaly cancellation conditions in terms of the general charges are written below:
\begin{align}
{\rm U}(1)_X \otimes \left[ {\rm SU}(3)_C \right]^2&\ :&
			2\tilde{x}_q - \tilde{x}_u - \tilde{x}_d &\ =\  0~, \nonumber \\
{\rm U}(1)_X \otimes \left[ {\rm SU}(2)_L \right]^2&\ :&
			3\tilde{x}_q + \tilde{x}_\ell &\ =\  0~, \nonumber \\
{\rm U}(1)_X \otimes \left[ {\rm U}(1)_Y \right]^2&\ :&
			\tilde{x}_q - 8\tilde{x}_u - 2\tilde{x}_d + 3\tilde{x}_\ell - 6\tilde{x}_e &\ =\  0~, \nonumber \\
\left[ {\rm U}(1)_X \right]^2 \otimes {\rm U}(1)_Y &\ :&
			{\tilde{x}_q}^2 - {\tilde{2x}_u}^2 + {\tilde{x}_d}^2 - {\tilde{x}_\ell}^2 + {\tilde{x}_e}^2 &\ =\  0~, \nonumber \\
\left[ {\rm U}(1)_X \right]^3&\ :&
			{6\tilde{x}_q}^3 - {3\tilde{x}_u}^3 - {3 \tilde{x}_d}^3 + {2\tilde{x}_\ell}^3 - {\tilde{x}_\nu}^3 - {\tilde{x}_e}^3 &\ =\  0~, \nonumber \\
{\rm U}(1)_X \otimes \left[ {\rm grav.} \right]^2&\ :&
			6\tilde{x}_q - 3\tilde{x}_u - 3 \tilde{x}_d + 2\tilde{x}_\ell - \tilde{x}_\nu - \tilde{x}_e &\ =\  0~, 
\label{anom-f-1}
\end{align}
respectively. The Yukawa interactions between the fermions and the scalars $(H, \Phi)$ can be written following the  $\mathcal{G}_{\rm SM} \otimes$ $U(1)_X$ gauge symmetry as
\begin{equation}
{\cal L}^{\rm Yukawa} = - Y_u^{\alpha \beta} \overline{q_L^\alpha} H u_R^\beta
                                - Y_d^{\alpha \beta} \overline{q_L^\alpha} \tilde{H} d_R^\beta
				 - Y_e^{\alpha \beta} \overline{\ell_L^\alpha} \tilde{H} e_R^\beta
				- Y_\nu^{\alpha \beta} \overline{\ell_L^\alpha} H N_R^\beta-\frac{1}{2} Y_N^\alpha \Phi \overline{(N_R^\alpha)^c} N_R^\alpha + {\rm H.c.}~,
\label{LYk}   
\end{equation}
where $H$ is the SM Higgs doublet and we transform it into $\tilde{H}$ following $i  \tau^2 H^*$ where $\tau^2$ is the second Pauli matrix. Hence, using Eq.~\ref{LYk} and following charge neutrality, we express the following relations between the general $U(1)_X$ charges of the particles as 
\begin{eqnarray}
-\frac{1}{2} x_H^{} &=& - \tilde{x}_q + \tilde{x}_u \ =\  \tilde{x}_q - \tilde{x}_d \ =\  \tilde{x}_\ell - \tilde{x}_e=\  - \tilde{x}_\ell + \tilde{x}_\nu, \nonumber \\
2 x_\Phi^{}	&=& - 2 \tilde{x}_\nu~. 
\label{Yuk}
\end{eqnarray} 
The general $U(1)_X$ charges of the fermions can be obtained solving Eqs.~\eqref{anom-f-1} and \eqref{Yuk} which finally can be expressed using the scalar charges $x_H^{}$ and $x_\Phi^{}$ respectively. Simply the anomaly free charge assignment of the general $U(1)_X$ can be expressed in terms of a linear combination of two anomaly free scenarios, namely, $U(1)_Y$ of the SM and $B-L$ charges. Finally, we find that the left and right handed fremions under the general $U(1)_X$ scenario have different charges and hence they interact differently with the neutral, BSM gauge boson $Z^\prime$ in the model. We take $x_\Phi=1$ without the loss of generality which corresponds to the $U(1)_{ B-L}$ and $U(1)_R$ scenarios with $x_H=0$~\cite{Davidson:1979wr,Mohapatra:1980qe,Wetterich:1981bx,Masiero:1982fi,Buchmuller:1991ce} and $x_H=2$~\cite{Jung:2009jz,Nomura:2017tih,Nomura:2017ezy,Jana:2019mez,Seto:2020jal}, respectively. $U(1)_{B-L}$ is a vector like scenario where left and right handed fermions of same type are equally charged under the $U(1)$ extension. In case of $U(1)_R$ scenario, we find that left handed fermions do not interact with the $Z^\prime$.  
\begin{table}[t]
\begin{center}
\begin{tabular}{||c||ccc||rcr|c||c||c||c||c||c||c||}
\hline
\hline
            & SU(3)$_C^{}$ & SU(2)$_L^{}$ & U(1)$_Y^{}$ & \multicolumn{3}{c|}{U(1)$_X^{}$}&$-2$&$-1$&$-0.5$& $0$& $0.5$ & $1$ & $2$  \\
            &&& &&&&$U(1)_{\rm{R}}$& & &$U(1)_{B-L}$&&&  \\
\hline
\hline
&&&&&&&&&&&&&\\[-12pt]
&&&&&&&&&&&&&\\
$q_L^\alpha$    & {\bf 3}   & {\bf 2}& $\frac{1}{6}$ & $\tilde{x}_q$ 		& = & $\frac{1}{6}x_H^{} + \frac{1}{3}x_\Phi^{}$   &$0$&$\frac{1}{6}$&$\frac{1}{4}$&$\frac{1}{3}$&$\frac{5}{12}$&$\frac{1}{2}$&$\frac{1}{3}$\\
$u_R^\alpha$    & {\bf 3} & {\bf 1}& $\frac{2}{3}$ & $\tilde{x}_u$ 		& = & $\frac{2}{3}x_H^{} + \frac{1}{3}x_\Phi^{}$   &$-1$&$-\frac{1}{3}$&$0$&$\frac{1}{3}$&$\frac{1}{2}$&$1$&$\frac{5}{3}$\\
$d_R^\alpha$    & {\bf 3} & {\bf 1}& $-\frac{1}{3}$ & $\tilde{x}_d$ 		& = & $-\frac{1}{3}x_H^{} + \frac{1}{3}x_\Phi^{}$  &$1$&$\frac{2}{3}$&$\frac{1}{2}$&$\frac{1}{3}$&$\frac{1}{6}$&$0$&$-\frac{1}{3}$\\
\hline
\hline
&&&&&&&&&&&&&\\
$\ell_L^\alpha$    & {\bf 1} & {\bf 2}& $-\frac{1}{2}$ & $\tilde{x}_\ell$ 	& = & $- \frac{1}{2}x_H^{} - x_\Phi^{}$   &$0$&$-\frac{1}{2}$&$-\frac{3}{4}$&$-1$&$\frac{5}{4}$&$-\frac{3}{2}$&$-2$ \\
$e_R^\alpha$   & {\bf 1} & {\bf 1}& $-1$   & $\tilde{x}_e$ 		& = & $- x_H^{} - x_\Phi^{}$   &$1$&$0$&$-\frac{1}{2}$&$-1$&$-\frac{3}{2}$&$-2$&$-3$ \\
\hline
\hline
&&&&&&&&&&&&&\\
$N_R^\alpha$   & {\bf 1} & {\bf 1}& $0$   & $\tilde{x}_\nu$ 	& = & $- x_\Phi^{}$  &$-1$&$-1$&$-1$&$-1$&$-1$&$-1$&$-1$ \\
\hline
\hline
&&&&&&&&&&&&&\\
$H$         & {\bf 1} & {\bf 2}& $-\frac{1}{2}$  &  $\tilde{x}_H^{}$ 	& = & $-\frac{1}{2}x_H^{}$\hspace*{7.0mm} &$1$&$\frac{1}{2}$&$\frac{1}{4}$&$0$&$-\frac{1}{4}$&$-\frac{1}{2}$&$-1$ \\ 
$\Phi$      & {\bf 1} & {\bf 1}& $0$  &  $\tilde{x}_\Phi^{}$ 	& = & $2 x_\Phi^{}$ &$2$&$2$&$2$&$2$&$2$&$2$&$2$  \\ 
\hline
\hline
\end{tabular}
\end{center}
\caption{
Field content of general $U(1)_X$ extension of the SM in the minimal form with the charges of the particles before and after anomaly cancellation considering different benchmark values of $x_H^{}$ setting $x_\Phi^{}=1$. 
In this charge assignment, $x_H^{}=0$ and $-2$ are the $U(1)_{B-L}$ and $U(1)_{\rm{R}}$ scenarios. 
Among the chiral scenarios, $U(1)_{B-L}$ is a vector like case.
}
\label{tab1}
\end{table}
Using the general form of the $U(1)$ charges of the charged fermions, we notice that for $x_H=-1$ the $U(1)$ charge of the right handed electron $(e_R)$ becomes zero and, as a result, it has no direct interaction with the $Z^\prime$ whereas other fermions will interact with the $Z^\prime$, manifesting the chiral nature of the model. In similar fashion, we find that for $x_H=-0.5$ the general $U(1)_X$ charge of the right handed up-type quark $u_R$ is zero implying no direct interaction with the $Z^\prime$, whereas other fermions will have nonzero general $U(1)_X$ charges allowing direct interactions with the $Z^\prime$. Similar behavior could be observed when $x_H=1$ where general $U(1)_X$ charge of right handed down type quark $(d_R)$ is zero resulting into no direct interaction with the $Z^\prime$. Detailed charge assignments for these combinations of $x_H$ and $x_\Phi$ are given in Tab.~\ref{tab1}. We consider two more scenarios with $x_H=0.5$ and $2$, setting $x_\Phi=1$, where all the charged fermions interact with the $Z^\prime$ manifesting the chiral behavior of the model.

The scalar sector of this scenario can be explored by introducing the renormalizable potential of this model and that can be given by
\begin{align}
  V \ = \ m_H^2(H^\dag H) + \lambda_H^{} (H^\dag H)^2 + m_\Phi^2 (\Phi^\dag \Phi) + \lambda_\Phi^{} (\Phi^\dag \Phi)^2 + \lambda_{\rm mix} (H^\dag H)(\Phi^\dag \Phi)~,
\end{align}
where $H$ and $\Phi$ can be separately approximated in the analysis of scalar potential, by taking $\lambda_{\rm mix}$ to be very small. 
After the breaking of general $U(1)_X$ and electroweak symmetries, the scalar fields develop vacuum expectation values (VEVs) as follows 
\begin{align}
  \braket{H} \ = \ \frac{1}{\sqrt{2}}\begin{pmatrix} v\\0 
  \end{pmatrix}~, \quad {\rm and}\quad 
  \braket{\Phi} \ =\  \frac{v_\Phi^{}}{\sqrt{2}}~,
\end{align}
where $v=246$ GeV is marked as the electroweak scale VEV at the potential minimum and $v_\Phi^{}$ is a free parameter. After the general $U(1)_X$ symmetry is broken, the mass of the BSM neutral gauge boson can be evolved setting $x_\Phi=1$ as  
\begin{equation}
 M_{Z^\prime}^{}=  2 g_X^{}  v_\Phi^{}~,
\end{equation}
in the limit of $v_\Phi^{} \gg v$. Here, $g_X$ is the general $U(1)_X$ coupling and the $Z^\prime$ mass is a free parameter.
\subsection{Case-II}
We consider another scenario where the SM is extended by a general $U(1)_X$ gauge group with three generations of RHNs. The field content of the model is given in Tab.~\ref{tab2}. The general U$(1)$ charges of the charged fermion fields are same for all generations, $\alpha (=1, 2, 3)$ where $\alpha$ is the generation index. The general $U(1)_X$ charges of the fields are written as $\tilde{x}_f$ before anomaly cancellation and $f$ stands for the three generations of quarks $(q_L^\alpha, u_R^\alpha, d_R^\alpha)$ and leptons $(\ell_L^\alpha, e_R^\alpha)$ respectively. We introduce two $SU(2)$ doublet Higgs fields $(H_{1,2})$ where one is the SM like $(H_1)$ and the other one is the BSM $(H_2)$ Higgs. Due to different general $U(1)_X$ charge assignments $H_1$ does not couple with the BSM fermions. We introduce three SM-singlet scalar fields $(\Phi_{1,2,3})$ which are differently charged under general $U(1)_X$ gauge group. Due to the general $U(1)_X$ gauge symmetry, first two generations of the RHNs have charge $-4$ each and the third generation RHN has $+5$ charge under the general $U(1)_X$ gauge group~\cite{Montero:2007cd}. The RHNs with semi non-universal $U(1)$ charges in this model help to cancel gauge and mixed gauge-gravity anomalies. As the result, we call this model an alternative general $U(1)_X$ scenario. Following the gauge and mixed gauge-gravity anomaly cancellation conditions, we relate the general $U(1)_X$ charges of the charged fermions as
\begin{table}[t]
\begin{center}
\begin{tabular}{||c||ccc||rcl|c||c||c||c||c||c||c||}
\hline
\hline
            & SU(3)$_C$ & SU(2)$_L$ & U(1)$_Y$ & \multicolumn{3}{c|}{$U(1)_X$}&$-2$&$-1$&$-0.5$& $0$& $0.5$ & $1$ & $2$  \\
            &&& &&&&U$(1)_{\rm{R}}$& & &B$-$L&&&  \\
\hline
\hline
&&&&&&&&&&&&&\\[-12pt]
&&&&&&&&&&&&&\\
$q_L^\alpha$    & {\bf 3}   & {\bf 2}& $\frac{1}{6}$ & $\tilde{x}_q$ 		& = & $\frac{1}{6}x_H^{} + \frac{1}{3}$   &$0$&$\frac{1}{6}$&$\frac{1}{4}$&$\frac{1}{3}$&$\frac{5}{12}$&$\frac{1}{2}$&$\frac{1}{3}$\\
$u_R^\alpha$    & {\bf 3} & {\bf 1}& $\frac{2}{3}$ & $\tilde{x}_u$ 		& = & $\frac{2}{3}x_H^{} + \frac{1}{3}$   &$-1$&$-\frac{1}{3}$&$0$&$\frac{1}{3}$&$\frac{1}{2}$&$1$&$\frac{5}{3}$\\
$d_R^\alpha$    & {\bf 3} & {\bf 1}& $-\frac{1}{3}$ & $\tilde{x}_d$ 		& = & $-\frac{1}{3}x_H^{} + \frac{1}{3}$  &$1$&$\frac{2}{3}$&$\frac{1}{2}$&$\frac{1}{3}$&$\frac{1}{6}$&$0$&$-\frac{1}{3}$\\
\hline
\hline
&&&&&&&&&&&&&\\
$\ell_L^\alpha$    & {\bf 1} & {\bf 2}& $-\frac{1}{2}$ & $\tilde{x}_\ell$ 	& = & $- \frac{1}{2}x_H^{} - 1$   &$0$&$-\frac{1}{2}$&$-\frac{3}{4}$&$-1$&$\frac{5}{4}$&$-\frac{3}{2}$&$-2$ \\
$e_R^\alpha$   & {\bf 1} & {\bf 1}& $-1$   & $\tilde{x}_e$ 		& = & $- x_H^{} - 1$   &$1$&$0$&$-\frac{1}{2}$&$-1$&$-\frac{3}{2}$&$-2$&$-3$ \\
\hline
\hline
&&&&&&&&&&&&&\\
$N_R^{1,2}$   & {\bf 1} & {\bf 1}& $0$   & $\tilde{x}_\nu$ 	& = & $-4$  &$-4$&$-4$&$-4$&$-4$&$-4$&$-4$&$-4$ \\
$N_R^3$   & {\bf 1} & {\bf 1}& $0$   & $\tilde{x}_\nu^{\prime}$ 	& = & $5$  &$5$&$5$&$5$&$5$&$5$&$5$&$5$ \\
\hline
\hline
&&&&&&&&&&&&&\\
$H_1$         & {\bf 1} & {\bf 2}& $-\frac{1}{2}$  &  $\tilde{x}_{H_{1}}^{}$ 	& = & $-\frac{x_H^{}}{2}$ &$1$&$\frac{1}{2}$&$\frac{1}{4}$&$0$&$-\frac{1}{4}$&$-\frac{1}{2}$&$-1$ \\ 
$H_2$         & {\bf 1} & {\bf 2}& $-\frac{1}{2}$  &  $\tilde{x}_{H_{2}}^{}$ 	& = &  $-\frac{1}{2} x_{H}^{}+3$ &$4$&$\frac{7}{2}$&$\frac{13}{2}$&$3$&$\frac{11}{4}$&$\frac{5}{2}$&$2$ \\ 
$\Phi_1$      & {\bf 1} & {\bf 1}& $0$  &  $ \tilde{x}_{\Phi_{1}}^{}$ 	& = & $+8$ &$+8$&$+8$&$+8$&$+8$&$+8$&$+8$&$+8$  \\ 
$\Phi_2$      & {\bf 1} & {\bf 1}& $0$  &  $\tilde{x}_{\Phi_{2}}^{}$ 	& = & $-10$ &$-10$&$-10$&$-10$&$-10$&$-10$&$-10$&$-10$  \\ 
$\Phi_3$      & {\bf 1} & {\bf 1}& $0$  &  $\tilde{x}_{\Phi_{3}}^{}$ 	& = & $-3$ &$-3$&$-3$&$-3$&$-3$&$-3$&$-3$&$-3$  \\ 
\hline
\hline
\end{tabular}
\end{center}
\caption{
Field content of the general $U(1)_X$ extension of the SM in the minimal form with the  charges of the particles before and after anomaly cancellation considering different benchmark values of $x_H^{}$. 
Here, $x_H^{}=0$ is an alternative $B-L$ case, which is purely vector like structure given as a reference in this article.
}
\label{tab2}
\end{table}
\begin{align}
{\rm U}(1)_X \otimes \left[ {\rm SU}(3)_C \right]^2&\ :&
			2\tilde{x}_q - \tilde{x}_u - \tilde{x}_d &\ =\  0~, \nonumber \\
{\rm U}(1)_X \otimes \left[ {\rm SU}(2)_L \right]^2&\ :&
			3\tilde{x}_q + \tilde{x}_\ell &\ =\  0~, \nonumber \\
{\rm U}(1)_X \otimes \left[ {\rm U}(1)_Y \right]^2&\ :&
			\tilde{x}_q - 8\tilde{x}_u - 2\tilde{x}_d + 3 \tilde{x_\ell} - 6\tilde{x_e} &\ =\  0~, \nonumber \\
\left[ {\rm U}(1)_X \right]^2 \otimes {\rm U}(1)_Y &\ :&
			{\tilde{x}_q}^2 - {2\tilde{x}_u}^2 + {\tilde{x}_d}^2 - {\tilde{x}_\ell}^2 + {\tilde{x}_e}^2 &\ =\  0~, \nonumber \\
\left[ {\rm U}(1)_X \right]^3&\ :&
			3({6\tilde{x}_q}^3 - {3\tilde{x}_u}^3 - {3\tilde{x}_d}^3 + {2\tilde{x}_\ell}^3-{\tilde{x}_e}^3) - 2 \tilde{x}_\nu^3-\tilde{x}_\nu^{\prime^3}   &\ =\  0~, \nonumber \\
{\rm U}(1)_X \otimes \left[ {\rm grav.} \right]^2&\ :&
			3(6 \tilde{x}_q - 3\tilde{x}_u - 3\tilde{x}_d + 2 \tilde{x}_\ell- \tilde{x}_e)-2 \tilde{x}_\nu-\tilde{x}_\nu^{\prime}  &\ =\  0~, 
\label{anom-f-2}
\end{align}
We find that due to general $U(1)_X$ charges the SM charged fermions interact differently with the $Z^\prime$, manifesting the chiral nature of the model. The second Higgs doublet $H_2$ interacts with the SM lepton doublet $(\ell^\alpha_L)$ and first two generations of the RHNs $(N_R^{1,2})$ due to the general $U(1)_X$ symmetry. Hence, the Dirac Yukawa mass term for $N_{R}^{1,2}$ can be generated. On the other hand, the corresponding Majorana mass term for $N_{R}^{1,2}$ can be generated from the Dirac Yukawa coupling with $\Phi_1$ followed by the general $U(1)_X$ symmetry breaking. The third generation of the RHN, $N_R^3$, has no Dirac Yukawa coupling involving any of the doublet Higgs fields being prohibited by the general $U(1)_X$ charge assignments. Therefore, it does not participate in the neutrino mass generation mechanism at the tree level, however, it can have Yukawa interaction with the $\Phi_2$ which further generates a Majorana mass term for $N_R^3$ after the general $U(1)_X$ symmetry breaking. Finally, we write the Yukawa interaction among the BSM sector as 
\bea
-\mathcal{L} _{\rm int}& \ \supset \ & \sum_{\alpha=1}^{3} \sum_{\beta=1}^{2} Y_{1}^{\alpha \beta} \overline{\ell_L^\alpha} H_2 N_R^\beta+\frac{1}{2} \sum_{\alpha=1}^{2} Y_{2}^{\alpha}  \Phi_1 \overline{(N_R^\alpha)^{c}} N_R^\alpha
+\frac{1}{2} Y_{3} \Phi_2 \overline{(N_R^3)^{c}} N_R^3+ \rm{H. c.}~,
\label{ExoticYukawa}
\eea 
taking $Y_2$ being diagonal without the loss of generality. As in the previous case, we can solve the gauge and mixed gauge-gravity anomalies to estimate the charges of the SM particles in Tab.~\ref{tab2}.

The scalar potential of this scenario can be given by   
\bea
  V&\ =\ &
m_{H_1}^2 (H_1^\dagger H_1) + \lambda_{H_1}  (H_1^\dagger H_1)^2 + m_{H_2}^2 (H_2^\dagger H_2) + \lambda_{H_2}  (H_2^\dagger H_2)^2 \nonumber \\
&& + m_{\Phi_1}^2 (\Phi_1^\dagger \Phi_1) + \lambda_1  (\Phi_1^\dagger \Phi_1)^2 
+ m_{\Phi_2}^2 (\Phi_2^\dagger \Phi_2) + \lambda_2   (\Phi_2^\dagger \Phi_2)^2 \nonumber \\
&&+ m_{\Phi_3}^2 (\Phi_3^\dagger \Phi_3) + \lambda_3   (\Phi_3^\dagger \Phi_3)^2 
+ ( \mu \Phi_3 (H_1^\dagger H_2) + {\rm H.c.} )  \nonumber \\
&&+ \lambda_4 (H_1^\dagger H_1) (H_2^\dagger H_2)+ \lambda_5 (H_1^\dagger H_2) (H_2^\dagger H_1) +\lambda_6 (H_1^\dagger H_1) (\Phi_1^\dagger \Phi_1)\nonumber \\
&&+ \lambda_7 (H_1^\dagger H_1) (\Phi_2^\dagger \Phi_2)+ \lambda_8 (H_1^\dagger H_2) (\Phi_3^\dagger \Phi_3) +\lambda_9 (H_2^\dagger H_2) (\Phi_1^\dagger \Phi_1)  \nonumber \\
&&+ \lambda_{10} (H_1^\dagger H_1) (\Phi_2^\dagger \Phi_2)+ \lambda_{11} (H_1^\dagger H_2) (\Phi_3^\dagger \Phi_3)+  \lambda_{12} (\Phi_1^\dagger \Phi_1) (\Phi_2^\dagger \Phi_2) \nonumber \\
&&+ \lambda_{13} (\Phi_2^\dagger \Phi_2) (\Phi_3^\dagger \Phi_3)+ \lambda_{14} (\Phi_3^\dagger \Phi_3) (\Phi_1^\dagger \Phi_1)~.
\label{HiggsPotential-2}
\eea
Choosing suitable parametrization for the scalar fields in this scenario to develop their respective VEVs, we can write 
\bea
  \braket{H_1} \ = \  \frac{1}{\sqrt 2}\left(  \begin{array}{c}  
    v_{h_1} \\
    0 \end{array}
\right)~,   \; 
  \braket{H_2} \ = \   \frac{1}{\sqrt{2}} \left(  \begin{array}{c}  
    v_{h_2}\\
    0 \end{array}
\right)~,  
  \braket{\Phi_1} \ = \  \frac{v_{1}}{\sqrt{2}}~,  \; 
  \braket{\Phi_2} \ = \  \frac{v_{2}}{\sqrt{2}}~,  \; 
  \braket{\Phi_3} \ = \  \frac{v_{3}}{\sqrt{2}}~,~~~~ 
\eea   
with the condition, $\sqrt{v_{h_1}^2 + v_{h_2}^2} = 246~{\rm GeV}$. In this alternative general $U(1)_X$ extension of the SM, we consider negligibly small scalar quartic couplings among SM scalar doublet fields $H_{1,2}$ and SM-singlet scalar fields $\Phi_{1,2,3}$. As a result, this ensures higher order mixing between the RHNs after the general $U(1)_X$ breaking to be very strongly suppressed. In Eq.~\eqref{HiggsPotential-2}, we may consider $0 < m_{\rm mix}^2 = \mu v_3/\sqrt{2} \ll m_{\Phi_3}^2$ which further leads to $v_{h_{2}} \sim m_{\rm mix}^2 v_{h_{1}}/m_{\Phi_{3}}^2 \ll v_{h_{1}}$~\cite{Ma:2000cc}.

Due to the presence of the general $U(1)_X$ gauge symmetry, the doublet scalar sector $H_{1,2}$ and singlet scalar sector $\Phi_{1,2,3}$ interact only through the coupling $\Phi_3 (H_1^\dagger H_2)+{\rm H.c.}$, however, this coupling has no significant effect to determine the VEVs $(v_{1,2,3})$ of the singlet scalar fields $(\Phi_{1,2,3})$, because there is already one collider constraint present in the form  of $v_1^2 + v_2^2+ v_3^2 \gg v_{h_1}^2 + v_{h_2}^2$; one find that $\sqrt{v_1^2 + v_2^2+ v_3^2}$ should be typically larger than around $1$ TeV from various constraints in the light $Z'$ case~\cite{Bauer:2018onh} and constraint from dilepton resonance in heavy $Z'$ case~\cite{CMS:2021ctt,ATLAS:2019erb} since the value is related to the $Z'$ boson mass and new gauge coupling. Therefore we arrange the parameters of the scalar potential in a way so that the VEVs of $\Phi_{1,2,3}$ will be almost same following $\mu < v_1$ whereas $\Phi_3$ can be considered as a spurion field which generate the mixing between $H_{1,2}$ in Eq.~(\ref{HiggsPotential-2}). Once the $\Phi_3$ acquires the VEV we get the mixing mass term between $H_{1,2}$ as $m_{\rm mix}= \frac{\mu v_3}{\sqrt{2}}$ which resembles the potential of two Higgs doublet model, however, due to the presence of the general $U(1)_X$ symmetry the SM-singlet fields $\Phi_{1,2,3}$ do not mix. As a result there are two existing physical Nambu-Goldstone (NG) bosons originating from the SM-singlet scalars. Due to the tiny quartic couplings and gauge couplings, the SM-singlet scalars become decoupled from the SM thermal bath in the early universe. Additionally, we consider that the singlet scalars are heavier than the neutral BSM gauge boson $Z^\prime$ preventing its decay into the NG bosons. The breaking of general U(1) gauge symmetry helps $Z^\prime$ to acquire the mass which copuld be given by 
\bea
 M_{Z^\prime} = g_X \sqrt{64 v_{1}^2+ 100 v_{2}^2+ 9v_3^2 +\frac{1}{4} x_H^2 v_{h_{1}}^2 + \left(-\frac{1}{2} x_H +3\right)^2  v_{h_{2}}^2}
\simeq g_X \sqrt{64 v_{1}^2+ 100 v_{2}^2+ 9 v_{3}^2}~.
\label{masses-Alt}   
\eea 
which is a free parameter and the general $U(1)_X$ gauge coupling $g_X$ is also a free parameter. Due to the general $U(1)_X$ gauge structure $H_2$ only couples with $N_R^{1,2}$ making this case a neutrinophilic two Higgs Doublet Model (2HDM) framework \cite{Ma:2000cc,Wang:2006jy,Gabriel:2006ns,Davidson:2009ha,Haba:2010zi}. 
\subsection{\texorpdfstring{$Z'$}{Z'} interactions with the fermions}
After the anomaly cancellation conditions are imposed, we notice that $Z^\prime$ in the above models can interact with the left and right handed SM fermions differently manifesting chiral nature of the models. Fixing $x_\Phi=1$ in Case-I we find that the chiral nature is the same as the Case-II. Therefore we write the interactions between the fermions with the $Z^\prime$ in the following as
\bea
\mathcal{L}_{\rm int} = -g_X (\overline{f}\gamma_\mu q_{f_{L}^{}}^{} P_L^{} f+ \overline{f}\gamma_\mu q_{f_{R}^{}}^{}  P_R^{} f) Z_\mu^\prime~,
\label{Lag1}
\eea
where $P_{L(R)}^{}= (1 \pm \gamma_5)/2$,  $q_{f_{L}^{}}^{}$ and $q_{f_{R}^{}}^{}$ are the corresponding general U$(1)$ charges of the left handed $(f_L)$ and right handed $(f_R)$ fermions. Hence we write vector coupling $(c_V=\frac{q_{f_L}+q_{f_R}}{2})$ and axial vector coupling $(c_A=\frac{q_{f_L}-q_{f_R}}{2})$ for the SM fermions following the charge assignments of the Cases-I and II in Tab.~\ref{tab-3} fixing $x_\Phi^{}=1$.
\begin{table}
    \begin{center}
\begin{tabular}{| *{15}{c|} }
    \hline
 & \multicolumn{7}{c|}{Vector coupling $(c_V)$}
            & \multicolumn{7}{c|}{Axial-vector coupling $(c_A)$}
            \\
    \hline
    \hline
SM fermions &$x_H=-2$&$-1$&$-0.5$&$0$&$0.5$&$1$&$2$&$x_H=-2$&$-1$&$-0.5$&$0$&$0.5$&$1$&$2$\\
    \hline
    \hline
Charged lepton $(\ell^\alpha)$  &$-\frac{3}{4}x_H^{}-1= \frac{1}{2}$& $-\frac{1}{4}$& $-\frac{5}{8}$& $-1$& $-\frac{11}{8}$ &$-\frac{7}{4}$  &$-\frac{5}{2}$&$\frac{1}{4} x_H^{}=-\frac{1}{2}$&$-\frac{1}{4}$&$-\frac{1}{8}$&0&$\frac{1}{8}$&$\frac{1}{4}$&$\frac{1}{2}$ \\
    \hline
SM-like neutrino $(\nu^\alpha_L)$ &$\frac{1}{4} x_H^{}+\frac{1}{2}=0$&$\frac{1}{4}$     &$\frac{3}{8}$     &$\frac{1}{2}$   & $\frac{5}{8}$ &$\frac{3}{4}$  &$1$&$\frac{1}{4} x_H^{}+\frac{1}{2}=0$&$\frac{1}{4}$&$\frac{3}{8}$& $\frac{1}{2}$  &$\frac{5}{8}$&$\frac{3}{4}$&$1$ \\
    \hline
    \hline
up-type quarks $(q_u^\alpha)$  &$\frac{5}{12} x_H^{}+\frac{1}{3}=-\frac{1}{2}$& $-\frac{1}{12}$    & $\frac{1}{8}$    &$\frac{1}{3}$& $\frac{13}{24}$ & $\frac{3}{4}$ &$\frac{7}{6}$&$-\frac{1}{4} x_H^{}=-\frac{1}{2}$&$-\frac{1}{4}$&$-\frac{1}{8}$&0&$\frac{1}{8}$&$\frac{1}{4}$&$\frac{1}{2}$ \\
    \hline
down-type quarks $(q_d^\alpha)$ &$-\frac{1}{12} x_H^{}+\frac{1}{3}=-\frac{1}{2}$& $-\frac{1}{4}$    & $-\frac{1}{8}$    &$\frac{1}{3}$& $\frac{7}{24}$ &$\frac{1}{4}$ &$\frac{1}{6}$&$-\frac{1}{4} x_H^{}=-\frac{1}{2}$&$-\frac{1}{4}$&$-\frac{1}{8}$&0&$\frac{1}{8}$&$\frac{1}{4}$&$\frac{1}{2}$ \\  
    \hline  
    \hline
\end{tabular}
\end{center}
\caption{Vector and axial vector couplings in general $U(1)_X$ scenarios for the couplings between SM fermions and $Z^\prime$. In B$-$L case considering $x_H=0$ and $x_\Phi=1$, the axial vector couplings for the charged fermions vanish. 
}
\label{tab-3}
\end{table}
Hence, we notice that interactions between SM fermions and $Z^\prime$ are chiral in nature in general $U(1)_X$ extension of the SM. The partial decay width of $Z^\prime$ into different fermions can be calculated using Eq.~\eqref{Lag1} and we write down the expression as 
\begin{align}
\label{eq:width-ll}
    \Gamma(Z' \to \bar{f} f)
    &= N_C^{} \frac{M_{Z^\prime}^{} g_{X}^2}{24 \pi} \left[ \left( q_{f_L^{}}^2 + q_{f_R^{}}^2 \right) \left( 1 - \frac{m_f^2}{M_{Z^\prime}^2} \right) + 6 q_{f_L^{}}^{} q_{f_R^{}}^{} \frac{m_f^2}{M_{Z^\prime}^2} \right]~,
\end{align}    
where $m_f$ is the mass of different SM fermions and $q_{L, R}$ are the functions of $x_H$. Here, $N_C^{}$ is the color factor being $1$ for the SM leptons and $3$ for the SM quarks. The light neutrinos $(\nu_L)$ are considered to be massless due to their tiny mass and putting $q_{f_R}=0$ in Eq.~\ref{eq:width-ll}, we obtain the partial decay width of $Z^\prime$ into a pair of one generation light neutrinos as 
\begin{align}   
\label{eq:width-nunu}
    \Gamma(Z' \to \nu \nu)
    = \frac{M_{Z^\prime}^{} g_{X}^2}{24 \pi} q_{f_L^{}}^2~,
\end{align} 
were $q_{f_L^{}}^{}$ is a function of $x_H$. The $Z^\prime$ gauge boson can decay into a pair of heavy Majorana neutrinos if $Z^\prime$ is heavier than twice the mass of the heavy neutrinos. The corresponding partial decay width into single generation of heavy neutrino pair can be written as
\begin{align}
\label{eq:width-NN}
    \Gamma(Z^\prime \to N_R^\alpha N_R^\alpha)
    = \frac{m_{Z'}^{} g_{X}^2}{24 \pi} q_{N_R^{}}^2 \left( 1 - 4 \frac{m_N^2}{M_{Z^\prime}^2} \right)^{\frac{3}{2}}~,
\end{align}
with $q_{N_R^{}}^{}$ is the general $U(1)_X$ charge of the heavy neutrinos which could be found from Tabs.~\ref{tab1} and \ref{tab2}, respectively and $m_N^{}$ is the mass of the heavy neutrinos. If we consider that the RHNs are heavier than the half of the $M_{Z^\prime}$ then the decay of $Z^\prime$ into a pair of RHN is kinematically forbidden. We find that the $U(1)_X$ charges of the fermions in Cases-II are same as those in Case-I with $x_\Phi=1$. As a result we can utilize same bounds for both the cases.  
\subsection{Neutrino mass}
The Yukawa interactions given in Eqs.~\ref{LYk} and \ref{ExoticYukawa} lead us to the generation of neutrino mass mechanism. The general $U(1)_X$ symmetry breaking generates the Majorana mass term for the three (first two) generations of the RHNs in Case-I (II) where BSM scalar $\Phi_{(1)}$ is involved. The Dirac mass term is generated after the electroweak symmetry breaking where SM Higgs doublet $H_{1(2)}$ is involved in Case-I(II). The corresponding Dirac and Majorana mass terms are written in Tab.~\ref{tab4}.  
\begin{table}[t]
\begin{center}
\begin{tabular}{|c|c|c|}
\hline\hline
      Models  & Majorana mass& Dirac mass  \\ 
                       & $(m_N)$& $(m_D)$\\
\hline\hline
Case-I&$m_{N_R^\alpha}=\frac{Y_N^\alpha}{\sqrt{2}} v_\Phi$&$m_{D}^{\alpha \beta}=\frac{Y_\nu}{\sqrt{2}} v_1$\\
&&\\
Case-II &$m_{N_R^{1,2}}=\frac{Y_{2}^{1,2}}{\sqrt{2}} v_1$&$m_D^{1,2}=\frac{Y_{1}^{1,2}}{\sqrt{2}}v_{h_2}$\\
\hline\hline
\end{tabular}
\end{center}
\caption{Dirac and Majorana masses in the neutrino sector. In Case-II we use the collider constraints to set $(v_1^2 + v_2^2+ v_3^2) \gg (v_{h_{1}}^2 + v_{h_{2}}^2)$ and in this case first two generations are participating in the neutrino mass generation mechanism.}
\label{tab4}
\end{table} 
Finally the light neutrino mass is generated by the seesaw mechanism to explain the origin of tiny neutrino mass term and flavor mixing. Following Tab.~\ref{tab4} the generic formula for the neutrino mass matrix can be written as 
\begin{equation}
   m_\nu= \begin{pmatrix} 0&m_D^{}\\m_D^T&m_N^{} \end{pmatrix}.
\label{num-1}
\end{equation}
Diaginalizing the neutrino mass matrix we find the light neutrino mass eigenvalues to be $`-m_D^{} m_N^{-1} m_D^T$'. In Case-II $N_R^{1,2}$ will generate the neutrino mass, on the other hand, at the tree level $N_R^3$ will not participate in the neutrino mass generation and it can be considered as a potential DM candidate in some scenarios. The neutrino mass generation mechanism and dark matter physics are not main motivations of this work, however, we provide a simple outline in this paper for completeness because these general $U(1)_X$ scenarios can generate neutrino mass at the tree level from the so-called seesaw mechanism which is an important aspect for studying such scenarios.  

\section{Calculation of the constraints on the chiral gauge couplings}
\label{sec:calculation}
The chiral $Z'$ gauge boson interacts with the SM fermions, and the couplings depend on the $U(1)_X$ charge of the SM Higgs doublet.
Therefore, experiments for scattering measurements of the SM particles can search the chiral $Z'$ gauge boson by measuring deviations of the scattering cross section from the SM value.
In this section, we show the scattering cross sections contributed by the chiral $Z'$ gauge boson and methods to calculate constraints on the $U(1)_X$ gauge coupling.
In this work, we consider FASER$\nu(2)$, SND$@$LHC, NA64 and MuonE as the experiments for scattering measurement. In addition, we estimate the constraints from $\nu-$electron, $\nu-$nucleon, electron and proton beam dump, long-lived gauge boson searches, respectively.  

\subsection{Scattering cross section contributed by chiral \texorpdfstring{$Z'$}{Z'} 
at FASER$\nu(2)$, SN$D$@LHC, NA64 and MuonE}

In this subsection, we summarize scattering cross sections via chiral $Z'$ interactions at FASER$\nu(2)$, SND@LHC, NA64 and MuonE experiments. 

\subsubsection{Prospects for FASER\texorpdfstring{$\nu(2)$}{v(2)} and SND@LHC}

\begin{figure}[htb]
\includegraphics[width=0.495\textwidth]{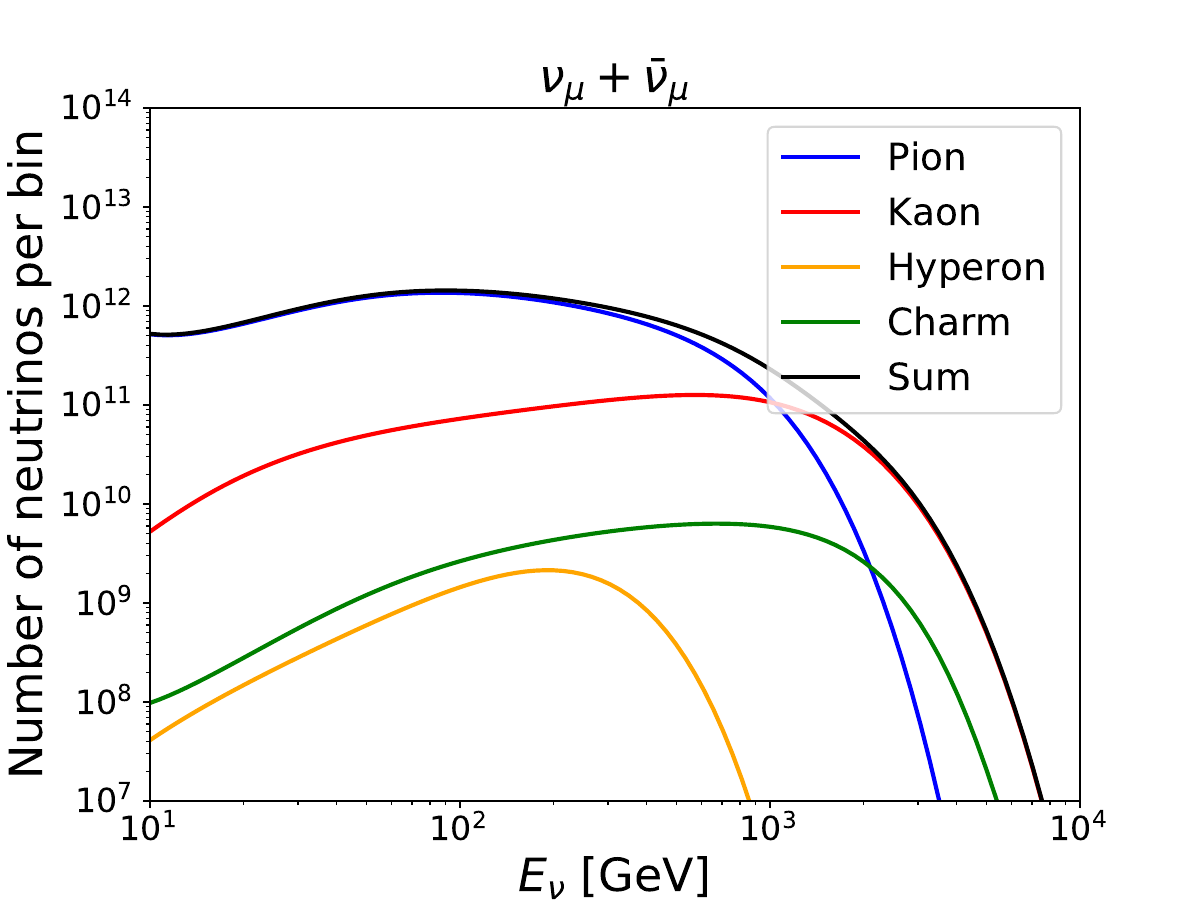}
\includegraphics[width=0.495\textwidth]{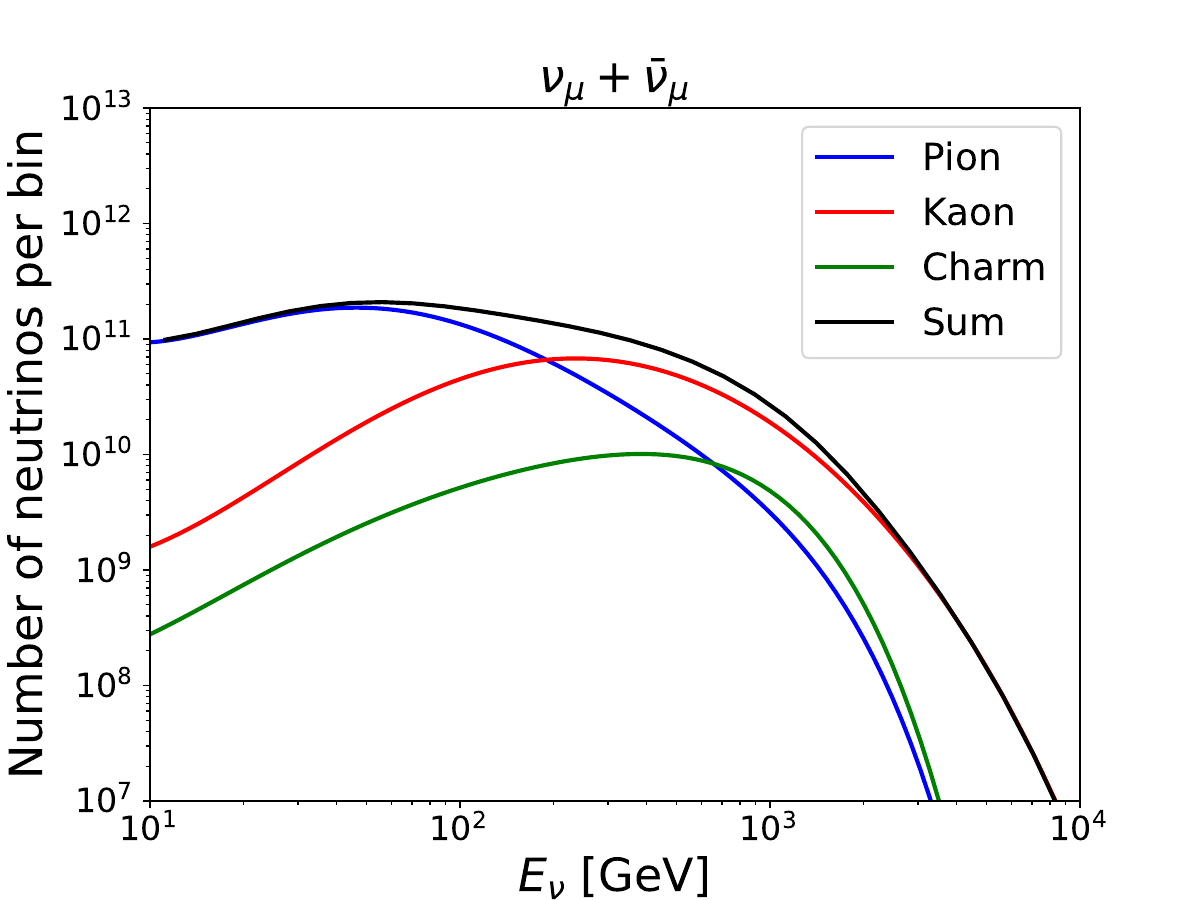}
\caption{Number of muon neutrino pass through the FASER$\nu$ (left panel) and SND@LHC (right panel) detectors. Thirty energy bins are defined uniformly on the logarithmic scale in [10,$10^{4}$] GeV.}
\label{fig:vmuflux}
\end{figure}

The existence of a light $Z^{\prime}$ affects the neutral-current deep-inelastic scattering at the LHC far-forward detectors. 
We study the constraints from the FASER$\nu$, FASER$\nu$2 and SND@LHC experiments in this subsection. 
The numbers of neutrinos that pass through the FASER$\nu$ and SND@LHC detectors have been simulated in Refs.~\cite{FASER:2019dxq,Kling:2021gos}. It has been found that the muon neutrino from pions and kaons decay is most abundant. The energy distributions of which are shown in Fig.~\ref{fig:vmuflux}. 
The corresponding neutrino flux at the FASER$\nu$2 detector can be obtained by rescaling the flux at the FASER$\nu$, assuming the neutrino distributes uniformly on the detector surface.

\begin{figure}[htb]
\includegraphics[width=0.475\textwidth]{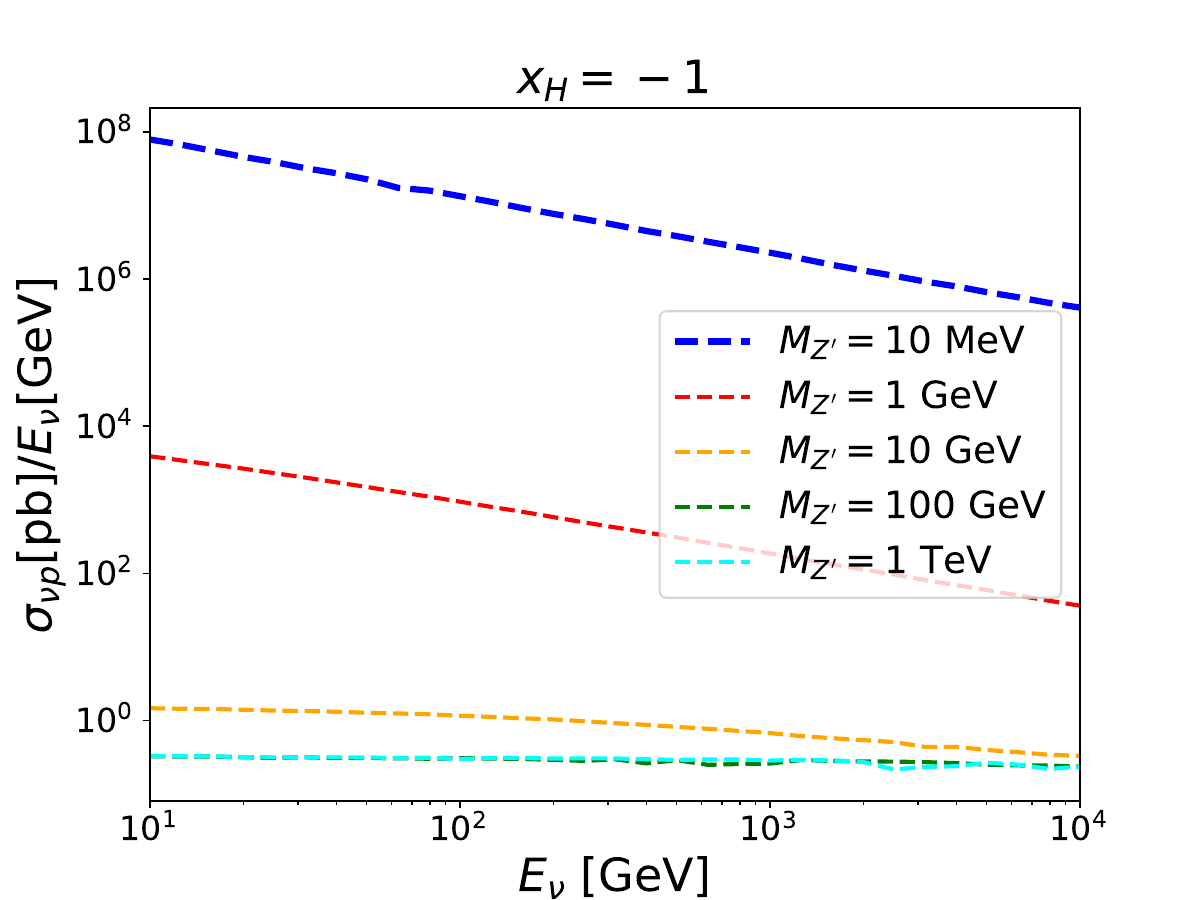}
\includegraphics[width=0.475\textwidth]{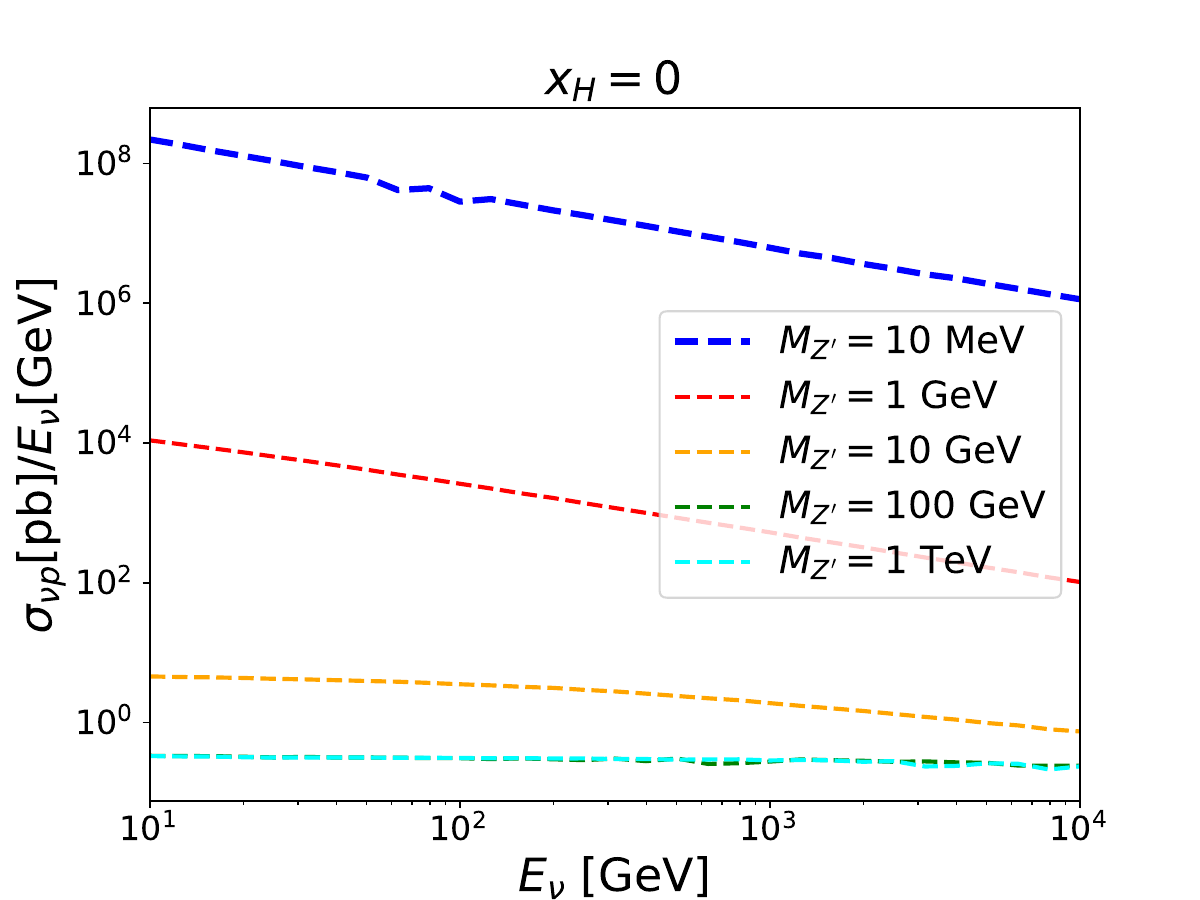}\\
\includegraphics[width=0.475\textwidth]{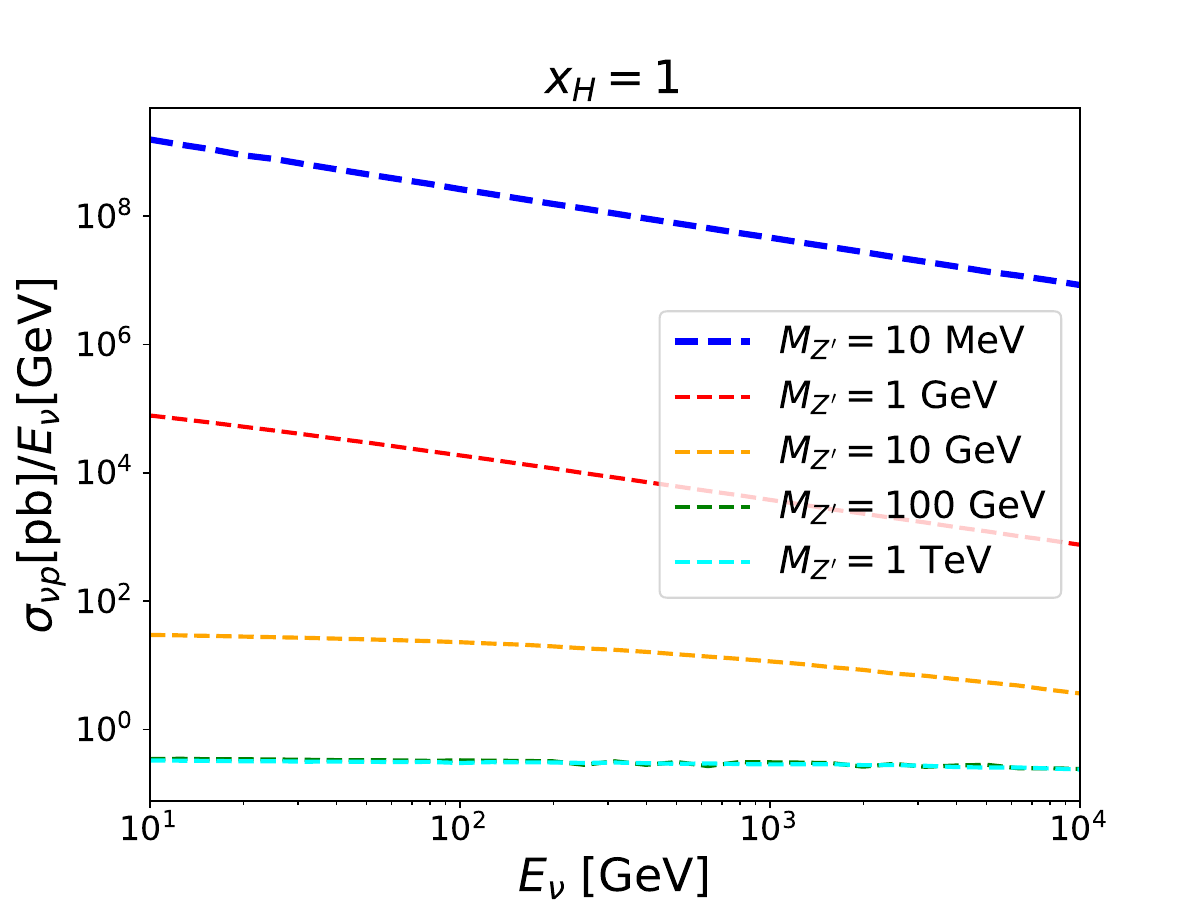}
\caption{The neutrino-proton scattering cross section $\sigma_{\nu p}$ divided by the incoming neutrino energy $E_{\nu}$. The gauge coupling $g_X$ is set to the unity. }
\label{fig:xsec/E}
\end{figure}

The MG5\_aMC@NLO package is used to calculate the fixed target deep-inelastic neutrino-proton scattering cross-section at the leading order. 
Assuming a benchmark detector made of tungsten target, the nCTEQ15FullNuc\_184\_74 set~\cite{Kovarik:2015cma} as implemented in LHAPDF6~\cite{Buckley:2014ana} is employed as the proton parton distribution function.
The partonic collision energy is taken as the factorization and renormalization scales in our simulation. 
The neutrino-proton scattering cross section $\sigma_{\nu p}$ is related to $g_X$, $x_{H}$, $Z^{\prime}$ mass as well as the incoming neutrino energy $E_{\nu}$. In Fig.~\ref{fig:xsec/E}, we present the relation between $E_{\nu}$ and $\sigma_{\nu p}/E_{\nu}$ for a few sets of $x_{H}$ and $M_{Z^{\prime}}$. The $g_X$ is fixed to the unity, since the change of $g_X$ can only lead to a total rescaling. 
The $Z^{\prime}$ contribution is slightly increased with increasing $x_{H}$ from $-1$ to 1. 
The dependence on the $M_{Z^{\prime}}$ is more complicated due to the interference effects between the $Z$ and $Z^{\prime}$ bosons. The $Z^{\prime}$ contribution is negligible for $M_{Z^{\prime}} \gtrsim 100$ GeV, and the cross-section becomes identical to the SM one in this region. 
Given the neutrino-nucleus cross section ($\sigma_{\nu N}$, which is $184 \times \sigma_{\nu p}$, mass number of Tungsten is approximately 184), one can estimate the probability of a neutrino interacting with the detector as 
\begin{align}
P= \frac{\sigma_{\nu N} \times ~\text{Number of Nuclei}}{\text{Detector Area}} = \frac{\sigma_{\nu N}}{A} \frac{m_{\rm det}}{m_{N}}~,~
\end{align}
where $N$ is the target nucleus, $A$ is the detector area, and $m_{N}$ is the mass of the target nucleus. 
The relevant detector configurations are listed in Tab.~\ref{tab:dets}. 

\begin{table}[h!]
\begin{center}
\begin{tabular}{|c||c|c|c|} \hline
Detector & $A$  & $m_{\rm det}$ & Integrated Luminosity \\ \hline
FASER$\nu$    & 25 cm $\times$ 25 cm & 1.2 tons  &  150 fb$^{-1}$ \\
FASER$\nu$2  & 50 cm $\times$ 50 cm & 10 tons   & 3000 fb$^{-1}$ \\
SND@LHC      & 39 cm $\times$ 39 cm &  0.8 tons  & 150 fb$^{-1}$ \\ \hline
\end{tabular}
\caption{\label{tab:dets} Detector configurations. }
\end{center}
\end{table}

\begin{sidewaysfigure}
\centering
\includegraphics[width=0.33\textwidth]{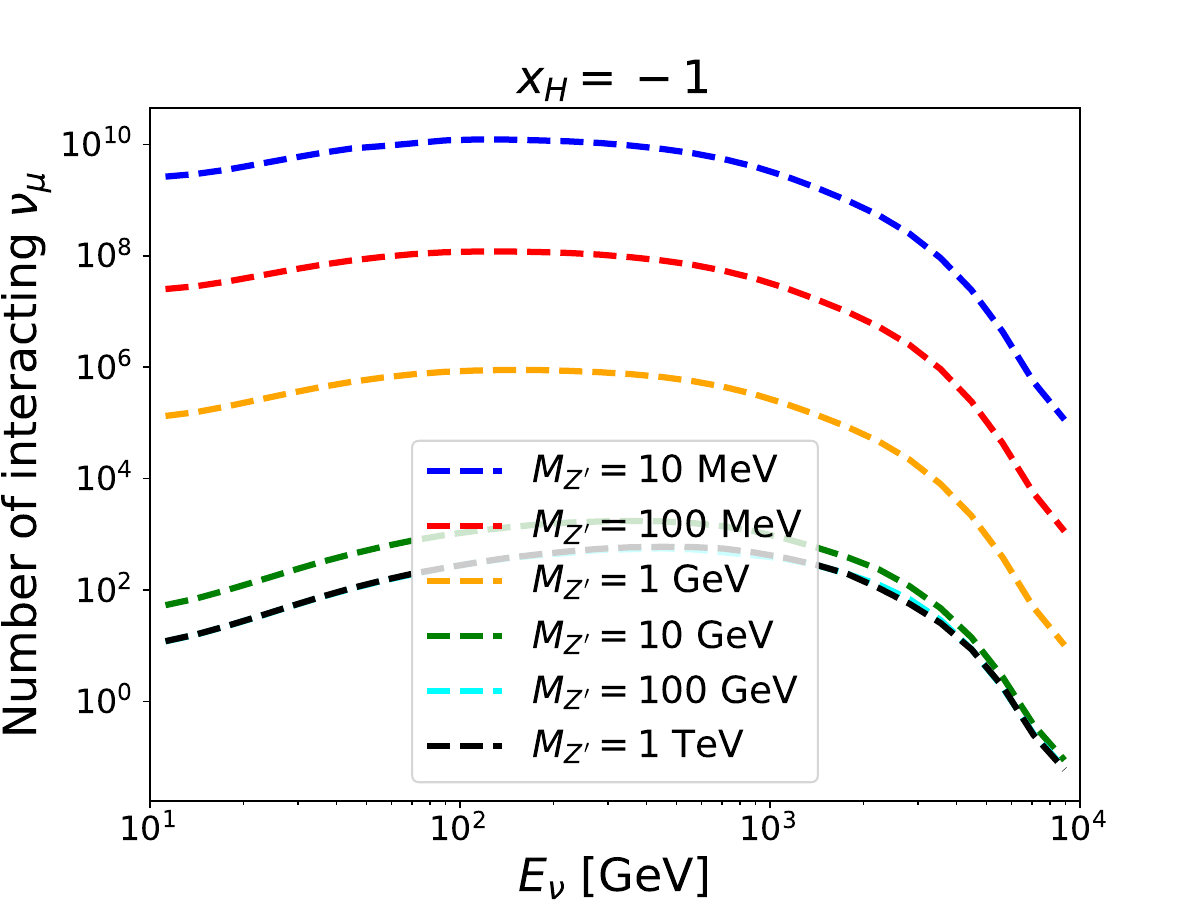}
\includegraphics[width=0.33\textwidth]{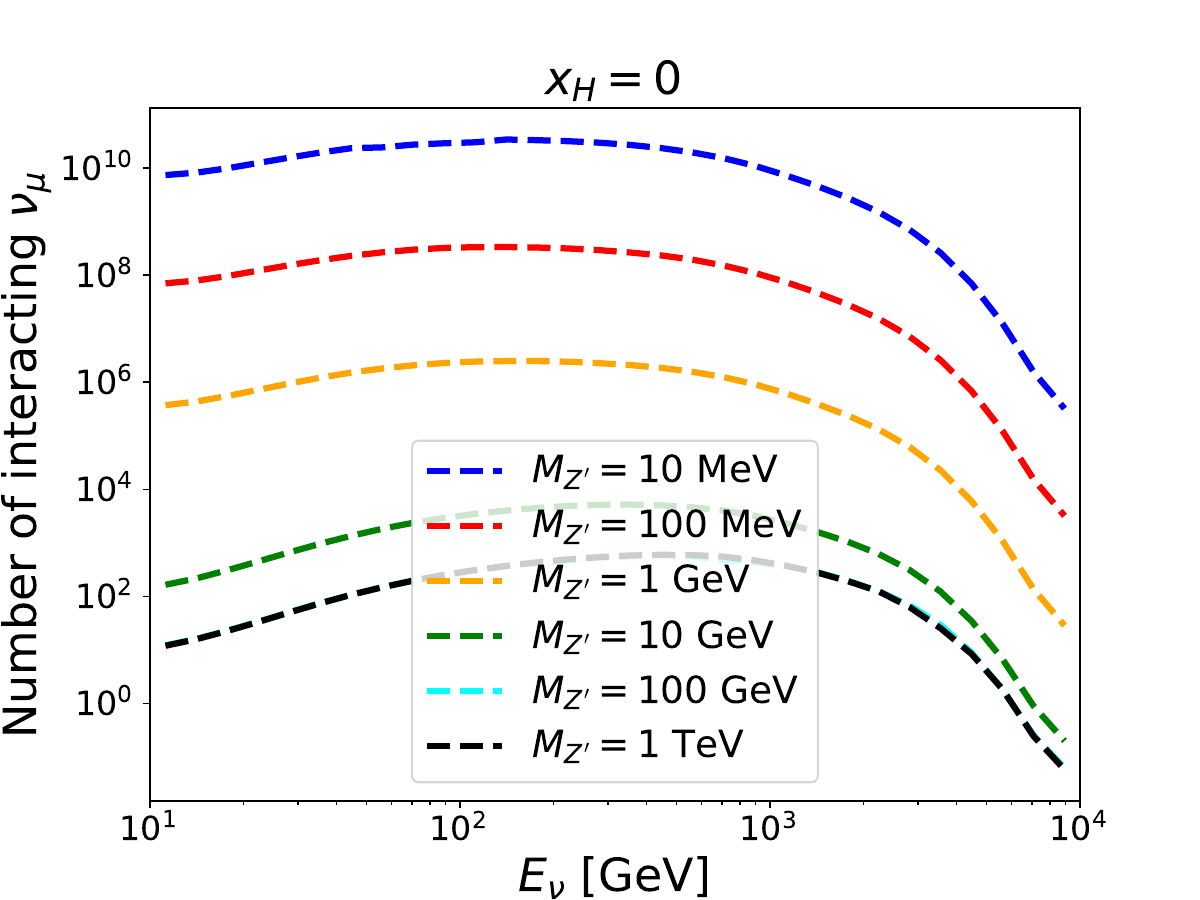}
\includegraphics[width=0.33\textwidth]{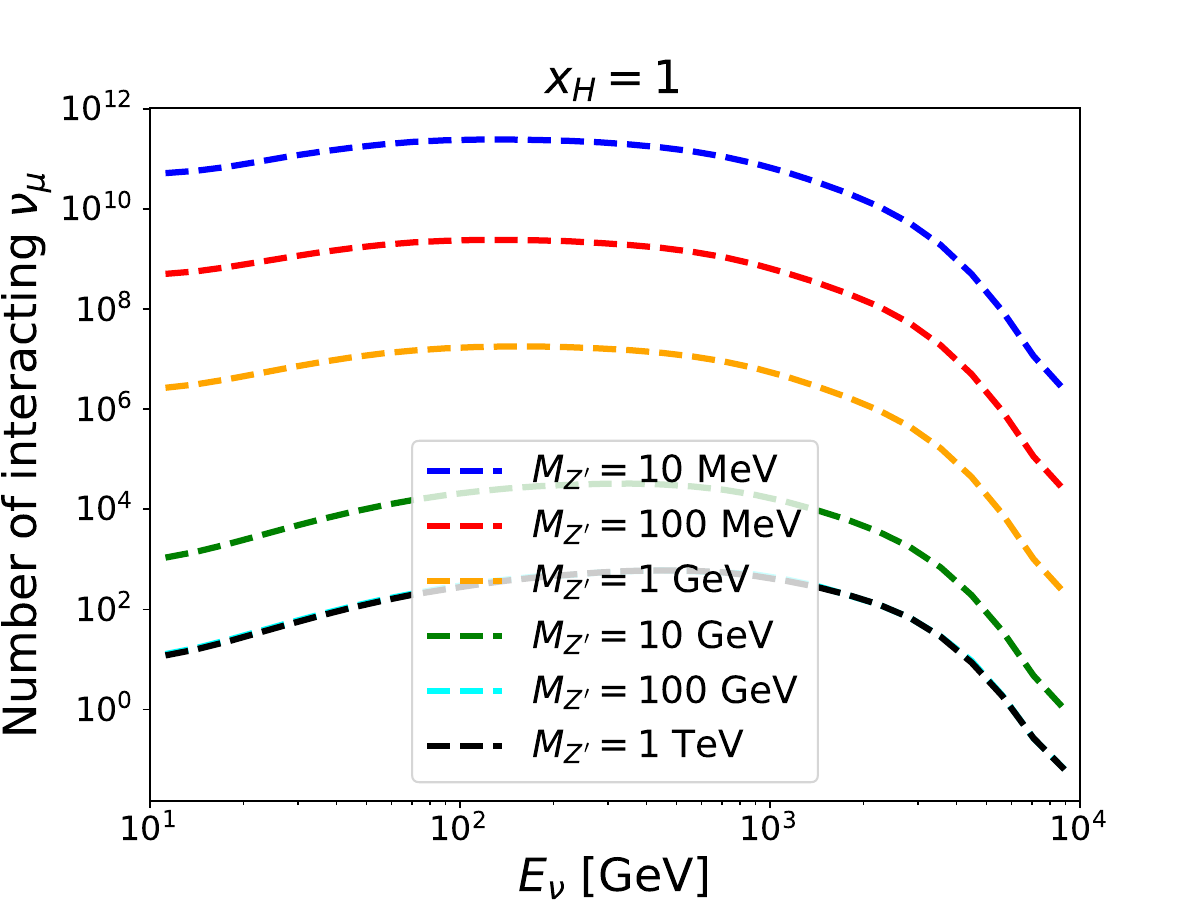}
\includegraphics[width=0.33\textwidth]{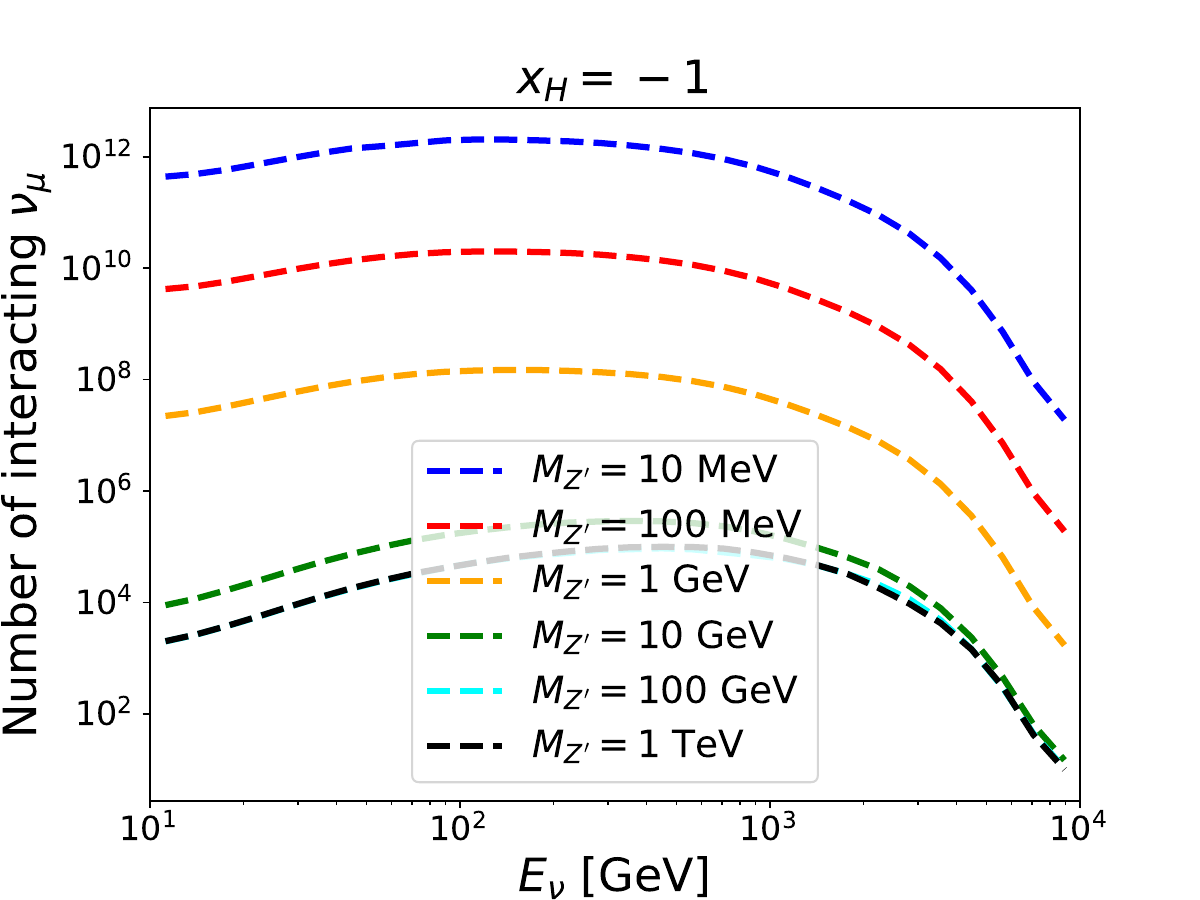}
\includegraphics[width=0.33\textwidth]{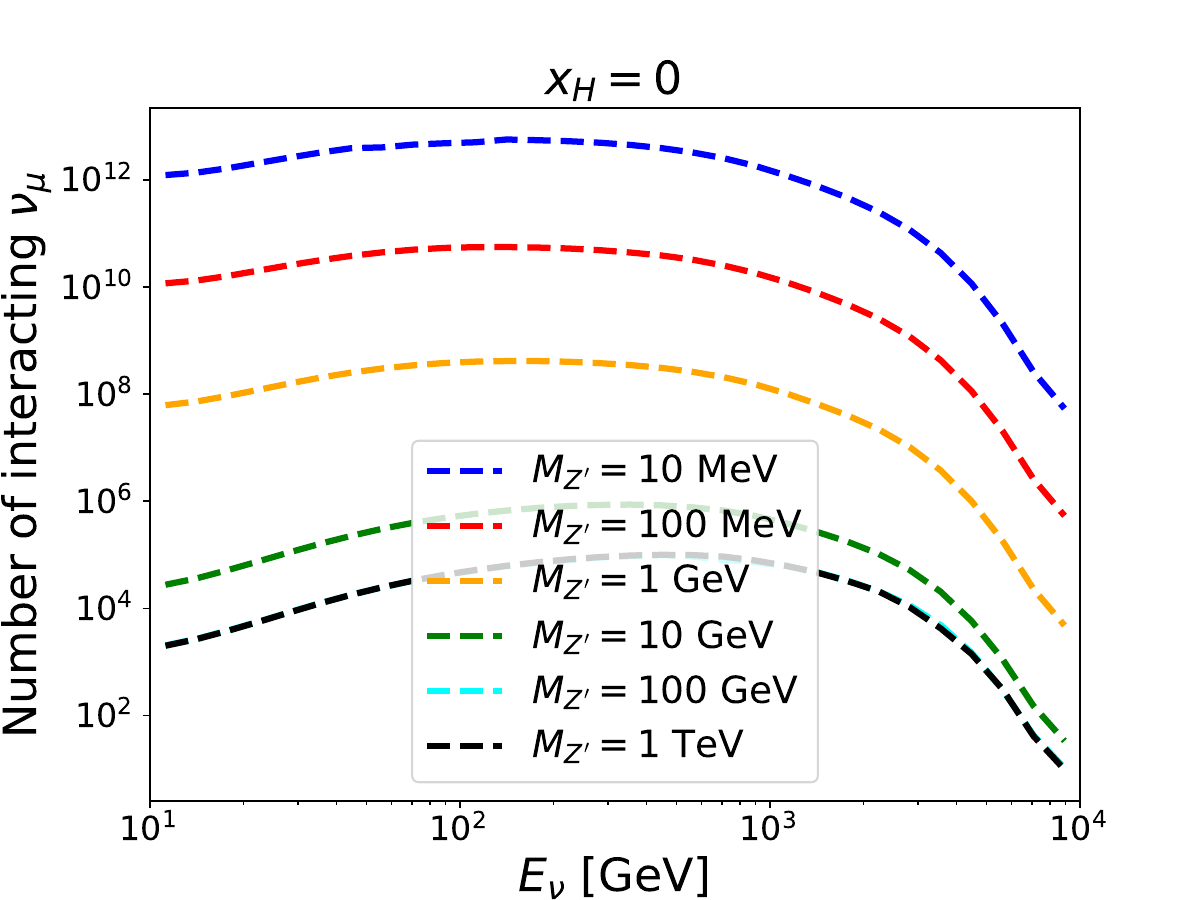}
\includegraphics[width=0.33\textwidth]{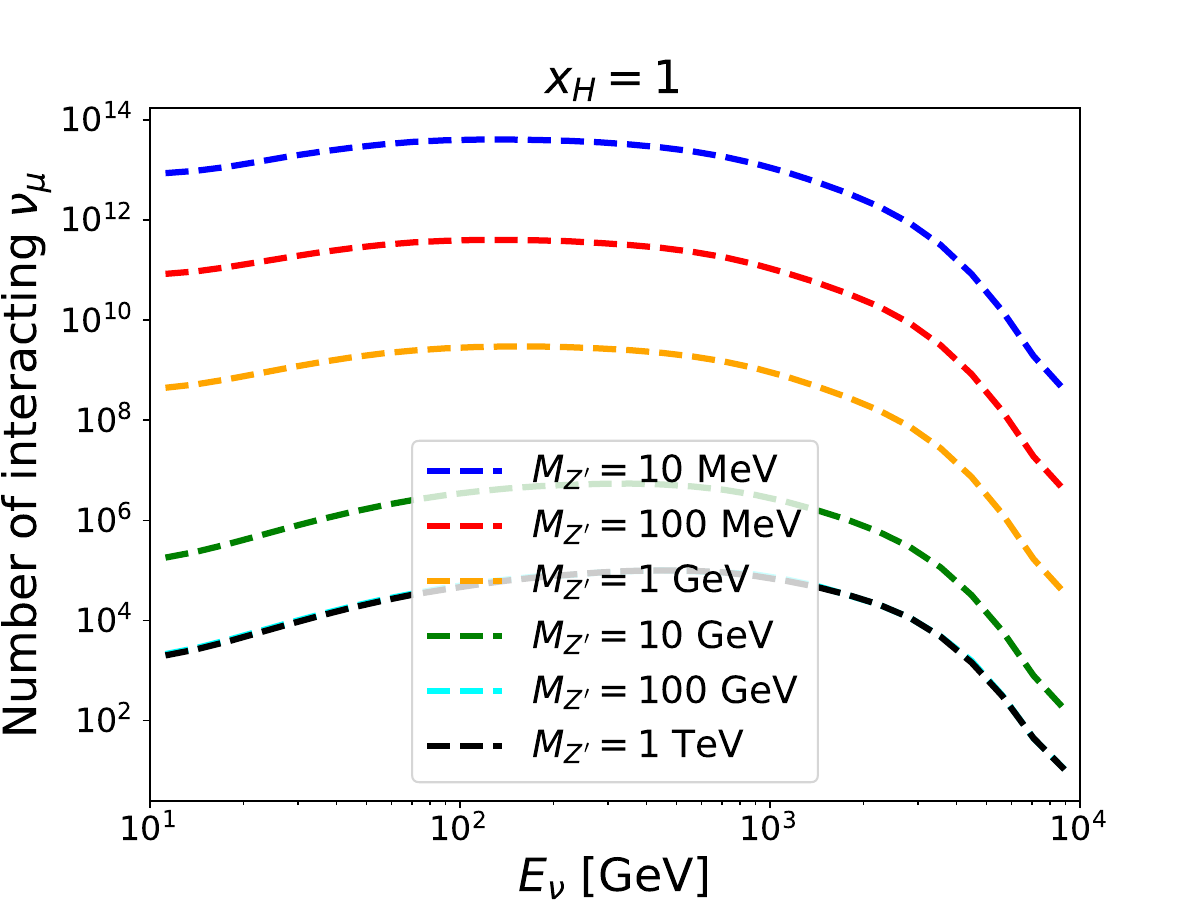}
\includegraphics[width=0.33\textwidth]{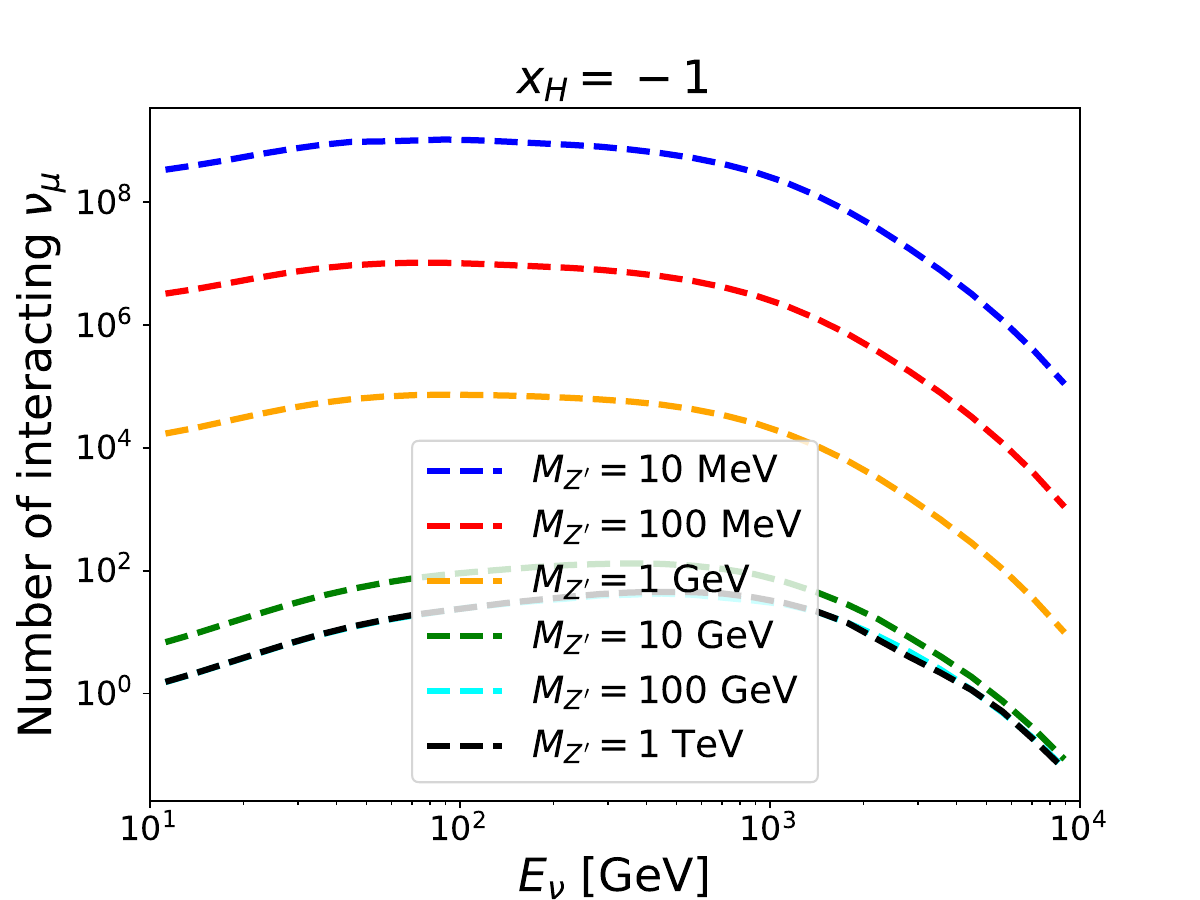}
\includegraphics[width=0.33\textwidth]{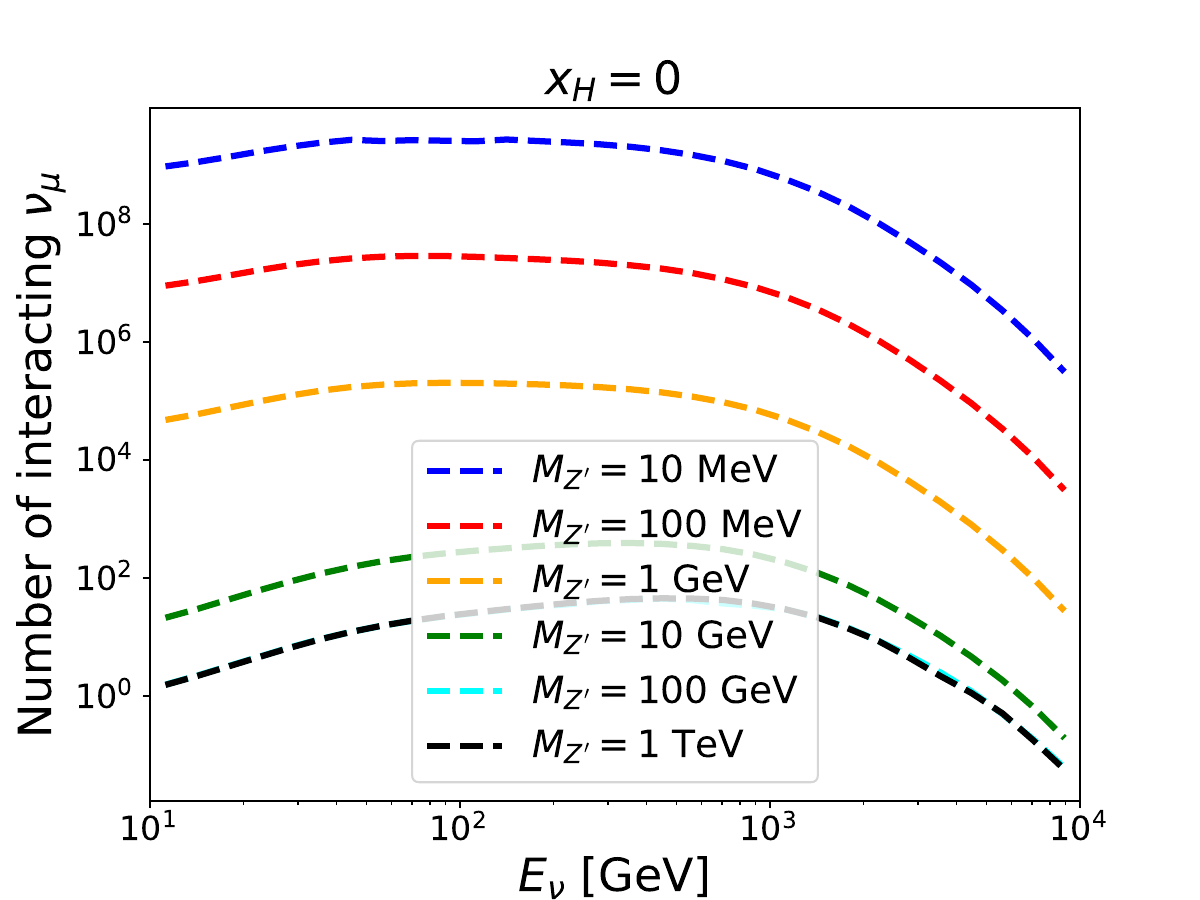}
\includegraphics[width=0.33\textwidth]{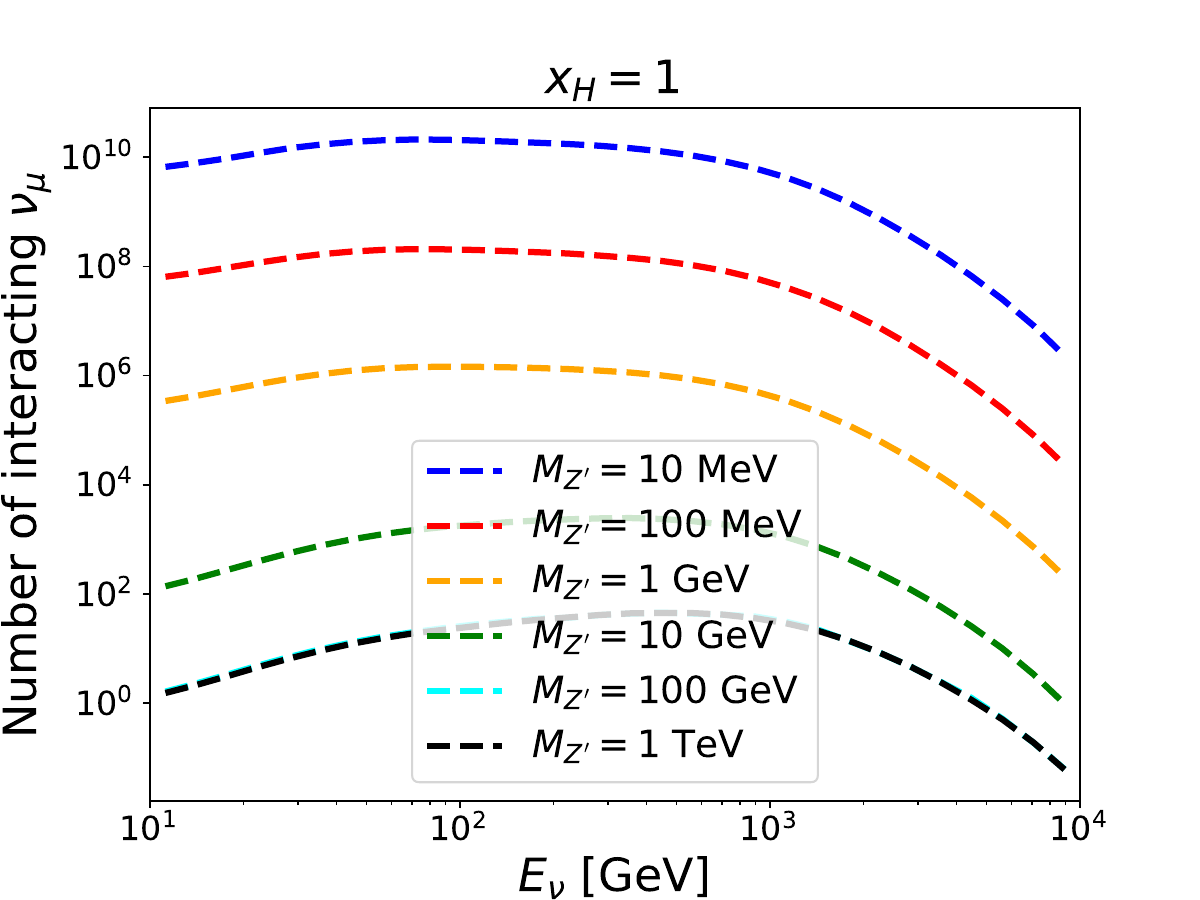}
\caption{The muon neutrino spectra at the FASER$\nu$ (top panels), FASER$\nu$2 (middle panels) and SND@LHC (bottom panels) detectors. }
\label{fig:nnudet}
\end{sidewaysfigure}

For given $g_X$, $x_{H}$ and $M_{Z^{\prime}}$, the number of interacting neutrinos in each $E_{\nu}$ bin can be calculated by the products of the number of neutrinos passing through the detector and the interaction probability. 
For illustration, energy spectra for the interacting muon neutrino at the FASER$\nu$, FASER$\nu$2 and SND@LHC detectors are shown in Fig.~\ref{fig:nnudet}. 
The $g_X$ is taken to be the unity. Three different values of $x_{H}$ and six different values of $M_{Z^{\prime}}$ are used. 
The number of interacting neutrinos is increased with increasing $x_{H}$ and decreasing $M_{Z^{\prime}}$. 
Moreover, we can observe that the number at the FASER$\nu$ detector is around one order of magnitude larger than that at the SND@LHC detector, while it is around two orders of magnitude below that at the FASER$\nu$2 detector. 

We follow a similar strategy as proposed in Ref.~\cite{Cheung:2021tmx} to estimate the sensitivity reach in the parameter space of $g_X$ and $M_{Z^{\prime}}$ for our model. 
Only the total numbers of interacting neutrinos are used in defining the $\chi^{2}$ measure ({\it i.e.} the shape of $E_{\nu}$ spectra are not of concern): 
\begin{align}
\chi^{2} = \min_{\alpha} \left[ \frac{N^{\nu_{e}}_{\rm BSM} -(1+\alpha) N^{\nu_{e}}_{\rm SM}}{N^{\nu_{e}}_{\rm BSM}} + \frac{N^{\nu_{\mu}}_{\rm BSM} -(1+\alpha) N^{\nu_{\mu}}_{\rm SM}}{N^{\nu_{\mu}}_{\rm BSM}} + \frac{N^{\nu_{\tau}}_{\rm BSM} -(1+\alpha) N^{\nu_{\tau}}_{\rm SM}}{N^{\nu_{\tau}}_{\rm BSM}} +(\frac{\alpha}{\sigma_{\rm norm}})^{2} \right]~,
\end{align}
where $N_{\rm BSM}$ and $N_{\rm SM}$ are the number of interacting neutrinos of each flavor in our model and in the SM model. 
The systematic uncertainties ($\sigma_{\rm norm}$) in each neutrino flavor are assumed to be the same and only one nuisance parameter $\alpha$ is used. 
The $\chi^{2}$ value is obtained by minimizing over the $\alpha$. The 95\% confidence level sensitivity reach corresponds to $\chi^{2} = 3.84$.

\subsubsection{Constraints from the NA64}
The fixed-target experiment NA64 at the CERN SPS~\cite{NA64:2016oww,NA64:2017vtt} aims to search for the $Z^{\prime}$ that is produced through the bremsstrahlung process in the high-energy electron beam colliding with heavy nuclei
\begin{align}
e^{-} Z \to e^{-} Z Z^{\prime}; ~~ Z^{\prime} \to \nu \nu
\end{align}
where the $Z^{\prime}$ is decaying invisibly. 
The partial width of each $Z^{\prime}$ decay channel is calculated by the DARKCAST package~\cite{Ilten:2018crw,Baruch:2022esd}, assuming that only vector interactions exist. The results are translated into those in our chiral model by using the same method as we proposed in Ref.~\cite{Asai:2022zxw}. 

\begin{figure}[htb]
\includegraphics[width=0.495\textwidth]{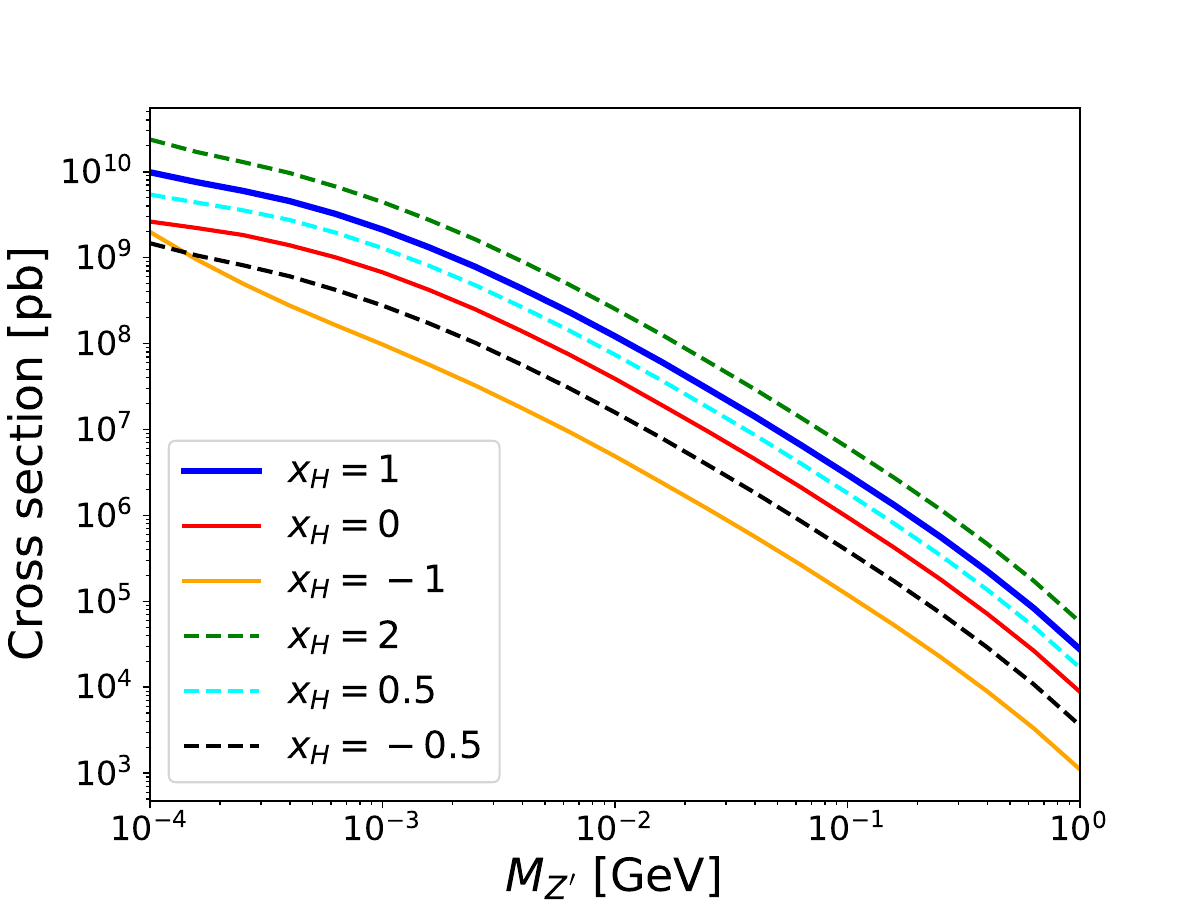}
\includegraphics[width=0.495\textwidth]{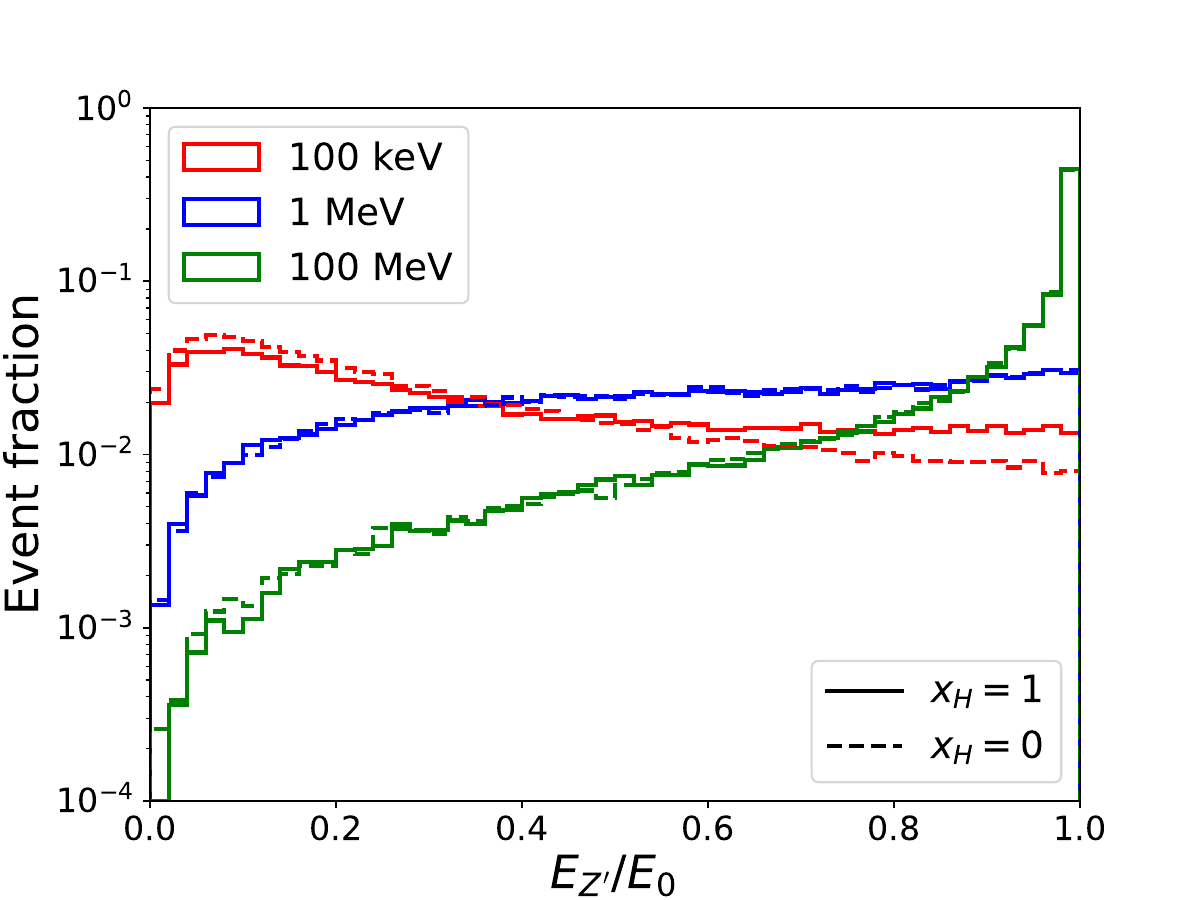}
\caption{Left panel: the total cross section for the bremsstrahlung $Z^{\prime}$ production. Right panel: normalized $E_{Z^{\prime}}/ E_{0}$ distribution for different $Z^{\prime}$ masses. $E_{0}=100$ GeV is the electron beam energy. 
}
\label{fig:xsec}
\end{figure}

The production cross section and the energy spectrum of the $Z^{\prime}$ is simulated by the MG5\_aMC@NLO package.
Since the target nucleus is lead at the NA64 experiment, the nCTEQ15\_208\_82 set in LHAPDF6 is employed as the proton PDF. 
The total cross section is proportional to $g_X^{2}$. Taking $g_X=0.2$ and incoming electron beam energy $E_{0}=100$ GeV, we present the electron-proton scattering cross sections with respect to varying $x_{H}$ and $M_{Z^{\prime}}$ in the left panel of Fig.~\ref{fig:xsec}. It can be observed that the cross-section is increased with increasing $x_{H}$ and decreasing $M_{Z^{\prime}}$. 
In the right panel of Fig.~\ref{fig:xsec}, the normalized bremsstrahlung $Z^{\prime}$ spectra are shown. 
The shape of $Z^{\prime}$ spectrum is highly dependent on the $Z^{\prime}$ mass while it is almost irrelevant to the $x_{H}$ value. Some analytic discussions about the feature of the spectrum are conducted in Ref.~\cite{Gninenko:2017yus}. 

Having calculated the differential bremsstrahlung $Z^{\prime}$ production cross section ($d \sigma_{Z^{\prime}} / d E_{Z^{\prime}}$) as well as the $Z^{\prime}$ decay branching ratios for all of the channels, we are able to calculate the number of signal events at the NA64 experiment as follows~\cite{NA64:2022yly}
\begin{align}
n_{Z^{\prime}}(g_X, x_{H} ,M_{Z^{\prime}}) = \int_{0.5 E_{e}}^{E_{e}} C \frac{d \sigma_{Z^{\prime}}}{d E_{Z^{\prime}}} \left[ \text{Br}(Z^{\prime} \to \nu \nu) +\sum_{l} \text{Br}(Z^{\prime} \to l^{+} l^{-}) \exp(-\frac{L_{\rm ECAL} + L_{\rm HCAL}}{L_{Z^{\prime}}}) \right], \label{eq:nevtna64}
\end{align} 
where $C$ is related to the detector parameters. The $L_{Z^{\prime}}= c \tau_{Z^{\prime}} E_{Z^{\prime}}/ M_{Z^{\prime}}$, $L_{\rm ECAL}$ and $L_{\rm HCAL}$ are the $Z^{\prime}$ decay length, electromagnetic calorimeter length, and hadronic calorimeter length, respectively. 
The Ref.~\cite{NA64:2022yly} uses data with $3.22 \times 10^{11}$ electrons on target collected during 2016–2021 runs at the NA64 experiment and obtains the 90\% C.L. exclusion limits for the $U(1)_{B-L}$ model. 
The corresponding bounds for our chiral model with different $x_{H}$ can be obtained by requiring the number of signal events $n_{Z^{\prime}}(g_X, x_{H}, M_{Z^{\prime}})$ in Eq.~\ref{eq:nevtna64} to be the same as that in the B$-$L case. 

\subsubsection{Constraints from the MUonE}
MUonE~\cite{Abbiendi:2016xup,Abbiendi:2677471} is an experiment at CERN aiming to measure elastic scatterings between the 150\,GeV $\mu^+$ beam and target being electrons in beryllium atoms and determining a contribution of hadronic vacuum polarization to the muon anomalous magnetic moment with a method~\cite{CarloniCalame:2015obs}.
The existence of the chiral $Z^\prime$ changes the cross section of the elastic scattering between positive muon and electron, and therefore, the chiral $Z^\prime$ can be searched by estimating a deviation of the scattering cross section from the SM prediction.

The elastic scattering cross section between the positive muon and electron is given by
\begin{equation}
    \frac{d\sigma(\mu^+e^-)}{dT} =
    \left. \frac{d\sigma(\mu^+e^-)}{dT} \right|_{\rm SM}
    \left. + \frac{d\sigma(\mu^+e^-)}{dT} \right|_{Z'}
    \left. + \frac{d\sigma(\mu^+e^-)}{dT} \right|_{\rm Int}~,
\end{equation}
where $T$ denotes the electron recoil energy, and the first, second, and third terms in the right-hand side (RHS) correspond to the contributions from the only SM interaction, only $Z'$ one and interference between the SM and $Z'$ ones. 
The purely SM contribution can be given by
\begin{align}
    \left. \frac{d\sigma(\mu^+e^-)}{dT} \right|_{\rm SM} =
    \frac{\pi \alpha_{\rm EM}^2}{(E_\mu^2 - m_\mu^2) m_e^2 (T - m_e)^2} \left\{ 2 E_\mu m_e (E_\mu - T + m_e) - (T - m_e) (2 m_e^2 + m_\mu^2 - m_e T) \right\}~,
\end{align}
where $\alpha_{\rm EM}$ denotes the fine structure constant, and $E_\mu$ stands for the energy of the positive muon.
The contributions from $Z'$ exchanging diagram and interference between the SM and $Z'$ are estimated by 
\begin{align}
    \left. \frac{d\sigma(\mu^+e^-)}{dT} \right|_{Z'} =&
    \frac{g_X^4 m_e}{128 \pi M_{Z^\prime}^4 (E_\mu^2 - m_\mu^2) (M_{Z^\prime}^2 + 2 m_e T - m_e^2)^2} \nonumber \\
    &\quad \times \left[
    (\tilde{x}_e + \tilde{x}_\ell)^4 \left\{ M_{Z^\prime}^4 \left( 2 E_\mu (E_\mu - T + m_e) + T^2 - 3 m_e T - m_\mu^2 + 2 m_e^2 \right)  \right\}
    \right. \nonumber \\
    &\qquad \quad + 2 (\tilde{x}_e + \tilde{x}_\ell)^3 (\tilde{x}_e - \tilde{x}_\ell) \left\{ M_{Z^\prime}^4 \left( 2 E_\mu^2 - m_e T - m_\mu^2 + m_e^2 \right) \right\} \nonumber \\
    &\qquad \quad + 2 (\tilde{x}_e + \tilde{x}_\ell)^2 (\tilde{x}_e - \tilde{x}_\ell)^2 \left\{ M_{Z^\prime}^4 \left( 2 E_\mu (E_\mu + T - m_e) - (T - m_e)^2 \right) \right. \nonumber \\
    &\hspace{50mm} \left. + 2 m_e m_\mu^2 M_{Z^\prime}^2 (T - m_e) + 2 m_e^2 m_\mu^2 (T - m_e)^2 \right\} \nonumber \\
    &\qquad \quad + 2 (\tilde{x}_e + \tilde{x}_\ell) (\tilde{x}_e - \tilde{x}_\ell)^3 \left\{ M_{Z^\prime}^4 \left( 2 E_\mu^2 + m_e T + m_\mu^2 - m_e^2 \right) + 4 m_e m_\mu^2 M_{Z^\prime}^2 (T - m_e) \right. \nonumber \\
    &\hspace{50mm} \left. + 4 m_e^2 m_\mu^2 (T - m_e)^2 \right\} \nonumber \\
    &\qquad \quad
    + (\tilde{x}_e - \tilde{x}_\ell)^4 \left\{ M_{Z^\prime}^4 \left( 2 E_\mu (E_\mu - T + m_e) + T^2 - m_e T + m_\mu^2 \right) + 4 m_e m_\mu^2 M_{Z^\prime}^2 (T - m_e) \right. \nonumber \\
    &\hspace{50mm} \left. \left. + 4 m_e^2 m_\mu^2 (T - m_e)^2 \right\} \right]~, \\
    \left. \frac{d\sigma(\mu^+e^-)}{dT} \right|_{\rm Int} =&
    \frac{\alpha_{\rm EM} g_X^2}{8 (E_\mu^2 - m_\mu^2) (T - m_e) (M_{Z^\prime}^2 + 2 m_e T - m_e^2)} \nonumber \\
    &\quad \times \left[ (\tilde{x}_e + \tilde{x}_\ell)^2 \left\{ 2 E_\mu \left( E_\mu - T + m_e \right) + T^2 - 3 m_e T - m_\mu^2 + 2 m_e^2 \right\} \right. \nonumber \\
    &\qquad \quad \left. + (\tilde{x}_e + \tilde{x}_\ell) (\tilde{x}_e - \tilde{x}_\ell) \left( 2 E_\mu^2 - m_e T - m_\mu^2 + m_e^2 \right) + (\tilde{x}_e - \tilde{x}_\ell)^2 \left( T - m_e \right) \left( 2 E_\mu - T + m_e \right) \right].~~
\end{align}
 
\begin{figure}[htb]
\includegraphics[width=0.49\textwidth]{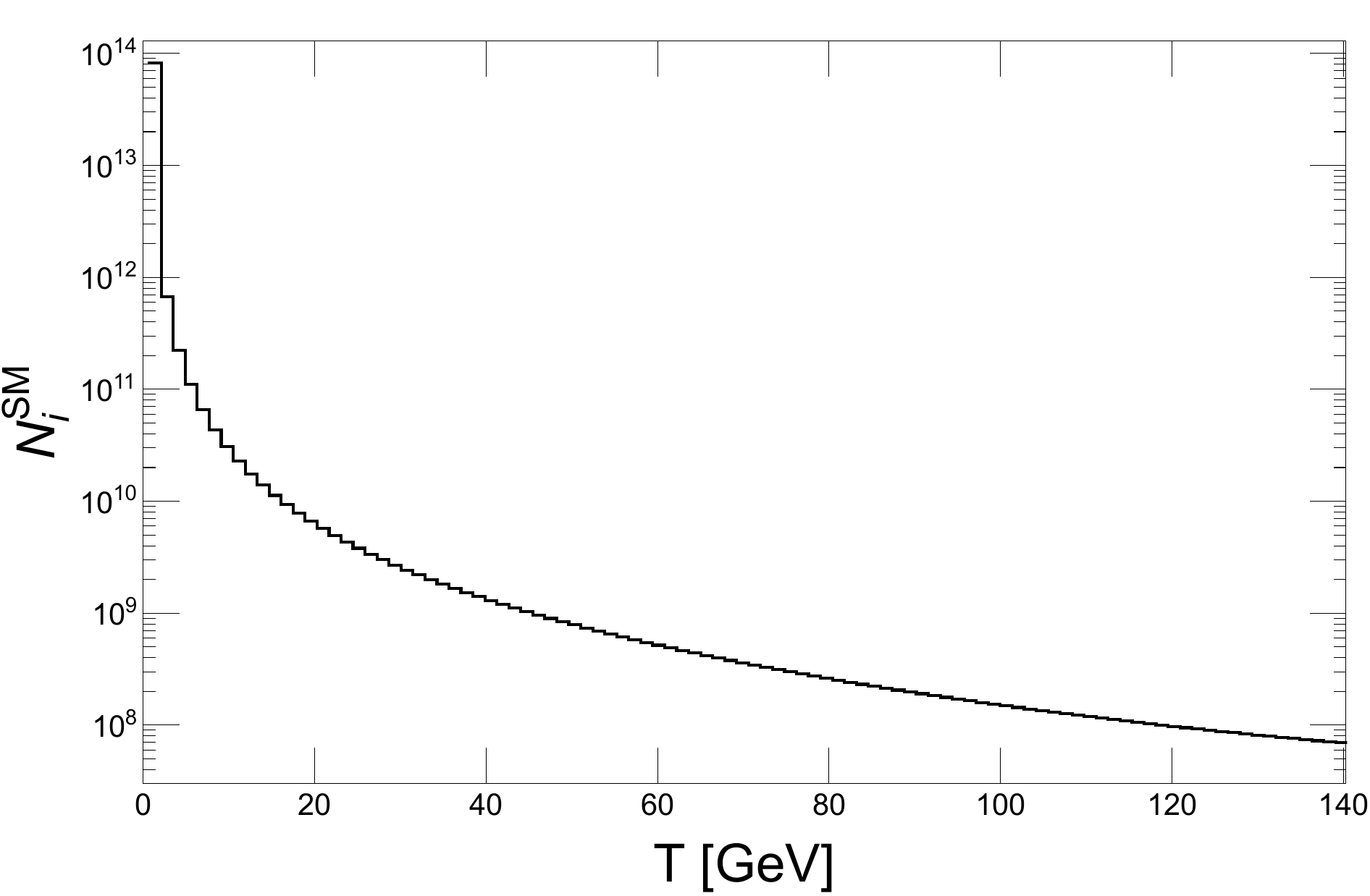}
\includegraphics[width=0.49\textwidth]{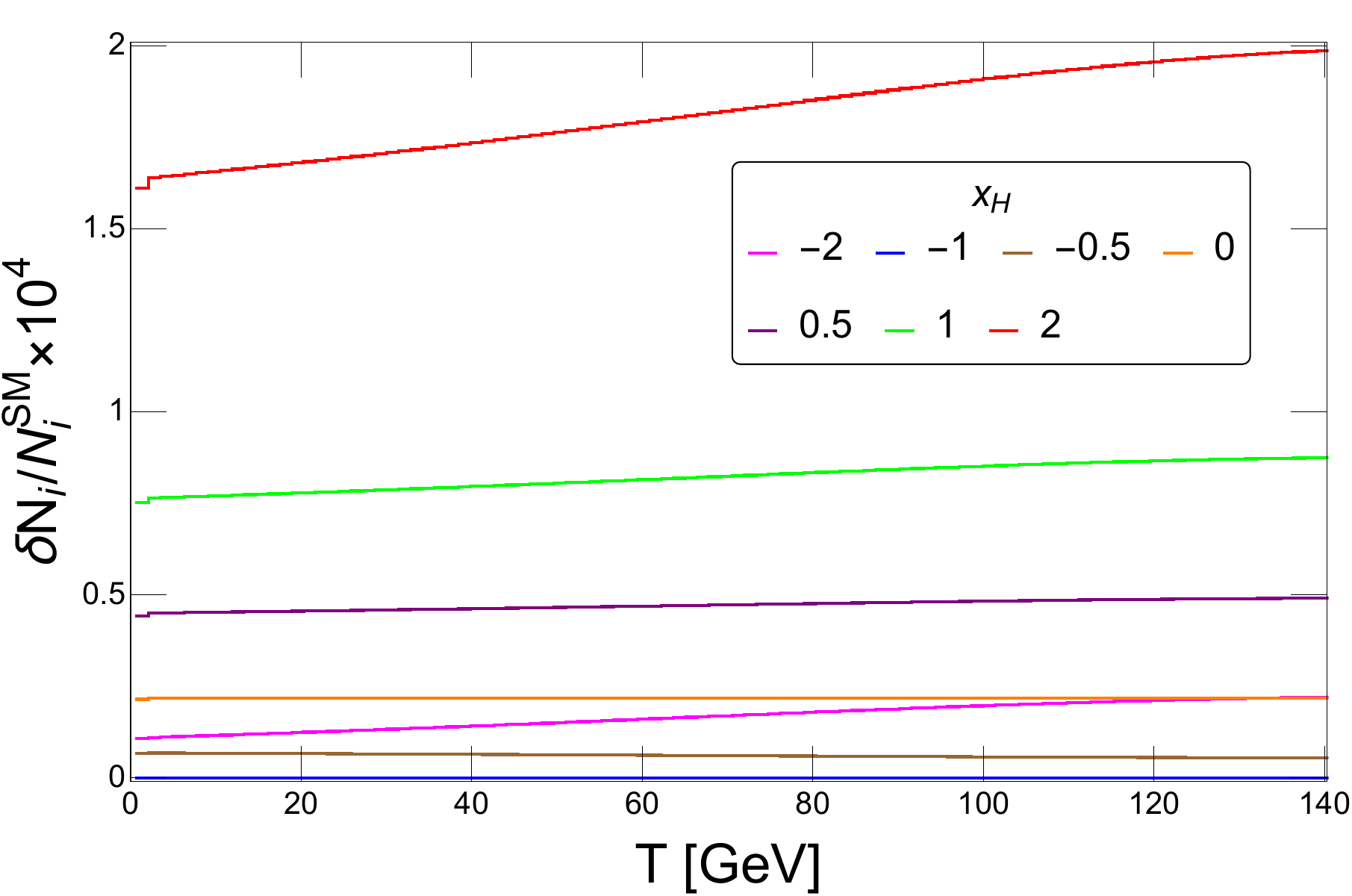}\\
\includegraphics[width=0.49\textwidth]{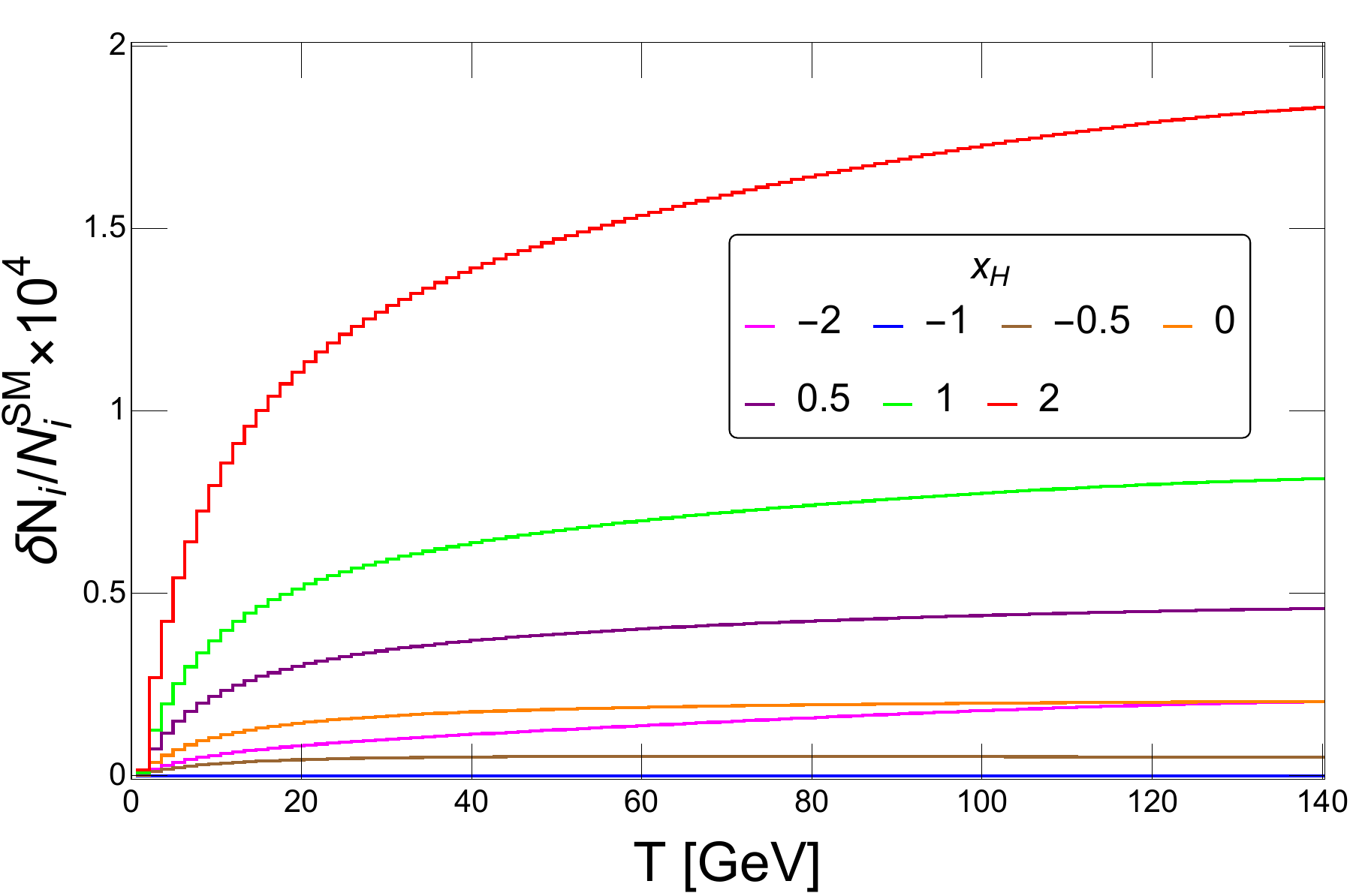}
\includegraphics[width=0.49\textwidth]{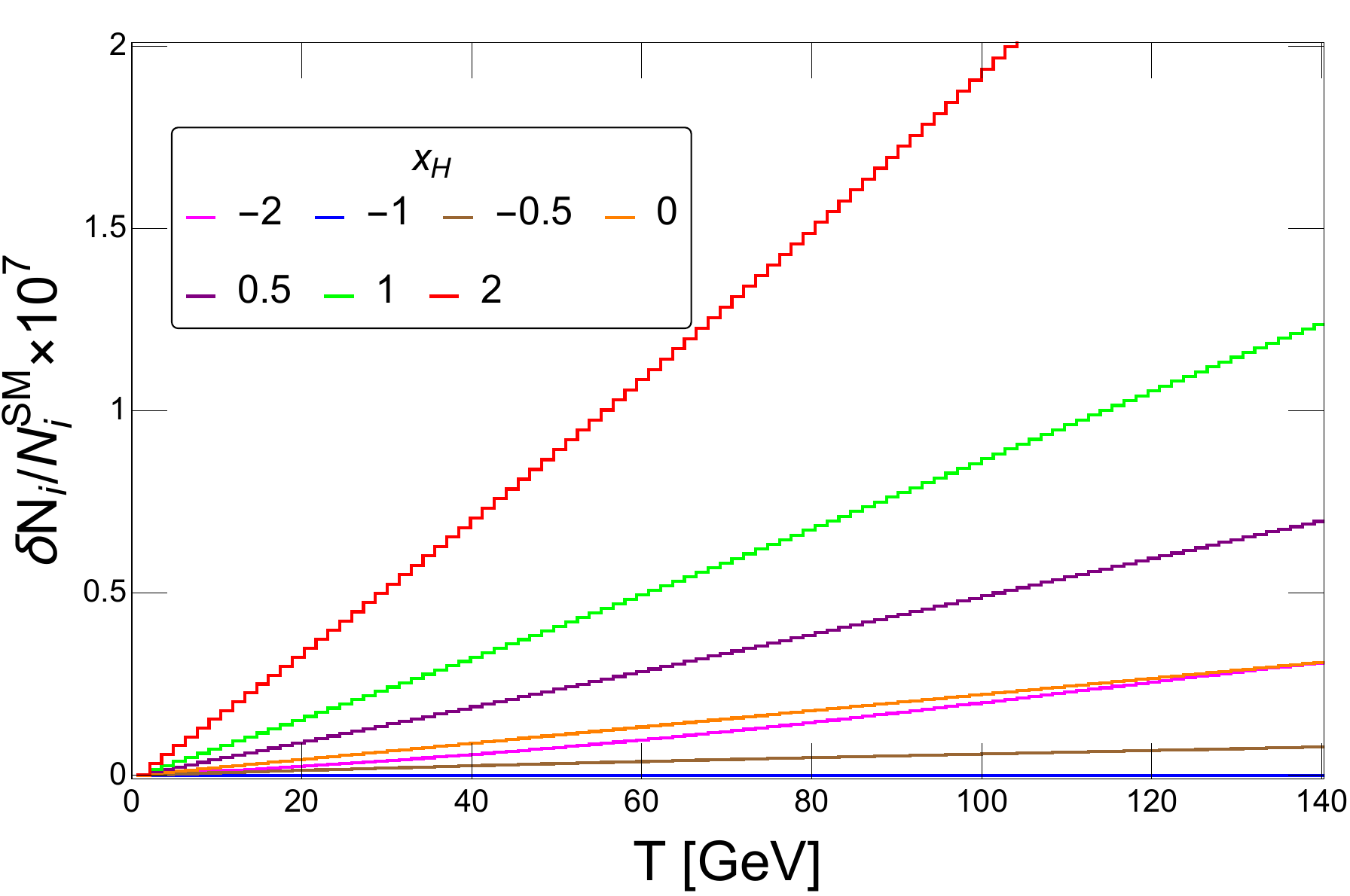}
\caption{
Top-left panel~:~the distribution of the number of events contributed only by the SM particles in MUonE.
Top-right, bottom-left, and bottom-right panels~:~the distributions of the deviation of the number of events in MUonE from that contributed only by the SM particles~:~$\delta N_i/N_i^{\rm SM} \equiv (N_i^{{\rm SM}+Z'} - N_i^{\rm SM}) / N_i^{\rm SM}$ for $(M_{Z'}/{\rm GeV}, g_X) = (10^{-3}, 10^{-3}), (10^{-1}, 10^{-3}), (10, 10^{-3})$, respectively.
}
\label{fig:event_muone}
\end{figure}

The number of elastic scattering signals in $i$-th bin of the electron recoil energy ($T_i < T < T_i + \Delta T$) is evaluated by~\cite{Dev:2020drf}
\begin{align}
\label{eq:eventnum_muone}
    N_i =
    \mathcal{L} \int_{T_i}^{T_i + \Delta T} \dd T \frac{\dd \sigma(\mu^+e^-)}{\dd T} \Theta(T) \Theta\left( T_{\rm max}(E_\mu) - T\right )~,
\end{align}
where $\mathcal{L}$ is the integrated luminosity, and $\mathcal{L} = 150$\,fb$^{-1}$ for MUonE. 
In Eq.~\eqref{eq:eventnum_muone}$, \Theta(T)$ is the Heaviside step function, and $T_{\rm max}(E_\mu)$ stands for the maximal value of the electron recoil energy determined by
\begin{align}
   T_{\rm max}(E_\mu) = 
   \frac{2 m_e (E_\mu^2 - m_\mu^2)}{2 E_\mu m_e + m_e^2 + m_\mu^2}~.
\end{align}
In Fig.~\ref{fig:event_muone}, the number of events contributed only by the SM particles and the deviation of the number of events in chiral $Z'$ model from that in the SM are shown.

In this paper, we evaluate the sensitivity of MUonE to the chiral $Z'$ gauge boson by the $\chi^2$-function, following Ref.~\cite{Dev:2020drf}.
The $\chi^2$-function is calculated by
\begin{align}
    \chi^2 =
    \sum_{i=1}^{100} \frac{(N_i - N_i^{\rm SM})^2}{\sigma_{{\rm stat},i}^2 + \sigma_{{\rm sys},i}^2}~,
\end{align}
where $\sigma_{{\rm stat},i} = \sqrt{N_i}$ stands for the statistical uncertainty, and $\sigma_{{\rm sys},i} = 10^{-5} N_i$ stands for the systematic uncertainty at the level of 10\,ppm~\cite{Abbiendi:2016xup}.
The 95\% confidence level sensitivity is obtained by solving $\chi^{2} = 3.84$.

\subsection{Constraints from proton/electron beam dump experiments}

For constraints from proton beam dump experiments NOMAD and CHARM, we obtain bound curves by rescaling the bounds of U(1)$_{B-L}$ case given in ref.~\cite{Bauer:2018onh}. 
The upper bound on $\{m_{Z'}^{}, g_X^{} \}$ plane is approximately derived applying scaling~\cite{Ilten:2018crw,Chakraborty:2021apc}%
\begin{equation}
\label{eq:upper}
\tau_{Z'}^{}(g_{B-L}^{\rm max}) \sim \tau_{Z'}^{}(g_X^{\rm max}, x_H^{}, x_\Phi^{})~,
\end{equation} 
where $g_{B-L}^{}$ denotes the gauge coupling in the U(1)$_{B-L}$ case, and $\tau_{Z'}^{}$ is the lifetime of the $Z^\prime$.
The lower bound is also scaled by applying
\begin{equation}
\label{eq:lowerP}
g_X^{\rm low} \sim g_{B-L}^{\rm low} \sqrt{ \frac{{\rm BR}(M \to Z'_{B-L} \gamma)~ {\rm BR}(Z'_{B-L} \to e^+ e^-) \tilde \tau_{Z'}^{} }{{\rm BR}(M \to Z' \gamma)~ {\rm BR}(Z' \to e^+ e^-) \tilde \tau_{Z'_{B-L} }} }~,
\end{equation}
where $\tilde \tau$ is lifetime with gauge coupling being unity, and
$Z'$ is produced via meson decay with 
$M = \pi^0$ for MONAD and $M = \eta$ for CHARM dominantly. 
The ratio of meson decay branching ratio thus represents the ratio of $Z'$ production cross sections.
Here, meson decay branching ratio is estimated using the 
method given in ref.~\cite{Ilten:2018crw}. In addition to that we use the same rescaling technique to estimate the constraints on $\{M_{Z'}, g_X \}$ plane from the FASER \cite{FASER:2023zcr} and NA62 \cite{NA62:2023qyn} experiments for different $U(1)_X$ charges where the ratio of $Z'$ production cross section between dark photon and $U(1)_X$ cases is estimated by calculating bremsstrahlung process as a good approximation using method in ref.~\cite{Asai:2022zxw}. 

For proton beam dump experiment $\nu$-cal, the bremsstrahlung process dominantly produces $Z'$ boson. 
In this work we use the excluded region given in ref.~\cite{Asai:2022zxw} on $\{M_{Z'}, g_X \}$ space where we take into account chiral structure of the $Z'$ interactions in estimating the $Z'$ production cross section.

For electron beam dump constraints from Orsay and KEK,  
The bound curves are obtained by rescaling the bounds of U(1)$_{B-L}$ case as for the proton beam dump one. 
We derive the constraint on the upper region on $\{M_{Z'}^{}, g_X^{} \}$ plane approximately by scaling with Eq.~\eqref{eq:upper}, which is the same as the proton beam dump case.
On the other hand, the constraint on the lower region is estimated by~\cite{Chakraborty:2021apc}
\begin{equation}
\label{eq:bremsstrahlung}
g_X^{\rm low} \sim g_{B-L}^{\rm low} \sqrt{ \frac{2 {\rm BR}(Z'_{B-L} \to e^+ e^-) \tilde \tau_{Z'}^{} }{(5x_H^{2}/4 + 3x_H x_\Phi + 2x_\Phi^{2}) {\rm BR}(Z' \to e^+ e^-) \tilde \tau_{Z'_{B-L}}^{}} }~,
\end{equation}
where $Z'$ is considered to be produced via bremsstrahlung process.

For electron beam dump constraints from E137 and E141 experiments,  
we use the results in ref.~\cite{Asai:2022zxw} for excluded region on $\{M_{Z'}, g_X \}$ in which chiral structure of the $Z'$ interactions is taken into account in estimating the $Z'$ production cross section via bremsstrahlung process. 

\subsection{Constraints from electron-(anti)neutrino scattering in neutrino experiments}

Here we discuss constraints on $\{M_{Z'}, g_X \}$ from electron-(anti)neutrino scattering processes that are tested by neutrino experiments: BOREXINO, TEXONO, GEMMA, CHARM-II and the J-PARC Sterile Neutrino Search at the J-PARC Spallation Neutron Source(JSNS$2$). 
To obtain the constraints we estimate the electron-(anti)neutrino scattering cross sections under the existence of $Z'$ interactions. 
The differential cross section can be expressed as 
\begin{equation}
\label{eq:nu-e-scattering}
\frac{d \sigma (\nu  e)}{d T}  =
\left. \frac{d \sigma (\nu  e)}{d T} \right|_{\rm SM} + 
\left. \frac{d \sigma (\nu  e)}{d T} \right|_{Z'} + 
\left. \frac{d \sigma (\nu  e)}{d T} \right|_{\rm Int} 
\end{equation}
where $T$ denotes the electron recoil energy and the first, second and third terms in the RHS correspond to the contributions from the only SM interactions, the only $Z'$ interactions and interference between the SM and $Z'$ interactions. 
The terms in RHS are given as follows~\cite{Chakraborty:2021apc}.
The purely SM contribution can be written by
\begin{align}
\label{eq:nu-e-scattering-SM}
\left. \frac{d \sigma (\nu  e)}{d T} \right|_{\rm SM} & = \frac{2 G_F^2 m_e}{\pi E_\nu^2} \left( a^2_1 E^2_\nu + a_2^2 (E_\nu - T)^2 - a_1 a_2 m_e T \right), 
\end{align}
where $E_\nu$ is the energy of initial neutrino.
Here $a_1$ and $a_2$ are given by 
\begin{align}
a_1 &= \left\{\sin^2 \theta_W + \frac12, \ \sin^2 \theta_W, \ \sin^2 \theta_W - \frac12, \ \sin^2 \theta_W \right\} \ {\rm for} \ \{\nu_e e, \bar \nu_e e, \nu_\beta e, \bar \nu_\beta e\}, \\
a_2 &= \left\{\sin^2 \theta_W, \ \sin^2 \theta_W +  \frac12, \ \sin^2 \theta_W, \ \sin^2 \theta_W -\frac12 \right\} \ {\rm for} \ \{\nu_e e, \bar \nu_e e, \nu_\beta e, \bar \nu_\beta e\},
\end{align}
where $\beta = \{\mu, \tau \}$.
The contribution from $Z'$ exchanging diagram is estimated by 
\begin{align}
\left. \frac{d \sigma (\overset{(-)}{\nu}_\alpha e )}{d T} \right|_{Z'} & = \frac{g_X^4 (\tilde{x}_\ell)^2 m_e}{4 \pi E_\nu^2 (2m_e T + M^2_{Z'})} [(2 E^2_\nu - 2 E_\nu T + T^2)(b^2_1 +b^2_2) \pm 2 b_1 b_2(2 E_\nu - T)T - m_e T(b^2_1 - b_2^2)],
\end{align}
where $b_1 = \frac{\tilde{x}_\ell + \tilde{x}_e}{2}$ and $b_2 = \frac{\tilde{x}_\ell - \tilde{x}_e}{2}$ with $\tilde{x}_{\ell, e}$ from Table.~\ref{tab1}, and the negative sign of $\pm$ is for the process of anti neutrino.
The contributions from interference between the SM and $Z'$ are also written, depending on the process, as follows:
\begin{align}
\left. \frac{d \sigma (\nu_e e)}{d T} \right|_{\rm int}  = & \frac{G_F g_X^2 \tilde{x}_\ell m_e}{\sqrt{2} \pi E^2_\nu (2m_e T +M^2_{Z'}) } [2 E_\nu^2(b_1 + b_2) + (2 E^2_\nu - 2 E_\nu T + T^2)(b_1 c_1 + b_2 c_2)] \nonumber \\
& + T (2E_\nu - T)(b_1 c_2 + b_2 c_1) - m_e T(b_1 -b_2 + b_1 c_1 - b_2 c_2)],  \\
\left. \frac{d \sigma (\bar{\nu}_e e)}{d T} \right|_{\rm int}  = & \frac{G_F g_X^2 \tilde{x}_\ell m_e}{\sqrt{2} \pi E^2_\nu (2m_e T +M^2_{Z'}) } [2 (E_\nu - T)^2(b_1 + b_2) + (2 E^2_\nu - 2 E_\nu T + T^2)(b_1 c_1 + b_2 c_2)] \nonumber \\
& - T (2E_\nu - T)(b_1 c_2 + b_2 c_1) - m_e T(b_1 -b_2 + b_1 c_1 - b_2 c_2)],  \\
\left. \frac{d \sigma (\overset{(-)}{\nu}_\beta e )}{d T} \right|_{\rm int} = & \frac{G_F g_X^2 \tilde{x}_\ell m_e}{\sqrt{2} \pi E^2_\nu (2m_e T +M^2_{Z'}) } [(2E_\nu^2 -2 E_\nu T - T^2)2(b_1c_1 + b_2 c_1) \pm T(2 E_\nu - T)(b_1 c_2 + b_2 c_1)] \nonumber \\
& - m_e T(b_1 c_1 - b_2 c_2)],  
\end{align}
where $c_1 = -1/2 + 2 \sin^2 \theta_W$ and $c_2 = -1/2$.
We then estimate the differential cross sections and derive the constraints for each experiment in the following way.

\noindent
{\it \bf BOREXINO}: The cross section of $\nu_e$-$e$ scattering process is estimated by the experiment where $\langle E_\nu \rangle = 862$ keV and $T \simeq [270, 665]$ keV for $^7B_e$ solar neutrino. We require the cross section with $Z'$ interaction should not be more than $8\%$ above that of the SM prediction~\cite{Bellini:2011rx} to obtain the constraint on $\{M_{Z'}, g_X \}$. 

\noindent
{\it \bf TEXONO}: $\bar \nu_e$-$e$ scattering process is measured by the experiment using 187 kg of CsI(Tl) scintillating crystal array with 29882/7369 kg-day of reactor ON/OFF data with electron recoil energy of $T \simeq [3, 8]$ MeV. 
The $\chi^2$ value is estimated as 
\begin{equation}
\chi^2 = \sum_{\rm bin} \frac{(R_{\rm data} - R_{\rm th})^2}{\Delta R^2},
\label{eq:chi2-2}
\end{equation}
where $R_{\rm data}$ and $R_{\rm th}$ are the event ratios measured by the experiment and predicted by the cross section in Eq.~\eqref{eq:nu-e-scattering}, and $\Delta R$ is the experimental uncertainty, for each recoil energy bin taken from data in ref.~\cite{TEXONO:2009knm}. Here we also applied anti neutrino flux in the reference.
The constraint on $\{M_{Z'}, g_X \}$ plane is then obtained by  $\chi^2$ analysis with $90 \%$ C.L. 

\noindent
{\it \bf GEMMA}: $\bar{\nu}_e$-$e$ scattering is observed with 1.5 kg HPGe detector where energy of neutrino is $\langle E_\nu \rangle \sim 1$-$2$ MeV 
and flux is $2.7 \times 10^{13}$ cm$^{-2}$s$^{-1}$. 
The $\chi^2$ value is estimated applying the formula Eq.~\eqref{eq:chi2-2} for the data given in ref.~\cite{Beda:2010hk} with 13000 ON-hours and 3000 OFF-hours,
and we derive the upper limit curve on $\{M_{Z'}, g_X \}$ with $90 \%$ C.L.

\noindent
{\it \bf CHARM-II}: $\nu_\mu (\bar \nu_\mu)$-electron scattering is observed where $2677\pm82$ and $2752\pm88$ events are respectively obtained for $\nu_\mu$ and $\bar \nu_\mu$ cases.
The mean neutrino energies for $\nu_\mu$ and $\bar \nu_\mu$ are respectively $\langle E_{\nu_\mu} \rangle = 23.7$ GeV and $\langle E_{\bar \nu_\mu} \rangle = 19.1$ GeV, and the range of observed recoil energy is 3-24 GeV.
The  $\chi^2$ value is estimated using the formula Eq.~\eqref{eq:chi2-2} for the data given in ref.~\cite{CHARM-II:1993phx, CHARM-II:1994dzw}, 
and we derive the upper curve on $\{M_{Z'}, g_X \}$ with $90 \%$ C.L.

\noindent
{\it \bf JSNS$2$}: In the experiment, 3 GeV proton collides with mercury target producing pions giving neutrino beams. We consider $\nu_e$-$e$ and $\bar \nu_\mu$-$e$ scattering processes to obtain the constraint. We estimate number of scattering events applying $3.8 \times 10^{22}$ protons on target per year, 17 tons of a gadolinium(Gd)-loaded liquid-scintillator detector (LS) detector and neutrino fluxes given in ref.~\cite{Ajimura:2017fld}.
The corresponding $\chi^2$ is estimated by 
\begin{equation}
\chi^2 = \underset{\alpha}{\rm min} \left[ \frac{(N_{\rm th} - (1+ \alpha) N_{\rm SM})^2}{N_{\rm th}} + \left( \frac{\alpha}{\sigma_{\rm norm}} \right)^2 \right],
\end{equation}
where $N_{\rm th}$ and $N_{\rm SM}$ are the expected number of events in our models and in the SM for 1 year, $\sigma_{\rm norm}$ is the systematic uncertainty in the neutrino flux normalization, and $\alpha$ is nuisance parameter. 
Here $\sigma_{\rm norm}$ is assumed to be $5 \%$ as the reference value.
Then the future sensitivity on $\{M_{Z'}, g_X \}$ plane is estimated by requiring $\chi^2$ value to be less than that of $90 \%$ C.L.

\subsection{Limits from coherent neutrino-nucleus scattering}

An upper limit curve on $\{M_{Z'}, g_X \}$ is also obtained from coherent elastic neutrino-nucleus scattering (CE$\nu$NS) that is measured by COHERENT experiment with CsI and Ar targets~\cite{COHERENT:2021xmm,COHERENT:2020iec,COHERENT:2017ipa}.
Here we derive the curve by rescaling the limit curve for $U(1)_{B-L}$ case given in Refs.~\cite{Melas:2023olz,Cadeddu:2020nbr} by comparing number of events in $U(1)_{B-L}$ and other cases.
We estimate the number of events at COHERENT experiment adopting formulas in the reference, as discussed below.

Firstly, the differential cross section for CE$\nu$NS process is estimated by~\cite{Barranco:2005yy,Patton:2012jr} 
\begin{equation}
\frac{d \sigma_{\nu-N}}{d T} (E,T) = \frac{G_F^2 M}{\pi} \left(1 - \frac{M T}{2 E^2} \right) Q_{{\rm SM}+Z'}^2,
\end{equation} 
where $E$ is the initial neutrino energy, $T$ is the recoil energy,  $M$ is the mass of target nucleus and $Q_{{\rm SM}+Z'}$ is the factor coming from interactions including SM and $Z'$ gauge bosons.
In our models $Q_{{\rm SM}+Z'}$ is given by 
\begin{equation}
Q_{{\rm SM}+Z'} = \left( g^p_V(\nu_\ell) + 2 \epsilon^{u V}_{\ell \ell} + \epsilon^{d V}_{\ell \ell} \right) Z F_Z(|{\bf q}^2|) + \left( g^n_V(\nu_\ell) +  \epsilon^{u V}_{\ell \ell} + 2\epsilon^{d V}_{\ell \ell} \right) N F_N(|{\bf q}^2|),
\end{equation}
where $g^{p(n)}_V$ is the neutrino-proton(neutron) coupling in the SM, $Z(N)$ is the number of proton(neutron) in the target nucleus, and $F_{Z(N)}(|{\bf q}^2|)$ is the from factors of the proton(neutron) for the target nucleus.
The $\epsilon^{q V}_{\ell \ell}$ is effective coupling explicitly given by
\begin{equation}
\epsilon^{q V}_{\ell \ell} = \frac{g_X^2 \tilde{x}_\ell \tilde{x}_q}{\sqrt{2} G_F ({\bf q}^2 + M^2_{Z'})}.
\end{equation}
We adopt the values of $g_V^p (\nu_e) = 0.0401$, $g_V^p=0.0318$ and $g_V^n = - 0.5094$ for the neutrino-proton(neutron) coupling in the SM~\cite{Cadeddu:2020lky,Erler:2013xha}. 
The Helm parametrization~\cite{Helm:1956zz} is applied for the form factors $F_{Z(N)}(|{\bf q}^2|)$ using proton rms radii $\{R_{p} ({\rm Cs}), R_p ({\rm I}), R_p ({\rm Ar}) \}  = \{4.804, 4.749, 3.448\}$ [fm] and 
neutron rms radii $\{R_{n} ({\rm Cs}), R_n ({\rm I}), R_n ({\rm Ar}) \}  = \{5.01, 4.94, 3.55\}$ [fm]~\cite{Fricke:1995zz,Angeli:2013epw,Bender:1999yt}.

Then we adopt the neutrino fluxes for the CE$\nu$NS event rate in the experiment which depend on the neutrino fluxes produced from the Spallation Neutron Source (SNS) at the Oak Ridge National 
Laboratories. They are given by~\cite{COHERENT:2017ipa,COHERENT:2020iec}
\begin{align}
\frac{d N_{\nu_\mu}}{d E} &= \eta \delta \left( E - \frac{m_\pi^2 - m_\mu^2}{2 m_\pi} \right), \\
\frac{d N_{\nu_{\bar \mu}}}{d E} &= \eta \frac{64 E^2}{m_\mu^3}   \left( \frac34 - \frac{E}{m_\mu} \right), \\
\frac{d N_{\nu_e}}{d E} &= \eta \frac{192 E^2}{m^3_\mu} \left( \frac12 - \frac{E}{m_\mu} \right), 
\end{align}
where $\eta = r N_{\rm POT}/(4 \pi L^2)$ with $L$, $N_{\rm POT}$ and $r$ being respectively the distance between the source and the detector, 
the number of proton-on-target (POT), and the number of neutrinos per flavor that are produced for each POT.
For these values, we use $r = 9 \times 10^{-2}$, $N_{\rm POT} = 13.7 \times 10^{22}$ and $L = 27.5$ m for Ar detector, and 
$r = 0.08$, $N_{\rm POT} = 17.6 \times 10^{22}$ and $L = 19.5$ m for CsI detector, respectively.

Finally the theoretical number of events for each energy bin in the experiment is derived from
\begin{equation}
N_i = N (\mathcal{N}) \int^{T_{i+1}}_{T_i} d T A(T) \int_{E_{\rm min}}^{E_{\rm max}} d E \sum_{\nu = \nu_e, \nu_{\mu}, \nu_{\bar \mu}} \frac{d N_\nu}{d E} \frac{d \sigma_{\nu-N}}{d T} (E, T),
\end{equation}
where $i$ corresponds to each recoil energy bin,  $E_{\rm min(max)} = \sqrt{M T/2} (m_\mu/2)$, and $A(T)$ is the energy-dependent reconstruction efficiency.
We estimate upper limit of the coupling $g_X$ for each $Z'$ mass for different models by rescaling that of $U(1)_{B-L}$ case in ref.~\cite{Cadeddu:2020nbr} 
by comparing the number of events for upper limit of $g_X$ in $U(1)_{B-L}$ with the number of events in each model.


\subsection{Limits from collider experiments: LEP-II, CMS, LHCb and BarBar}

Here we briefly summarize our estimation of limit on $\{M_{Z'}, g_X \}$ from results of several collider experiments:

{\it (i) The limit from LEP-II}: We estimate upper limit curve on  $\{M_{Z'}, g_X \}$ from the results of LEP-II~\cite{ALEPH:2013dgf,ALEPH:2005ab} that measure $e^+ e^- \to \bar f f$ scattering cross sections at the $Z$ peak with $f$ being the SM fermions.
The scattering cross sections are estimated including $Z'$ exchanging diagram in addition to the SM processes. 
Then the value of cross section is compared with the observed value.
As a result, we consider constraints coming from the cross section of $e^+ e^- \to q \bar q$ process giving hadronic final state whose value is $\sigma = 41.544 \pm 0.037$ nb, and $e^+ e^- \to \ell^+  \ell^-$ process with $R_\ell = \Gamma_{had}/\Gamma_\ell =20.768 \pm 0.024$.
We thus obtain upper limit of $g_X$ as function of $m_{Z'}$ when we require the total cross section is within the 90\% C.L. of the observed value.

{\it (ii) The limit from dark photon search at the LHC experiments}: 
The bounds on $\{M_{Z'}, g_X \}$ are also obtained from the results of CMS~\cite{CMS:2023slr} and LHCb~\cite{LHCb:2019vmc} experiments searching for dark photon $A'$ which decays into $\mu^+ \mu^-$ pair.
 They provide us the bounds on the mass of dark photon $m_{A'}$ and kinetic mixing parameter $\epsilon$.
 The upper limit on $g_X$ as a function of $M_{Z'}$ can be estimated in use of following re-scaling 
 \begin{equation}
 g_X^{\rm max}(M_{Z'} = m_{A'}) = \epsilon^{\rm max}(m_{A'}) e \sqrt{\frac{\sigma(pp \to A') BR(A' \to \mu^+ \mu^-)}{\sigma(pp \to Z')BR(Z' \to \mu^+ \mu^-)} },
 \end{equation}
 where $\sigma(pp \to A'(Z') )$ is dark photon($Z'$) production cross section estimated by {\it CalcHEP3.5} \cite{Belyaev:2012qa} implementing relevant interactions,
  and $\epsilon^{\rm max} (m_{A'})$ is the upper limit of kinetic mixing parameter as a function of dark photon mass.

{\it (iii) The limit from BaBar experiment}:
 At the BaBar experiment, dark photon is searched for via the process $e^+ e^- \to A' \gamma$~\cite{BaBar:2014zli,BaBar:2017tiz}.
 They consider visible $A'$ which decays into $\{e^+e^-, \mu^+ \mu^-, \text{light mesons} \}$ final states and invisible $A'$ which decays into invisible final sate such as neutrinos.
 To estimate the bounds on $g_X$, we rescale the upper limit of gauge coupling in $U(1)_{B-L}$ case given in ref.~\cite{Bauer:2018onh} as function of $Z'$ mass. 
The rescaling formula for visible $Z'$ decay mode is 
 \begin{equation}
  g_X^{\rm max}(M_{Z'} ) = g_{B-L}^{\rm max}(M_{Z'})  \sqrt{\frac{\sigma(e^+e^- \to \gamma Z'_{B-L}) BR(Z'_{B-L} \to \text{visible states})}{\sigma(e^+ e^- \to \gamma Z')BR(Z' \to \text{visible states})} },
 \end{equation}
 where $Z'_{B-L}$ indicate $Z'$ boson in case of $U(1)_{B-L}$.
The rescaling formula for invisible $Z'$ decay mode is 
 \begin{equation}
  g_X^{\rm max}(M_{Z'} ) = g_{B-L}^{\rm max}(M_{Z'})  \sqrt{\frac{\sigma(e^+e^- \to \gamma Z'_{B-L}) BR(Z'_{B-L} \to \bar \nu \nu)}{\sigma(e^+ e^- \to \gamma Z')BR(Z' \to \bar \nu \nu) } },
 \end{equation}
 where all neutrino modes are taken into account.

\subsection{Limits from electron and muon $(g-2)$}

The $Z'$ boson contributes to electron and muon $(g-2)$, $\Delta a_{\mu, e}$, at one loop level.
Calculating one-loop diagram, we obtain the formula as~\cite{Jegerlehner:2009ry}
\begin{equation}
\Delta a_{\ell} = \frac{g_X^2}{8 \pi^2} \frac{m_\ell^2}{M_{Z'}^2} \int^1_0 dx \frac{(\tilde{x}_\ell + \tilde{x}_e)^2 2x^2(1-x) +(\tilde{x}_\ell - \tilde{x}_e)^2 \left(2x (1-x)(x-4) -4 x^3 \frac{m_\ell^2}{M_{Z'}^2}\right)}{(1-x) \left(1 -x \frac{m_\ell^2}{M_{Z'}^2} \right) + x \frac{m_\ell^2}{M_{Z'}^2} },
\end{equation}
 where $\ell = \{e, \mu\}$.
 As a reference we derive the region that accommodate experimental values of electron and muon $(g-2)$ in the models.
 For electron $(g-2)$, the experimentally obtained ranges are
 \begin{align}
\Delta a_e (^{133}\text{Cs}) & = -(8.8\pm 3.6)\times 10^{-13}~~\text{\cite{Parker:2018vye}}\,, \nonumber \\
\Delta a_e (^{87}\text{Rb}) &= (4.8 \pm 3.0)\times 10^{-13}~~\text{\cite{Morel:2020dww}}\,,
\end{align}
where the corresponding deviation from the SM are $2.4\sigma$ and $1.6\sigma$ respectively.
We consider $\Delta a_e$ constraint on $Z'$ interaction to satisfy either range of $\Delta a_e$ depending on the charge assignment.
For muon $(g-2)$ we apply the experimentally obtained range of~\cite{Muong-2:2021ojo} 
\begin{align}
\Delta a_\mu = (25.1\pm 5.9)\times 10^{-10},
\label{exp_dmu}
\end{align}
that corresponds to 4.2$\sigma$ deviation from the SM prediction.
We show the parameter region that satisfies the $\Delta a_\mu$ range in the models.

\section{Results and Discussions}
\label{sec:result}
\begin{figure}[h]
\begin{center}
\includegraphics[width=1.0\textwidth]{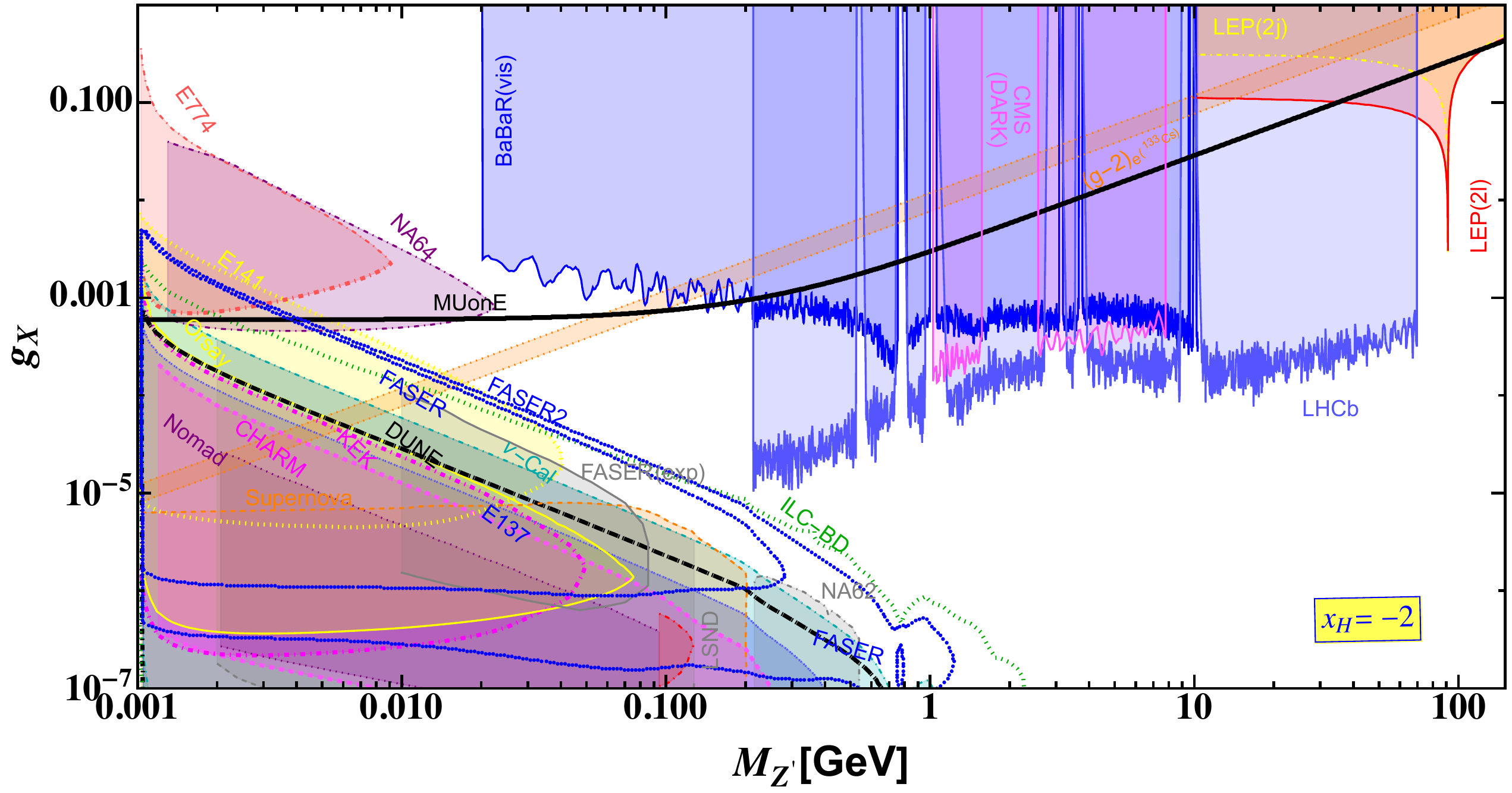}
\includegraphics[width=1.0\textwidth]{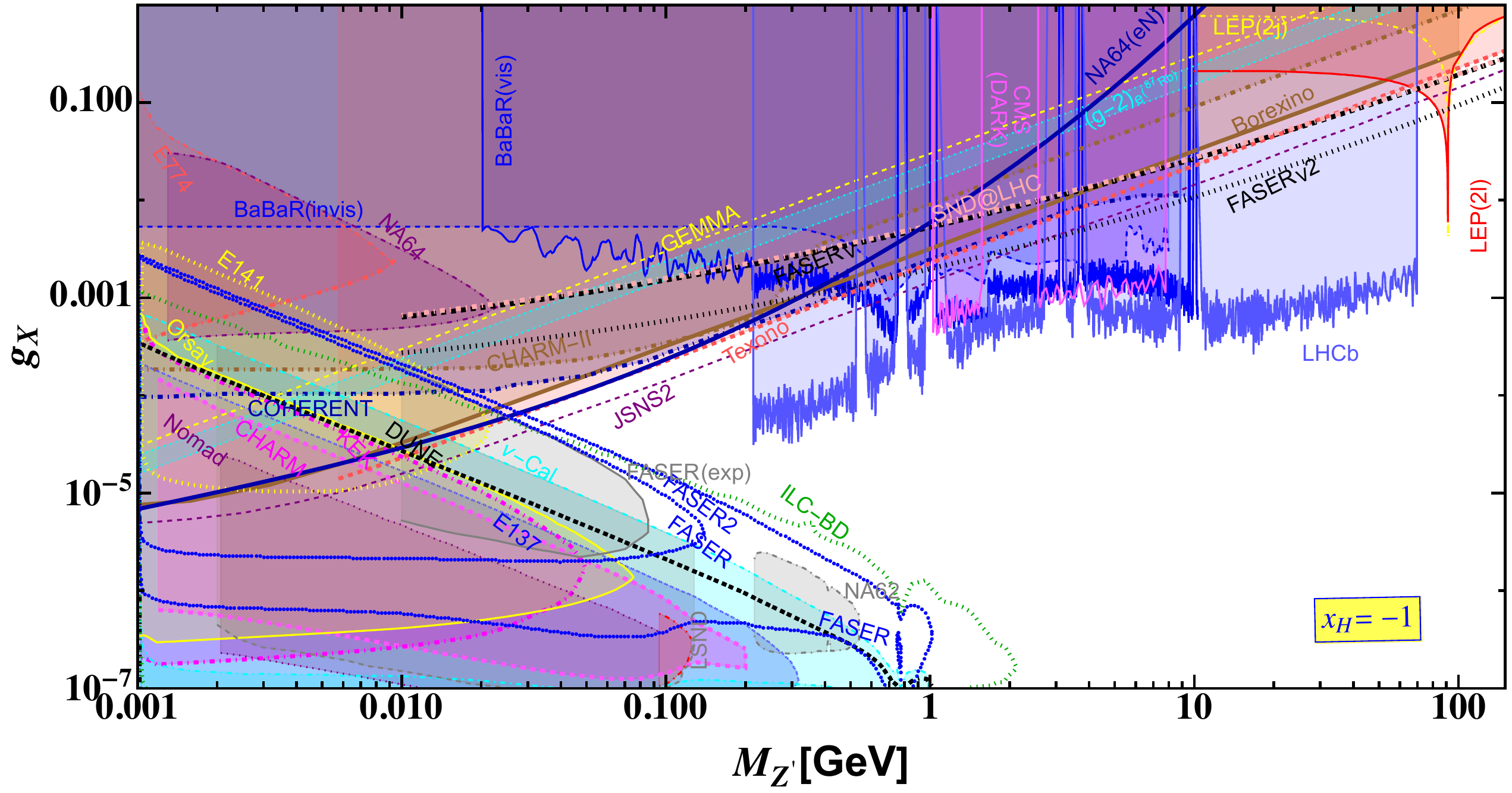}
\caption{Limits on $g_X^{}-M_{Z^\prime}^{}$ plane for $x_H^{} = -2$ (upper panel, $U(1)_R$ case) and $x_H=-1$ (lower panel) taking $x_\Phi^{}=1$ considering $10^{-3}$ GeV $\leq M_{Z^\prime}^{} \leq 150$ GeV. For $x_H=-2$ we show the region sensitive to the MuonE experiment. Recasting the data we compare the parameter region with those we estimated from LEP,  Dark photon searches at BaBaR, LHCb and CMS (CMS Dark) and different beam dump experiments at  Orsay, KEK, E137, CHARM, Nomad, $\nu-$cal, E141, E774, NA64, NA62, FASER and involving prospective bounds from FASER(2), DUNE and ILC-BD, respectively from theoretical analyses. For $x_H=-1$ we show the regions sensitive to FASER$\nu$, FASER$\nu$2, SND$@$LHC, NA64$(eN)$ and JSNS2 experiments respectively. We compare our results recasting limits from LEP, CHARM-II, GEMMA, BOREXINO, COHERENT, TEXONO, Dark photon search at BaBaR (vis and invis), LHCb and CMS (CMS Dark) and different beam dump experiments at Orsay, KEK, E137, CHARM, Nomad, $\nu-$cal, E141, E774, NA64 and involving prospective bounds from FASER(2), DUNE and ILC, respectively. In this case MUonE bound can not be calculated because $e_R$ does not interact with the $Z^\prime$.}
\label{fig:xHm2}
\end{center}
\end{figure}

\begin{figure}[h]
\begin{center}
\includegraphics[width=1.0\textwidth]{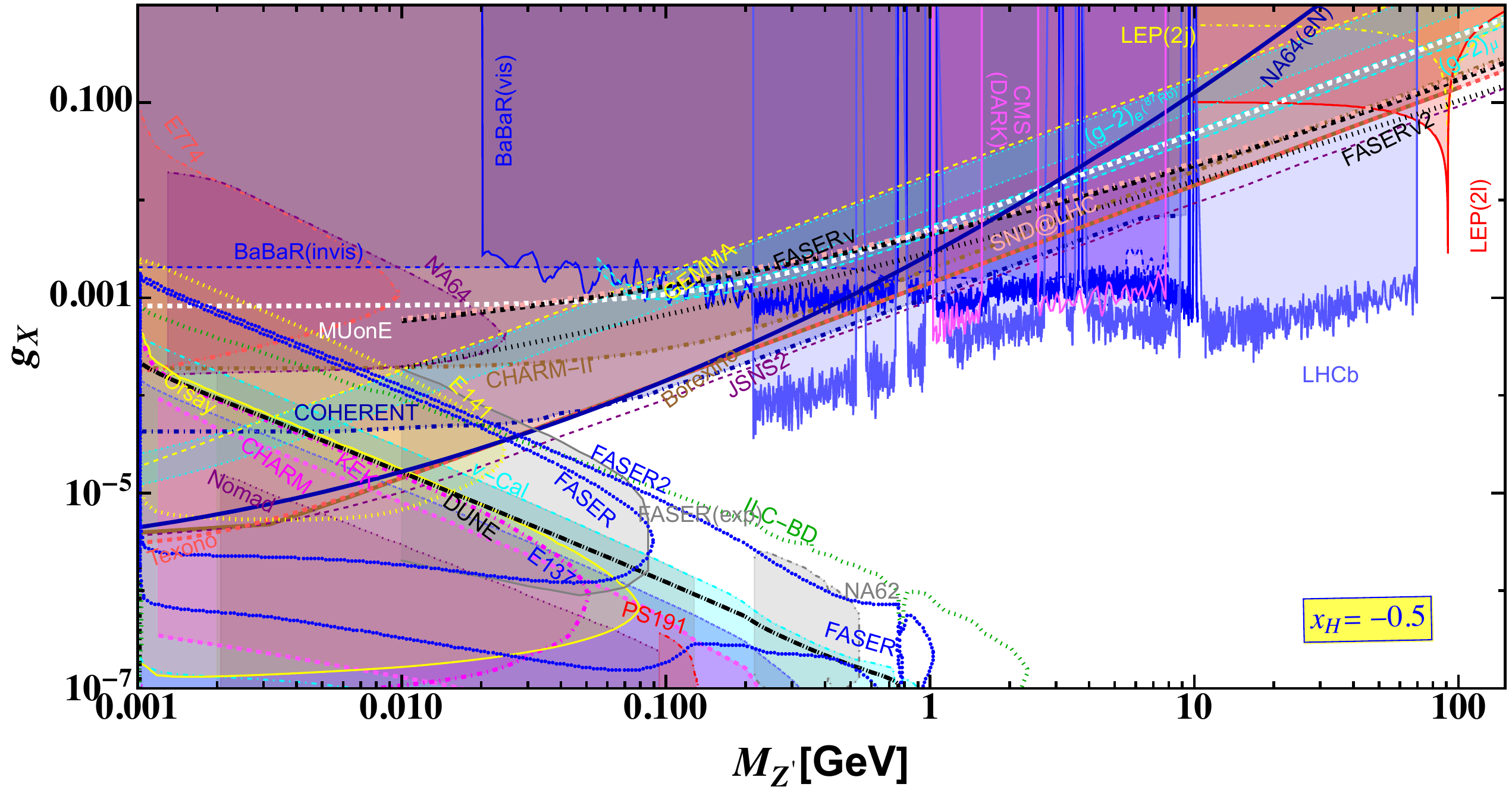}
\includegraphics[width=1.0\textwidth]{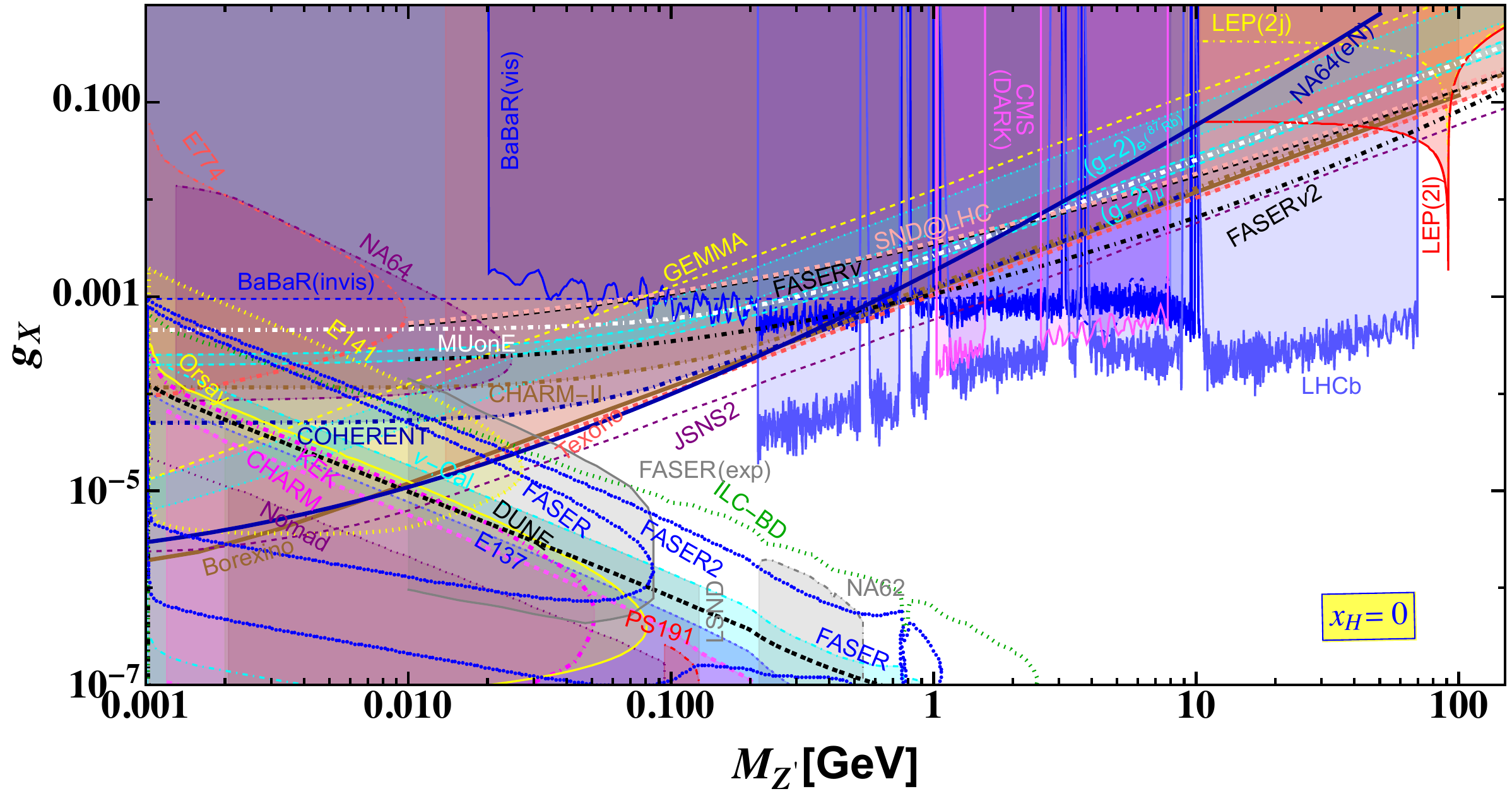}
\caption{Limits on $g_X^{}-M_{Z^\prime}^{}$ plane for $x_H^{} = -0.5$ (upper panel) and $x_H^{}=0$ (lower panel, B-L case) taking $x_\Phi^{}=1$ considering $10^{-3}$ GeV $\leq M_{Z^\prime}^{} \leq 150$ GeV showing the regions sensitive to FASER$\nu$, FASER$\nu$2, SND$@$LHC, NA64$(eN)$ and JSNS2 experiments. Recasting the existing results in our case, we compare parameter regions obtained from LEP, dark photon searches at BaBaR(vis and invis), LHCb and CMS(CMS Dark), scattering experiments CHARM-II, GEMMA, BOREXINO, COHERENT, TEXONO and different beam dump experiments from Orsay, KEK, E137, CHARM, Nomad, $\nu-$cal, E141, E774, NA64, NA62, FASER involving prospective bounds from FASER(2), DUNE and ILC(ILC-BD), respectively from theoretical analyses.}
\label{fig:xHm05}
\end{center}
\end{figure}

\begin{figure}[h]
\begin{center}
\includegraphics[width=1.0\textwidth]{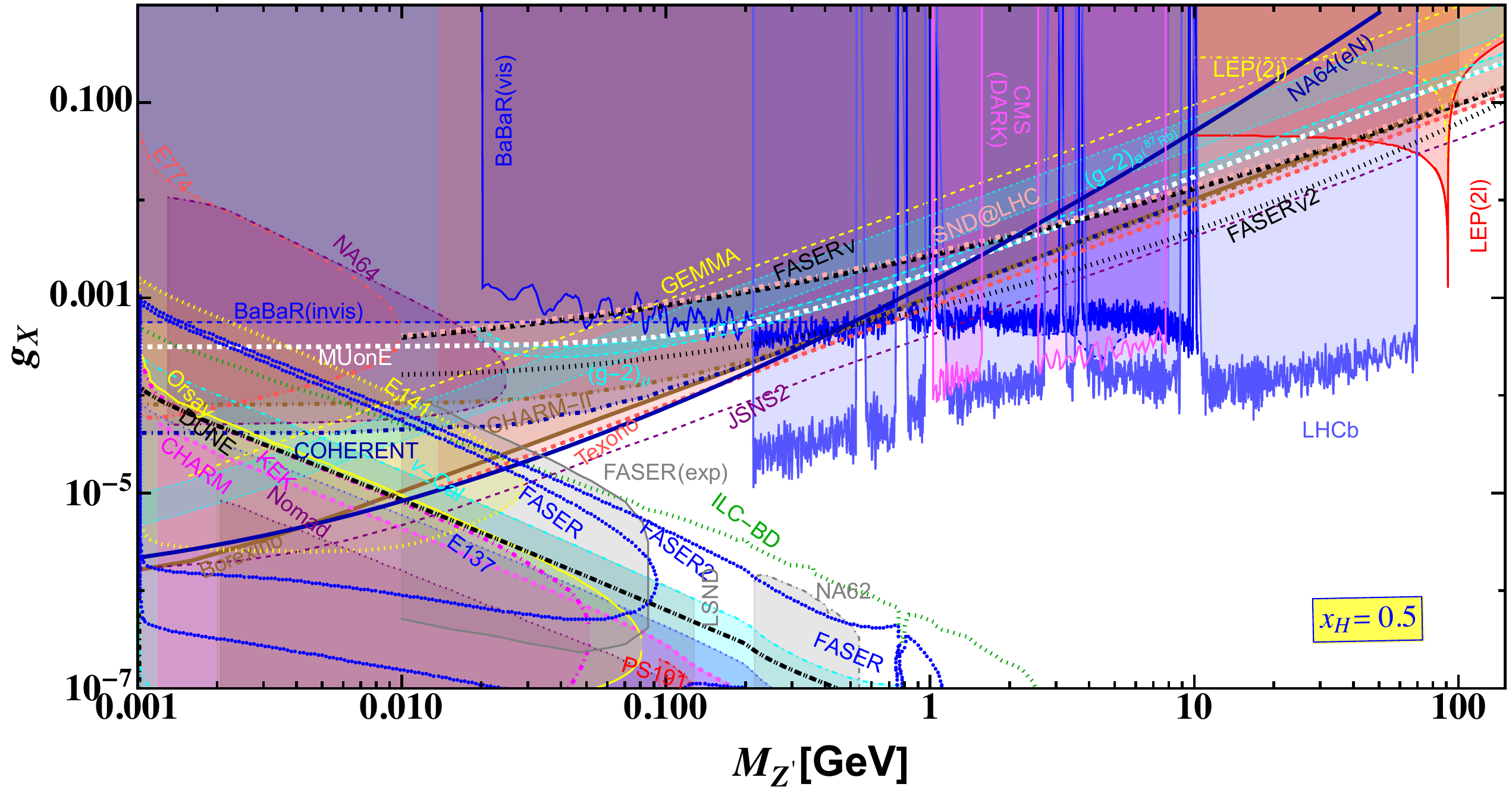}
\includegraphics[width=1.0\textwidth]{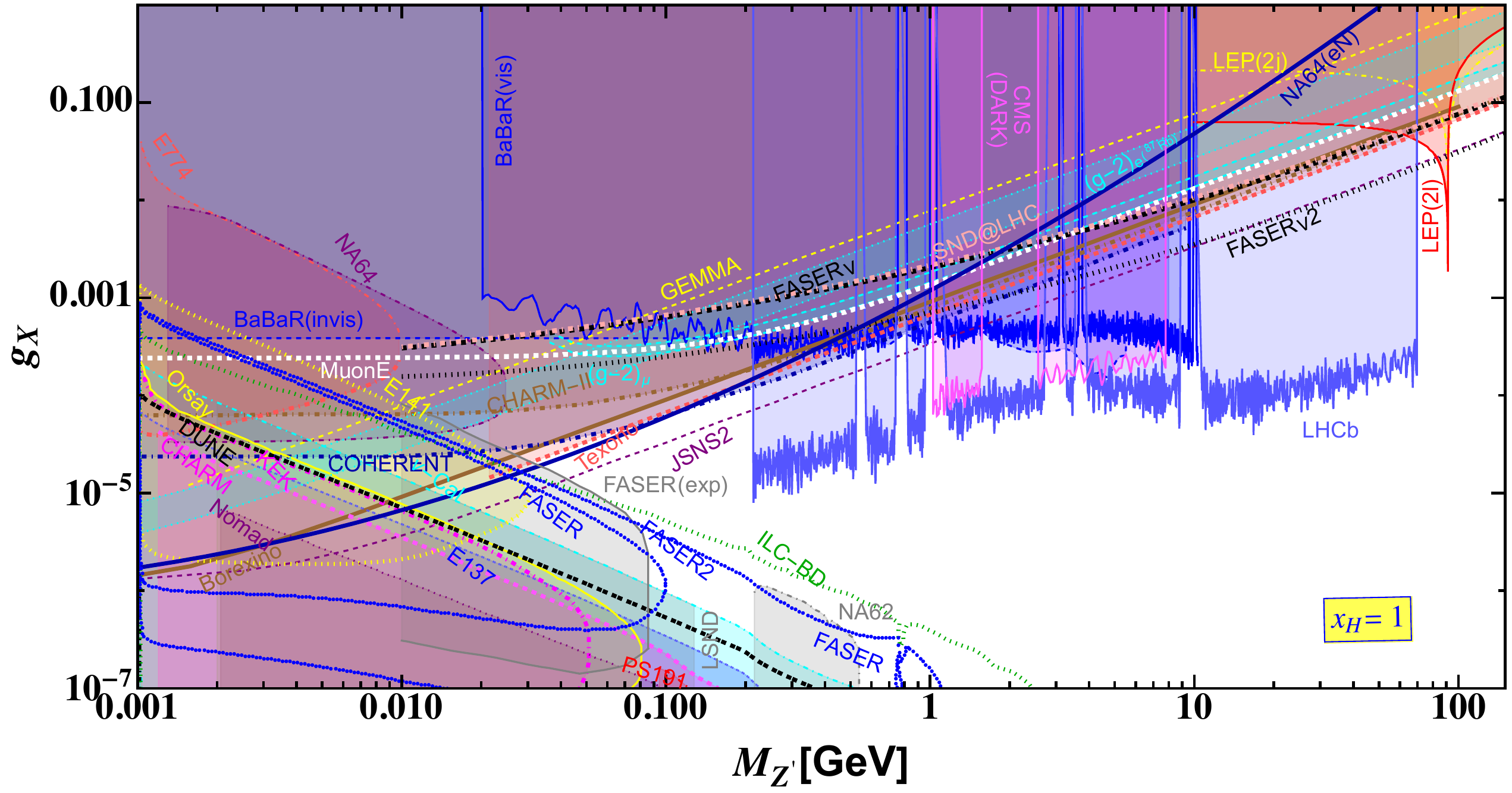}
\caption{Limits on $g_X^{}-M_{Z^\prime}^{}$ plane for $x_H^{} = 0.5$ (upper panel) and $x_H^{}=1$(lower panel) taking $x_\Phi^{}=1$ considering $10^{-3}$ GeV $\leq M_{Z^\prime}^{} \leq 150$ GeV showing the regions sensitive to FASER$\nu$, FASER$\nu$2, SND$@$LHC, NA64$(eN)$ and JSNS2 experiments. Recasting the existing results in our case, we compare parameter regions obtained from scattering experiments at LEP, CHARM-II, GEMMA, BOREXINO, COHERENT, TEXONO, Dark photon searches at BaBaR(vis and invis), LHCb and CMS(CMS Dark) experiments and different beam dump experiments at Orsay, KEK, E137, CHARM, Nomad, $\nu-$cal, E141, E774, NA64, NA62, FASER involving prospective theoretical bounds from FASER(2), DUNE and ILC(ILC-BD), respectively.}
\label{fig:xH05}
\end{center}
\end{figure}

\begin{figure}[t]
\begin{center}
\includegraphics[width=1.0\textwidth]{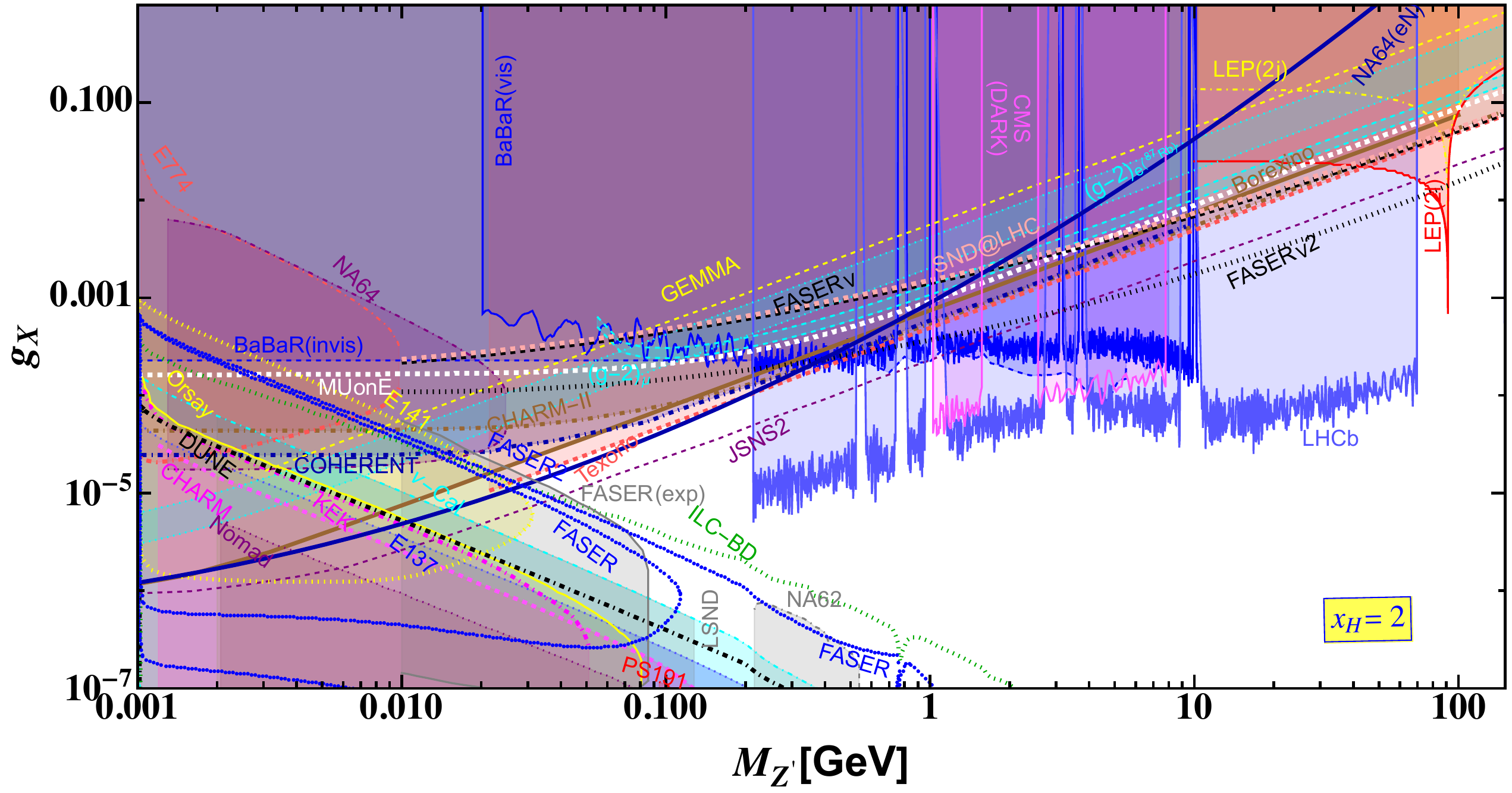}
\caption{Limits on $g_X^{}-M_{Z^\prime}^{}$ plane for $x_H^{} = 2$ taking $x_\Phi^{}=1$ considering $10^{-3}$ GeV $\leq M_{Z^\prime}^{} \leq 150$ GeV showing the regions sensitive to FASER$\nu$, FASER$\nu$2, SND$@$LHC, NA64$(eN)$ and JSNS2 experiments. Recasting the existing results in our case, we compare parameter regions obtained from the scattering experiments at LEP, CHARM-II, GEMMA, BOREXINO, COHERENT, TEXONO, dark photon searches at BaBaR(vis and invis), LHCb and CMS(CMS Dark) and different beam dump experiments at Orsay, KEK, E137, CHARM, Nomad, $\nu-$cal, E141, E774, NA64, NA62 and FASER and involving prospective bounds from FASER(2), DUNE and ILC(ILC-BD) respectively.}
\label{fig:xH2}
\end{center}
\end{figure}

We calculate the exclusion and future sensitivity regions for the chiral $Z^\prime$ gauge boson from neutrino-nucleon scattering measurements (FASER$\nu$(2), SND@LHC), missing energy search (NA64), muon-electron scattering measurements (MUonE), proton and electron beam dump experiments (Orsay, KEK, E137, CHARM, Nomad, $\nu-$cal, E141, E774, NA64), electron-neutrino scattering measurements (BOREXINO, TEXONO, JSNS2), coherent neutrino-nucleus scattering (COHERENT), and collider experiments (LEP-II, CMS, LHCb, BarBar). In Figs. \ref{fig:xHm2}-\ref{fig:xH2}, we show the exclusion and future sensitivity regions for $x_H^{} = -2, -1, -0.5, 0, 0.5, 1$, and $2$ considering $x_\Phi^{} = 1$ where the horizontal and vertical axes are $Z^\prime$ mass within the range $10^{-3}$ GeV $\leq M_{Z^\prime} \leq 150$ GeV and gauge coupling constant $g_X$, respectively. The shaded regions have already been excluded, and on the other hand, the curves without shaded regions are sensitivity curves by  future experiments. Along this line we compare the sensitivity lines from Supernova (only in case of $x_H=-2$, bounds for other $x_H$ can be found in \cite{Asai:2022zxw}) and beam dump experiments at FASER(2), DUNE and ILC (ILC-BD) from \cite{Asai:2022zxw}, respectively. The shaded regions in the figures are ruled out by respective experiments.   

We show the prospective sensitivity in the $g_X-M_{Z^\prime}$ plane obtained from the elastic $\mu^+-e^-$ scattering at the MUonE experiment in the upper panel of Fig.~\ref{fig:xHm2} for $x_H^{}=-2$ where left handed fermions do not interact with the $Z^\prime$. Bounds obtained from the MUonE experiment vary between $6\times 10^{-4} \leq g_X \leq 0.4$ for $10^{-3}$ GeV $\leq M_{Z^\prime}\leq 150$ GeV. It is found that MUonE bounds for $M_{Z^\prime} \leq 0.03$ GeV are ruled out by beam dump searches from NA64 and E774 respectively whereas these bounds for $M_{Z^\prime} \geq 0.175$ GeV are ruled out by estimated limits obtained from the dark photon searches at BaBaR, LHCb, CMS (CMS Dark), respectively except for a narrow window from the LHCb experiment around $M_{Z^\prime}\simeq 1$ GeV. Limits obtained from electron magnetic dipole moment for Cesium-133 is stronger than the limits obtained from $\mu-e$ scattering at MUonE for $0.0174$ GeV $\leq M_{Z^\prime} \leq 0.0894$ GeV with $1.4\times 10^{-4} \leq g_X \leq 7.16\times 10^{-4}$.  Following the analysis given in \cite{Asai:2022zxw} we find that prospective sensitivity from the beam dump experiments at FASER(2) can reach up to $5 \times 10^{-3}$ at $M_{Z^\prime}=10^{-3}$ GeV whereas ILC-BD reaches up to $2.5 \times 10^{-3}$ at that $M_{Z^\prime}$. However, these bounds are ruled out by the results from different beam dump searches from NA64, E774 and E141 whereas prospective reach from FASER(2) and ILC-BD can be improved for $M_{Z^\prime} \simeq 0.3$ GeV. Following the analysis in \cite{Asai:2022zxw} we find that Supernova bounds can be slightly stronger than the $\nu-$cal bounds around $M_{Z^\prime}\simeq 0.1$ GeV. Within $10^{-3}$ GeV $\leq M_{Z^\prime} \leq 0.18$ GeV the prospective bounds from DUNE (prospective DUNE sensitivity plots can be found in \cite{Asai:2022zxw} for $0.1$ GeV $\leq M_{Z^\prime} \leq 3$ GeV) is weaker than $\nu-$cal. We find that recent experimental observations from FASER (FASER-exp)\cite{FASER:2023zcr} and NA62 \cite{NA62:2023qyn} are represented by gray solid and dot-dashed lines and the corresponding excluded regions are shaded in gray. Most of these limits are well within the $\nu-$cal bounds, however, rest of them are just above the $\nu-$cal contour offering stronger constrains around $0.04$ GeV $\leq M_{Z^\prime} \leq 0.07$ GeV from FASER-exp and $0.25$ GeV $\leq M_{Z^\prime} \leq 0.55$ GeV from NA62 experiment respectively. The experimental limits from FASER experiment (FASER-exp) are weaker than the theoretically estimated lines for the FASER experiment represented by blue dotted contour for $M_{Z^\prime}> 0.04$ GeV. Recasting BaBaR(vis) data for dark photon searches we obtain a stringent bound for $M_{Z^\prime} \leq 0.175$ GeV. Recasting the dark photon searches at the LHCb and CMS experiments, we find that stringent bounds can be calculated for $0.175$ GeV $\leq M_{Z^\prime} \leq 70$ GeV and $1$ GeV $\leq M_{Z^\prime} \leq 8$ GeV respectively. Except for small windows in these ranges where $M_{Z^\prime}\simeq 10$ GeV, stringent bounds can be obtained from the visible final state in BaBaR(vis) experiment. For $M_{Z^\prime} \leq 0.19$ GeV we compare our results recasting the limits from different beam dump experiments including Nomad, CHARM, KEK, E774, E141, E137 and $\nu-$cal experiments with the prospective limits obtained from FASER(2), DUNE and ILC-BD. Furthermore dilepton and dijet searches also rule out the bounds for $M_{Z^\prime} \geq 70$ GeV. We also show that dilepton and dijet limits when $M_{Z^\prime}^{}$ is at the $Z-$pole giving a bound about $3\times10^{-3}$ when $x_H=-2$. Due to the structure of the $U(1)_R^{}$ scenario where left handed fermions do not interact with the $Z^\prime$, we find no other constrains in this scenario from experiments like $\nu-$electron, $\nu-$nucleon, etc where we can not explore the coupling between $\nu_L^{}$ and $Z^\prime$. 

Limits for $x_H=-1$ are shown in the lower panel of Fig.~\ref{fig:xHm2}. For this charge there is no coupling between $e_R$ and $Z^\prime$ resulting no direct bound from $\mu^+-e^-$ scattering at the MUonE experiment. However, for this value of $x_H$, we obtain prospective limits from FASER$\nu (2)$, SND$@$LHC, $\nu-$nucleus scattering at NA64 (NA64$(eN)$) where we compare recasting the bounds from the existing scattering experiments like GEMMA, BOREXINO, CHARM-II, TEXONO and COHERENT. Bounds obtained from the TEXNONO and COHERENT experiments are stronger than the other bounds obtained from the existing scattering experiments for the $Z^\prime$ roughly within $0.03$ GeV $\leq M_{Z^\prime} \leq 0.1$ GeV. The prospective bounds on the $U(1)_X$ gauge coupling varies between $7\times 10^{-6} \leq g_X \leq 1$ from the electron-nucleon scattering process in the NA64$(eN)$ experiment for $10^{-3}$ GeV $\leq M_{Z^\prime} \leq 10$ GeV. However, the bound is comparable with the bounds obtained from the TEXONO and COHERENT experiments for the $Z^\prime$ mass within $0.03$ GeV $\leq M_{Z^\prime} \leq 0.1$ GeV. Whereas bounds obtained from NA64$(eN)$ experiment is weak compared to the beam dump experiments for $M_{Z^\prime}^{} \leq 0.028$ GeV. Here we recast existing results from different beam dump experiments like Orsay, KEK, E137, CHARM, Nomad, $\nu-$cal, E141, E774, NA64, respectively for $x_H=-1$ to show complementarity with the other experiments. Prospective bounds on the $U(1)_X$ gauge coupling obtained from the beam dump experiments like FASER(2) and ILC-BD could be stringent compared to the estimated bounds from the existing beam dump experiments for $M_{Z^\prime} \geq 0.2$ GeV. Interestingly the NA64$(eN)$ line crosses ILC-BD and FASER2 lines at this mass point. Prospective limits obtained by performing the beam dump study at DUNE we obtain that those limits are weaker than the limits obtained by recasting the existing results from the $\nu-$cal experiment. We find that recent experimental observations from FASER (FASER-exp)\cite{FASER:2023zcr} and NA62 \cite{NA62:2023qyn} are represented by gray solid and dot-dashed lines and the corresponding excluded regions are shaded in gray. Most of these limits are well within the $\nu-$cal bounds, however, rest of them are just above the $\nu-$cal contour offering stronger constrains around $0.0225$ GeV $\leq M_{Z^\prime} \leq 0.08$ GeV from FASER-exp and $0.25$ GeV $\leq M_{Z^\prime} \leq 0.55$ GeV from NA62 experiment respectively. The FASER-exp contour almost covers the theoretical region shown by the blue dotted line for FASER experiment for $M_{Z^\prime}> 0.07$ GeV. Estimated prospective bounds obtained from $\nu-$nucleon scattering at FASER$\nu$, SND$@$LHC vary between $6.69\times 10^{-4} \leq g_X \leq 0.29$ and those in case of FASER$\nu2$ vary between $2.8 \times 10^{-4} \leq g_X \leq 0.14$ for $0.01$ GeV $\leq M_{Z^\prime} \leq 150$ GeV, respectively showing stronger prospective bounds for $91.2$ GeV $\leq M_{Z^\prime} \leq 150$ GeV. We recast dilepton and dijet results from LEP which provide most stringent limit as $g_X \simeq 5\times 10^{-3}$ at the $Z-$pole. We recast recent results of dark photon searches at the LHCb and CMS providing constrains for $0.2$ GeV $\leq M_{Z^\prime} \leq 70$ GeV and $1$ GeV $\leq M_{Z^\prime} \leq 8$ GeV respectively. Constrains on $g_X$ from LHCb vary between $3.12 \times 10^{-5} \leq g_X \leq 1.09\times 10^{-3}$ and those from CMS Dark vary between $4.0 \times 10^{-4} \leq g_X \leq 1.1 \times 10^{-3}$ leaving some narrow windows where BaBaR(vis) provides the strongest bound around $M_{Z^\prime} \simeq 10$ GeV as $g_X \simeq 7 \times 10^{-4}$. Studying $\nu-$electron scattering at JSNS2 we find that a prospective sensitivity for $g_X$ could reach as low as $2.78\times 10^{-5}$ and as high as $3 \times 10^{-3}$ for $M_{Z^\prime}=0.022$ GeV and $0.22$ GeV respectively which could be probed in future. JSNS2 provides stronger bound for $M_{Z^\prime}$ above the $Z-$pole. Recasting the data from the different beam dump experiments we find that prospective bounds obtained from JSNS2 could be weaker than some of these bounds for $M_{Z^\prime} \leq 0.025$ GeV.In addition to that we estimate bounds on the $U(1)_X$ coupling analyzing the electron $g-2$ data which belongs to the shaded region being ruled out by the scattering, dark photon search and beam dump experiments, respectively.

Limits for $x_H=-0.5$ are shown in the upper panel of Fig.~\ref{fig:xHm05}. In this case $u_R$ does not interact with the $Z^\prime$. We obtain the prospective sensitivity on $g_X$ by estimating $\mu^--e^+$ scattering process involving $Z^\prime$ contribution at MUonE experiment. Estimated limits are weaker than the projected sensitivities obtained from the $\nu-$nucleon scattering at the experiments like FASER$\nu$ and SND$@$LHC except for $M_{Z^\prime} \leq 0.73$ GeV. Studying $\nu-$nucleon scattering for the FASER$\nu2$ experiment we find that the prospective sensitivities could be stronger than the projections  of FASER$\nu$, SND$@$LHC and MUonE experiments respectively. The prospective limit obtained from the FASER$\nu2$ experiment could reach at $g_X\simeq 0.15$ for $M_{Z^\prime}$ going beyond the $Z-$pole. At the $Z-$pole we recast dilepton and dijet bounds from LEP experiment and find that the exclusion limit on the coupling could reach at $g_X\simeq 3\times 10^{-3}$. Limits estimated by recasting the COHERENT $\nu-$nucleus scattering are obtained to be strong for $0.066$ GeV $\leq M_{Z^\prime} \leq 0.213$ GeV where coupling varies between $9.0\times10^{-5} \leq g_X \leq 2.36\times10^{-4}$. We find that COHERENT limits are weaker than the limits obtained by the $e-$nucleus scattering at the NA64 experiment (NA64$(eN)$) for $M_{Z^\prime} \leq 0.07$ GeV while NA64$(eN)$ limits are weaker than the those from COHERENT, TEXONO and BOREXINO experiments beyond this $M_{Z^\prime}$. Recasting the Dark photon searches at the experiments like BaBRaR, LHCb and CMS we find that stringent limits come from LHCb within $0.21$ GeV $\leq M_{Z^\prime} \leq 70$ GeV whereas those from BaBaR(vis) could have severe bounds around $M_{Z^\prime}\simeq 10$ GeV and CMS Dark could have severe bound around $M_{Z^\prime} \simeq 1$ GeV and $3$ GeV, respectively. We obtain prospective sensitivity from the JSNS2 experiment studying $\nu-$electron scattering where the most stringent bound comes within the range $0.024$ GeV $\leq M_{Z^\prime} \leq 0.215$ GeV where the $U(1)_X$ gauge coupling varies between $2.46\times 10^{-5} \leq g_X \leq 1.97\times 10^{-4}$. JSNS2 could also produces a stringent prospective bound on $g_X$ for $M_{Z^\prime}$ beyond the $Z-$pole which is close to the prospective limit from FASER$\nu2$ for $M_{Z^\prime} \simeq 150$ GeV. The prospective sensitivity line from JSNS2 crosses over the bounds obtained from the prospective limits from the beam dump scenarios like FASER at $M_{Z^\prime}=0.0287$ GeV whereas FASER2 and ILC-BD lines at $M_{Z^\prime}=0.0326$ GeV and the corresponding couplings are $2.8\times 10^{-5}$ and $3.21\times 10^{-5}$, respectively. We find that recent experimental observations from FASER (FASER-exp)\cite{FASER:2023zcr} and NA62 \cite{NA62:2023qyn} are represented by gray solid and dot-dashed lines and the corresponding excluded regions are shaded in gray. Some parts of these limits are well within the $\nu-$cal bounds, however, rest of them are above the $\nu-$cal contour offering stronger constrains around $0.03$ GeV $\leq M_{Z^\prime} \leq 0.09$ GeV from FASER-exp and $0.225$ GeV $\leq M_{Z^\prime} \leq 0.5$ GeV from NA62 experiment respectively. The FASER-exp contour almost covers the theoretical region shown by the blue dotted line for $x_H=-0.5$. Recasting the data obtained from the existing results of E774, E137, E141, NA64, Orsay, KEK, Nomad and $\nu-$cal we find that prospective bounds obtained from DUNE for $M_{Z^\prime} \leq 0.15$ GeV are weaker than the bounds obtained from recasting the data of the $\nu-$cal experiment. We estimated bounds on $g_X$ from the electron and muon $g-2$ data. However, the limits are weaker compared to different existing and prospective scattering and dark photon experiments mentioned for this charge. Similar behavior is seen for the sensitivity line obtained after recasting the data from GEMMA experiment which is weak compared to existing beam dump, scattering and dark photon searches, respectively.

We estimate limits on $g_X$ for $x_H=0$ depending on $M_{Z^\prime}$ in the lower panel of Fig.~\ref{fig:xHm05}. We mention that $x_H=0$ is the well known B$-$L scenario. We estimate prospective limits at experiments like MUonE, FASER$\nu(2)$ and SND$@$LHC and find that these sensitivities belong to the shaded region below the $Z-$pole. However, above $Z-$pole the prospective limits from FASER$\nu2$ becomes stringent for $M_{Z^\prime} \leq 150$ GeV, the range of $Z^\prime$ mass under consideration. Recasting the dilepton and dijet search results from LEP we find the strongest bound at $Z-pole$ to be $0.0018$. Recasting available data, we find that limits obtained from GEMMA, CHARM-II, COHERENT also belong to the shaded region being weak compared to the bounds from different scattering experiments involving TEXONO and BOREXINO, beam dump experiments involving E774, E131, E141, NA64, Orsay, KEK, Nomad and $\nu-$cal, CHARM and dark photon search experiments involving BaBaR(visible and invisible modes), LHCb and CMS after we recast the available data from these experiments. We calculate prospective limits from $e-$nucleon scattering at the NA64 experiment which are slightly stronger than existing bounds obtained from the TEXONO experiment for $0.02$ GeV $\leq M_{Z^\prime}\leq 0.164$ GeV where limits on the $U(1)_X$ coupling can reach up to $2.0\times 10^{-5} \leq g_X \leq 7.2 \times 10^{-4}$. We show bounds obtained from the dark photon search experiments at BaBaR(visible and invisible), LHCb and CMS and found that LHCb provides a strong bound on the $U(1)_X$ gauge coupling for $0.21$ GeV $\leq M_{Z^\prime} \leq 70$ GeV whereas for narrow windows around $M_{Z^\prime} \simeq 10$ GeV where limits obtained from BaBaR are stringent and limits on $g_X$ are stringent around $M_{Z^\prime} \simeq 1$ GeV and 3 GeV from the CMS Dark, respectively. Prospective bounds from the JSNS2 experiment could be stronger than the other scattering experiments for $0.028$ GeV $\leq M_{Z^\prime} \leq 0.21$ GeV where constrains on the $U(1)_X$ coupling constant could be as strong as $1.7 \times 10^{-5} \leq g_X \leq 1.3 \times 10^{-4}$. Starting from beyond the $Z-$pole, prospective search reach from JSNS2 could be around $g_X=0.02$ for $M_{Z^\prime} \leq 150$ GeV. We compare our results with different prospective beam dump scenarios at DUNE, FASER(2) and ILC-BD. Doing that we find that prospective reach from JSNS2 could intersect the prospective beam dump lines from FASER(2) and ILC-BD at $\{M_{Z^\prime}, g_X\}=\{0.025~\rm GeV, 1.55 \times 10^{-5}\}, \{0.032~\rm GeV, 1.92 \times 10^{-5}\}, \{0.035~\rm GeV, 2.13 \times 10^{-5}\}$, respectively which could be probed in the near future. Prospective limits obtained from the DUNE are weaker than the $\nu-$cal bounds for $M_{Z^\prime} \leq 0.08$ GeV. We find that recent experimental observations from FASER (FASER-exp) \cite{FASER:2023zcr} and NA62 \cite{NA62:2023qyn} are represented by gray solid and dot-dashed lines and the corresponding excluded regions are shaded in gray. Some parts of these limits are well within the $\nu-$cal bounds, however, rest of them are above the $\nu-$cal contour offering stronger constrains around $0.03$ GeV $\leq M_{Z^\prime} \leq 0.08$ GeV from FASER-exp and $0.225$ GeV $\leq M_{Z^\prime} \leq 0.5$ GeV from NA62 experiment respectively. The FASER-exp contour covers the theoretical region shown by the blue dotted line for $0.01$ GeV $\leq M_{Z^\prime} \leq 0.085$ GeV, however, these are stronger than the theoretical limits.

The limits on $g_X-M_{Z^\prime}$ plane for $x_H=0.5$ has been shown in the upper panel of Fig.~\ref{fig:xH05} where all the fermions interact with the $Z^\prime$, however, left and right handed interactions with the $Z^\prime$ are different. We estimated prospective limits from the $\mu^+-e^-$ scattering at the MUonE experiment and compare it with the bounds recasting the data from other scattering experiment involving GEMMA, TEXONO, BOREXONO, CHARM-II and COHERENT. We find that bounds from TEXONO rules out the other results. We also estimate the bounds from the prospective search reaches at the experiments involving SND$@$LHC and FASER$\nu(2)$. We find that FASER$\nu2$ could provide a stringent limit beyond the $Z-$pole up to $M_{Z^\prime} \leq 150$ GeV where constrains on $g_X$ could reach around 0.1. Using the dilepton and dijet searches from LEP we recast those data in our scenario to obtain a strong limit on $M_{Z^\prime}$ at the $Z-$pole as $g_X=0.0012$. We find that electron-nucleon scattering in NA64$(eN)$ gives a strong bound on the $U(1)_X$ coupling as $2.22\times 10^{-5} \leq g_X \leq 1.17\times 10^{-4}$ for the $Z^\prime$ within $0.028$ GeV $\leq M_{Z^\prime} \leq 0.14$ GeV which is slightly stronger than the bounds obtained from TEXONO and COHERENT within that range of $M_{Z^\prime}$. We calculate bounds on $g_X$ recasting the data obtained from the dark photon searches at BaBaR, LHCb and CMS, respectively. Hence we find that LHCb provides a stringent bound for the $Z^\prime$ mass within $0.21$ GeV $\leq M_{Z^\prime} \leq 70$ GeV where limits on $g_X$ vary between $10^{-5} \leq g_X \leq 2 \times 10^{-4}$, while, around $M_{Z^\prime} \simeq 10$ GeV BaBaR provides stringent constraints whereas around $M_{Z^\prime}\simeq 1$ GeV and $3$ GeV CMS Dark provides stringent constraints on the $U(1)_X$ gauge coupling. We study $\nu-e$ scattering for JSNS2 experiment. JSNS2 shows prospective strongest bounds compared to other scattering experiments and dark photon searches within $0.031$ GeV $\leq M_{Z^\prime} \leq 0.21$ GeV where limits on the coupling vary between $1.32 \times 10^{-5} \leq g_X \leq 9.1 \times 10^{-5}$. Following the JSNS2 line we find that it may provide a strong bound on $g_X$ beyond the $Z-$pole which might reach up to $g_X\simeq 0.06$ for $M_{Z^\prime}\leq 150$ GeV. Limits on $g_X$ calculating electron and muon $g-2$ scenarios are found to be weak compared to the existing scattering, beam dump and dark photon search experiments. Recasting the data from the beam dump experiments involving NA64, E141, E137, CHARM, Nomad, $\nu-$cal, KEK, we find the limits on $g_X-M_{Z^\prime}$ plane shown by the shaded areas. We find that recent experimental observations from FASER (FASER-exp) \cite{FASER:2023zcr} and NA62 \cite{NA62:2023qyn} are represented by gray solid and dot-dashed lines and the corresponding excluded regions are shaded in gray. Some parts of these limits are well within the $\nu-$cal bounds, however, rest of them are above the $\nu-$cal contour offering stronger constrains around $0.035$ GeV $\leq M_{Z^\prime} \leq 0.09$ GeV from FASER-exp and $0.225$ GeV $\leq M_{Z^\prime} \leq 0.525$ GeV from NA62 experiment respectively. We also show the prospective sensitivities from FASER(2), DUNE and ILC-BD where sensitivity from DUNE is weaker compared to $\nu-$cal for $M_{Z^\prime} \leq 0.08$ GeV. The prospective sensitivity line from JSNS2 crosses the prospective FASER, FASER2 and ILC-BD lines at $\{M_{Z^\prime}, g_X\}= \{0.0305 \rm{GeV}, 1.29\times 10^{-5}\},~\{0.035 \rm{GeV}, 1.54\times 10^{-5}\},~\{0.0374 \rm{GeV}, 1.65\times 10^{-5}\}$, respectively. These limits could be probed by the scattering and beam dump experiments in future. 

In the lower panel of Fig.~\ref{fig:xH05} we show the constrains on the general $U(1)_X$ coupling for different $M_{Z^\prime}$ using $x_H=1$ where $d_R$ does not interact with the $Z^\prime$. We estimate prospective limits from $\mu^+-e^-$ scattering at the MUonE experimet and compare it with the bounds recasting the results from CHARM-II, TEXONO, BOREXINO, GEMMA, and COHERENT experiments, respectively. We find that comparing with all these limits MUonE is weak staying in the shaded region for $10^{-3}$ GeV $\leq M_{Z^\prime} \leq 150$ GeV. Prospective sensitivity obtained from the electron-nucleon scattering at the NA64 experiment are denoted by NA64$(eN)$ line. This provides a strongest prospective limit within $0.0257$ GeV $\leq M_{Z^\prime} \leq 0.0085$ GeV where $U(1)_X$ coupling could reach down to $1.423 \times 10^{-5} \leq g_X \leq 10^{-5}$. Recasting the TEXONO data we find that in this context strong constrains come for $0.0254$ GeV $\leq M_{Z^\prime} \leq 0.072$ GeV where limits on the gauge coupling vary within $1.72 \times 10^{-5} \leq g_X \leq 4.64 \times 10^{-5}$. We find that bounds obtained from TEXONO can be stronger than the limits obtained from recasting the dilepton and dijet searches at the LEP experiment beyond the $Z-$ pole but $ M_{Z^\prime} \leq 150$ GeV where limits on the gauge coupling could vary between $0.031 \leq g_X \leq 0.051$. In addition, we obtain that LEP bounds at $Z-$pole could reach at $g_X\simeq 0.0019$ from the dilepton and dijet searches. FASER$\nu2$ provides a prospective sensitivity beyond the $Z-$pole and below 150 GeV where limits could vary between $0.03 \leq g_X \leq 0.05$.  The fermions in the $U(1)_X$ scenario under consideration interact equally with the $Z^\prime$ making a generation independent nature of the model irrespective of $x_H$ which affects bounds from $g-2$ analysis. This is true for any value of $x_H$. We find that limits obtained from the dark photon searches at LHCb shows the strongest bounds on $U(1)_X$ gauge coupling within $8.41 \times 10^{-6} \leq g_X \leq 1.637 \times 10^{-4}$ for $0.21$ GeV $\leq M_{Z^\prime} \leq 70$ GeV. We find that dark photon search at CMS provides the strongest bounds on the gauge coupling for $M_{Z^\prime}$ around 1 GeVa nd 3 GeV respectively whereas same scenario appears from the BaBaR experiment for $M_{Z^\prime}$ around 10 GeV. Prospective limits estimated in the context of JSNS2 experiment are found to reach at $1.85 \times 10^{-4}\leq g_X \leq 1.95 \times 10^{-4}$ for $0.517$ GeV $\leq M_{Z^\prime} \leq 0.56$ GeV and beyond $Z-$pole, the bounds are comparable with the prospective ones from FASER$\nu2$. We find that recent experimental observations from FASER (FASER-exp) \cite{FASER:2023zcr} and NA62 \cite{NA62:2023qyn} are represented by gray solid and dot-dashed lines and the corresponding excluded regions are shaded in gray. Some parts of these limits are well within the $\nu-$cal bounds, however, rest of them are above the $\nu-$cal contour offering stronger constrains around $0.035$ GeV $\leq M_{Z^\prime} \leq 0.09$ GeV from FASER-exp and $0.225$ GeV $\leq M_{Z^\prime} \leq 0.525$ GeV from NA62 experiment respectively. Prospective limits from JSNS2 crosses the prospective sensitivity lines at the beam dump experiments involving FASER, FASER2 and ILC-BD at $\{M_{Z^\prime}, g_X \}= \{0.033 \rm{GeV}, 1.05 \times 10^{-5}\},~\{0.036 \rm{GeV}, 1.21 \times 10^{-5}\},~\{0.0375 \rm{GeV}, 1.30 \times 10^{-5}\}$, respectively which could be probed in future. We compare these bounds recasting the data from different beam dump experiments like E141, NA64, KEK, Orsay, CHARM, E137 and $\nu-$cal which are shown by different shaded regions. We find that prospective bounds obtained from DUNE are weaker than $\nu-$cal for $M_{Z^\prime} \leq 0.06$ GeV. 

We show the bounds from different experiments for $x_H=2$ in Fig.~\ref{fig:xH2}. This charge is another example where all fermions interact with the $Z^\prime$, however, their left and right handed counterparts interact differently with the $Z^\prime$. Being influenced by the charge assignment we find that prospective bounds on $g_X$ for different $M_{Z^\prime}$ obtained form the MUonE experiment are weaker compared to the bounds after recasting the data from TEXONO, BOREXINO, CHARM-II and COHERENT experiments. We estimate prospective bounds on $g_X$ for different prospective experiments like SND$@$LHC and FASER$\nu(2)$. We find that prospective searches from FASER$\nu 2$ are strong only beyond the $Z-$pole and up to $M_{Z^\prime} \leq 150$GeV. In this mass range the limits on $g_X$ vary between $0.014 \leq g_X \leq 0.025$. Recasting the dilepton and dijet searches from the LEP experiment we find that limit on $g_X$ at the $Z-$pole could be as stringent as $6.8 \times 10^{-4}$. We also find that electron-nucleon scattering in NA64 experiment shown by the line NA64$(eN)$ provides a strong bound on the gauge coupling around $1.1\times 10^{-5} \leq g_X \leq 1.1 \times 10^{-4}$ for $0.0261$ GeV $\leq M_{Z^\prime} \leq 0.21$ GeV. Studying $\nu-$electron scattering in context of JSNS2 experiment we estimate the prospective bounds on $g_X$ with respect to $M_{Z^\prime}$. The strongest future limits on $g_X$ can be estimated for $Z^\prime$ within $0.032$ GeV $\leq M_{Z^\prime} \leq 0.212$ GeV as $7.735 \times 10^{-6} \leq g_X \leq 5.13 \times 10^{-5}$ which crosses respective future sensitivity lines obtained from the beam dump experiments like FASER, FASER2 and ILC-BD  at $\{M_{Z^\prime}, g_X\}=\{0.033~\rm{GeV}, 7.855 \times 10^{-6}\},~\{0.0371~\rm{GeV}, 8.57\times 10^{-6}\},~\{0.0386~\rm{GeV}, 9.07 \times 10^{-6}\}$ which could be probed in future. We compare our results for $x_H=2$ recasting the bounds obtained from the existing results from the beam dump experiments like E141, NA64, KEK, Orsay, CHARM, E137 and $\nu-$cal respectively. The excluded regions are shown by different shades. We find that recent experimental observations from FASER (FASER-exp) \cite{FASER:2023zcr} and NA62 \cite{NA62:2023qyn} are represented by gray solid and dot-dashed lines and the corresponding excluded regions are shaded in gray. Some parts of these limits are well within the $\nu-$cal bounds, however, rest of them are above the $\nu-$cal contour offering stronger constrains around $0.035$ GeV $\leq M_{Z^\prime} \leq 0.095$ GeV from FASER-exp and $0.225$ GeV $\leq M_{Z^\prime} \leq 0.525$ GeV from NA62 experiment respectively.
We find that prospective bounds obtained from the beam dump scenario at DUNE for $M_{Z^\prime} \leq 0.04$ GeV are weaker than those obtained from $\nu-$cal experiment recasting the existing data. Dark photon searches from LHCb provides stronger limit for $0.21$ GeV $\leq M_{Z^\prime} \leq 70$ GeV and beyond $Z-$pole up to $M_{Z^\prime}=150$ GeV. We find that LHCb limits below $Z-$pole vary within $3.5\times 10^{-6} \leq g_X \leq 10^{-4}$. Stringent limits can be obtained from the dark photon search experiments at BaBaR and CMS Dark around $M_{Z^\prime}\simeq 10$ GeV and $M_{Z^\prime}\simeq 1$ GeV and $3$ GeV within narrow windows from LHCb experiment. Finally we comment that limits obtained from muon and electron $g-2$ studies are weaker than the scattering and beam dump experiments due to the generation independent nature of the fermionic couplings with $Z^{\prime}$.
\section{Conclusions}
\label{sec:summary}
In this paper we consider chiral scenarios where $Z^\prime$ interacts with the left and right handed fermions differently. We obtain that depending on $U(1)_X$ charges the interactions of the femions with the $Z^\prime$ get modified by manifesting chiral nature of the scenarios under consideration. Such interactions affect $Z^\prime$ mediated neutrino-electron, electron-nucleon, electron-muon scattering processes which could be probed at the experiments like FASER$\nu(2)$, SND$@$LHC, NA64$(eN)$, MUonE, JSNS2, dark photon searches at BaBaR, LHCb and CMS experiments respectively. Further we compare our results with dilepton, dijet searches from LEP, neutrino-nucleus coherent scattering at COHERENT experiment, electron-neutrino scattering experiments like BOREXINO, TEXONO, GEMMA and CHARM-II respectively. We compare our results studying visible and invisible final states at the BaBaR experiment. Finally we show complementarity with different beam dump experiments like $\nu-$cal, E137, E141, NA64, E774, Orsay, CHARM, KEK, Nomad and future experiments like FASER, FASER2 and ILC-BD. We find that the experimental results from FASER matches with the theoretical limits estimated for general $U(1)_X$ charges having some bounds stronger than our estimated ones for increasing $M_{Z^\prime}$. We have also shown the NA62 regions which covers some prospective regions which could be probed by FASER2 experiment in future. 

Analyzing different interactions we find that some prospective bounds at NA64$(eN)$, FASER$\nu2$, JSNS2 could be probed in future. Some of the existing experimental limits from LEP, TEXONO, BaBaR(visible), dark photon searches at LHCb and CMS show stringent upper limits on $g_X$ for the respective $Z^\prime$ mass. JSNS2 bounds crosses the future sensitivities estimated from FASER, FASER2 and ILC-BD which could also be verified in future, however, their cross-overs depend on $x_H$ which could be checked after the realistic experimental results will be available. Depending on $U(1)_X$ charge, we find that beam dump experiments like $\nu-$cal, E137, E141, NA64, E774, Orsay, CHARM, KEK, Nomad rules out the values of $U(1)_X$ coupling between $10^{-6} \leq g_X \leq 0.01$ depending on the $Z^\prime$ mass for $M_{Z^\prime} \leq 0.08$ GeV. Within the mass range of $Z^\prime$, we find that DUNE will provide a weaker bound from the beam dump scenario. We find that weaker limits obtained analyzing the $g-2$ data because in our model three generations of the fermions are equally coupled with the $Z^\prime$. We point out that limits from SND$@$LHC, FASER$\nu$ are weak compared to the other scattering experiments below the $Z-$pole. Finally, from our analysis it has been found that scattering experiment could probe lighter $Z^\prime$ between $0.02$ GeV $\leq M_{Z^\prime} \leq 0.2$ GeV which could be simultaneously probed by proposed beam dump experiments involving FASER, FASER2 and ILC-BD. We find that in case of $U(1)_R$ scenario MUonE could provide a stringent bound for $0.02$ GeV $\leq M_{Z^\prime} \leq 0.175$ GeV. On the other hand heavier $Z^\prime$ above the $Z-$pole but $M_{Z^\prime} \leq 150$ GeV which could be probed by high energy colliders experiments in future for the cases we considered except $x_H = -2$. Hence we conclude that studying $Z^\prime$ mediated interactions in addition to the SM processes, limits on general $U(1)_X$ couplings could be interesting to probe $\mathcal{G}_{\rm SM}\otimes U(1)_X$ scenario in future.  
\section*{Acknowledgments}
We thank Takashi Shimomura and Yuichi Uesaka for useful advice.
This work is supported by JSPS KAKENHI Grant Numbers JP21K20365 and JP23K13097 [KA], JP19K03860, JP19K03865 and JP23K03402 [OS], by the Fundamental Research Funds for the Central Universities [TN, JL], by the Natural Science Foundation of Sichuan Province under grant No. 2023NSFSC1329 and the National Natural Science Foundation of China under grant No. 11905149 [JL].
\bibliographystyle{utphys28mod} 
\bibliography{bibliography}
\end{document}